\documentclass[pdftex]{article}\usepackage[]{graphicx}\usepackage[]{color}
\makeatletter
\def\maxwidth{ %
  \ifdim\Gin@nat@width>\linewidth
    \linewidth
  \else
    \Gin@nat@width
  \fi
}
\makeatother

\definecolor{fgcolor}{rgb}{0.345, 0.345, 0.345}
\newcommand{\hlnum}[1]{\textcolor[rgb]{0.686,0.059,0.569}{#1}}%
\newcommand{\hlstr}[1]{\textcolor[rgb]{0.192,0.494,0.8}{#1}}%
\newcommand{\hlcom}[1]{\textcolor[rgb]{0.678,0.584,0.686}{\textit{#1}}}%
\newcommand{\hlopt}[1]{\textcolor[rgb]{0,0,0}{#1}}%
\newcommand{\hlstd}[1]{\textcolor[rgb]{0.345,0.345,0.345}{#1}}%
\newcommand{\hlkwa}[1]{\textcolor[rgb]{0.161,0.373,0.58}{\textbf{#1}}}%
\newcommand{\hlkwb}[1]{\textcolor[rgb]{0.69,0.353,0.396}{#1}}%
\newcommand{\hlkwc}[1]{\textcolor[rgb]{0.333,0.667,0.333}{#1}}%
\newcommand{\hlkwd}[1]{\textcolor[rgb]{0.737,0.353,0.396}{\textbf{#1}}}%

\usepackage{framed}
\makeatletter
\newenvironment{kframe}{%
 \def\at@end@of@kframe{}%
 \ifinner\ifhmode%
  \def\at@end@of@kframe{\end{minipage}}%
  \begin{minipage}{\columnwidth}%
 \fi\fi%
 \def\FrameCommand##1{\hskip\@totalleftmargin \hskip-\fboxsep
 \colorbox{shadecolor}{##1}\hskip-\fboxsep
     \hskip-\linewidth \hskip-\@totalleftmargin \hskip\columnwidth}%
 \MakeFramed {\advance\hsize-\width
   \@totalleftmargin\z@ \linewidth\hsize
   \@setminipage}}%
 {\par\unskip\endMakeFramed%
 \at@end@of@kframe}
\makeatother

\definecolor{shadecolor}{rgb}{.97, .97, .97}
\definecolor{messagecolor}{rgb}{0, 0, 0}
\definecolor{warningcolor}{rgb}{1, 0, 1}
\definecolor{errorcolor}{rgb}{1, 0, 0}
\newenvironment{knitrout}{}{} 

\usepackage{alltt}

\usepackage{amsmath}
\usepackage{amssymb}
\usepackage[round,sort]{natbib}
\usepackage{graphicx}

\newlength{\onecolfig}
\newlength{\twocolfig}
\date{\today}

\newcommand{\code}[1]{\texttt{#1}}

\usepackage{paralist}
\usepackage[colorlinks=true,urlcolor=blue,citecolor=blue,linkcolor=blue]{hyperref}
\hypersetup{pdftitle={Avoidable errors: Ebola models and forecasts}}
\usepackage{caption}

\IfFileExists{upquote.sty}{\usepackage{upquote}}{}
\begin{document}

\begin{flushleft}
\item {\Large
\textbf{Avoidable errors in the modeling of outbreaks of emerging pathogens, with special reference to Ebola}
}\\
\vspace{12pt}
Aaron A.~King$^{1,2,3,4,\ast}$\\
Matthieu Domenech de Cell\`es$^{1}$\\
Felicia M.~G.~Magpantay$^{1}$\\
Pejman Rohani$^{1,2,4}$\\
\vspace{12pt}
\textbf{1} Department of Ecology \& Evolutionary Biology, University of Michigan, Ann Arbor, Michigan, USA 48109\\
\textbf{2} Center for the Study of Complex Systems, University of Michigan, Ann Arbor, Michigan, USA 48109\\
\textbf{3} Department of Mathematics, University of Michigan, Ann Arbor, Michigan, USA 48109\\
\textbf{4} Fogarty International Center, National  Institutes of Health, Bethesda,  Maryland, USA 20892\\
$\ast$ E-mail: kingaa@umich.edu
\end{flushleft}

\begin{abstract}
As an emergent infectious disease outbreak unfolds, public health response is reliant on information on key epidemiological quantities, such as transmission potential and serial interval.
Increasingly, transmission models fit to incidence data are used to estimate these parameters and guide policy.
Some widely-used modeling practices lead to potentially large errors in parameter estimates and, consequently, errors in model-based forecasts.
Even more worryingly, in such situations, confidence in parameter estimates and forecasts can itself be far over-estimated, leading to the potential for large errors that mask their own presence.
Fortunately, straightforward and computationally inexpensive alternatives exist that avoid these problems.
Here, we first use a simulation study to demonstrate potential pitfalls of the standard practice of fitting deterministic models to cumulative incidence data.
Next, we demonstrate an alternative based on stochastic models fit to raw data from an early phase of 2014 West Africa Ebola Virus Disease outbreak.
We show not only that bias is thereby reduced, but that uncertainty in estimates and forecasts is better quantified and that, critically, lack of model fit is more readily diagnosed.
We conclude with a short list of principles to guide the modeling response to future infectious disease outbreaks.

\noindent
This is a preprint.
The definitive version is published as Proceedings of the Royal Society of London, Series B (published online 1 April 2015, \href{http://dx.doi.org/10.1098/rspb.2015.0347}{DOI:10.1098/rspb.2015.0347}).
\end{abstract}

\section{Introduction}

The success of model-based policy in response to outbreaks of bovine spongiform encephelopathy \citep{Anderson1996c} and foot-and-mouth disease \citep{Keeling2001a,Ferguson2001} established the utility of scientifically informed disease transmission models as tools in a comprehensive strategy for mitigating emerging epidemics.
Increasingly, the expectation is that reliable forecasts will be available in real time.
Recent examples in which model-based forecasts were produced within
weeks of the index case include severe acute respiratory syndrome
\citep[SARS;][]{Riley2003,Lipsitch2003a}, pandemic H1N1 influenza \citep{Fraser2009}, cholera in Haiti and Zimbabwe \citep{Tuite2011},  Middle East respiratory syndrome \citep[MERS;][]{Breban2013}, and lately, Ebola virus disease (EBVD) in West Africa \citep{Fisman2014,Chowell2014}.
In the early stages of an emerging pathogen outbreak, key unknowns include its transmission potential, the likely magnitude and timing of the epidemic peak, total outbreak size, and the durations of the incubation and infectious phases.
Many of these quantities can be estimated using clinical and household transmission data, which are, by definition, rare in the early stages of such an outbreak.
Much interest therefore centers on estimates of these quantities from incidence reports that accumulate as the outbreak gathers pace.
Such estimates are obtained by fitting mathematical models of disease transmission to incidence data.

As is always the case in the practice of confronting models with data, decisions must be made as to the structure of fitted models and the data to which they will be fit.
Concerning the first, in view of the urgency of policy demands and paucity of information, the simplest models are, quite reasonably, typically the first to be employed.
With even the simplest models, such as the classical susceptible-infected-recovered (SIR) model, the choice of data to which the model is fit can have significant implications for science and policy.
Here, we explored these issues using a combination of inference on simulated data and on actual data from an early phase of the 2013--2015 West Africa EBVD outbreak.
We find that some of the standard choices of model and data can lead to potentially serious errors.
Since, regardless of the model choice, all model-based conclusions hinge on the ability of the model to fit the data, we argue that it is important to seek out evidence of model misspecification.
We demonstrate an approach based on stochastic modeling that allows straightforward diagnosis of model misspecification and proper quantification of forecast uncertainty.

\section{Deterministic models fit to cumulative incidence curves: a recipe for error and overconfidence}

\begin{figure}
\begin{center}
\begin{knitrout}\scriptsize
\definecolor{shadecolor}{rgb}{0.969, 0.969, 0.969}\color{fgcolor}

{\centering \includegraphics[width=\linewidth]{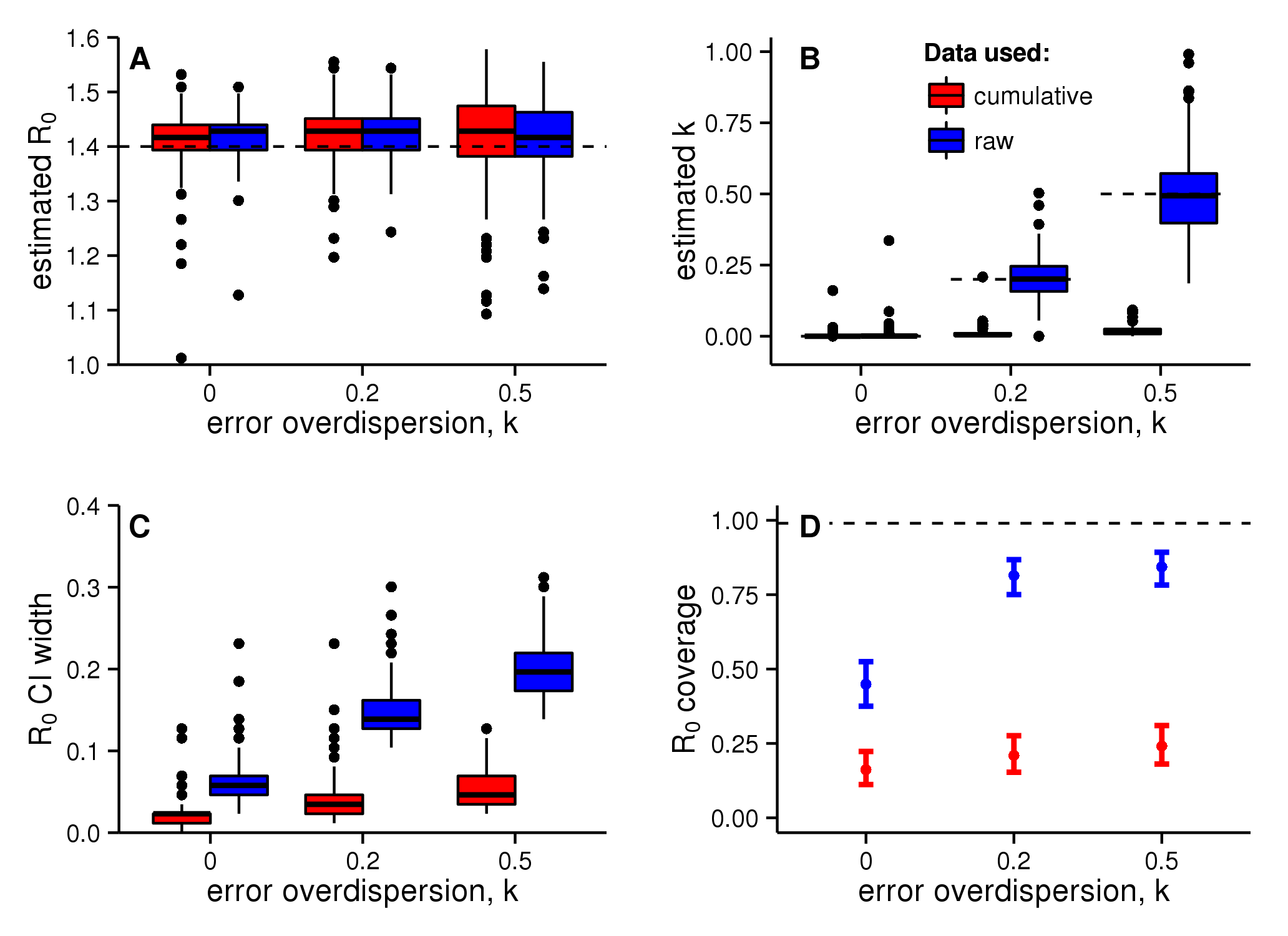} 

}

\end{knitrout}
\end{center}
\caption{\label{fig:tm}
  Results from simulation study fitting deterministic models to stochastically simulated data.
  500 simulated data sets of length 39~wk were generated by the stochastic model described in the Methods section at each of three levels of the measurement error overdispersion parameter, $k$.
  The deterministic model was fit to both raw (blue) and accumulated (red) incidence data.
  (A) Estimates of $R_0$.
  True value used in generating the data is shown by the dashed line.
  (B) Estimates of error overdispersion, $k$.
  (C) Widths of nominal $99\%$ profile likelihood confidence intervals (CI) for $R_0$.
  (D) Actual coverage of the CI, i.e., probability that the true value of $R_0$ lay within the CI.
  Ideally, actual coverage would agree with nominal coverage ($99\%$, dashed line).
}
\end{figure}

An inexpensive and therefore common strategy is to formulate deterministic transmission models and fit these to data using least squares or related methods.
These approaches seek parameters for which model trajectories pass as close to the data as possible.
Because, in such an exercise, the model itself is deterministic, all discrepancies between model prediction and data are in effect ascribed to measurement error.
Implicitly, the method of least squares assumes that these errors are independent and normally distributed, with a constant variance.
This assumption can be replaced without difficulty by more realistic assumptions of non-normal errors and, in particular, an error variance that depends on the mean.
As for the data to be fit, many have opted to fit model trajectories to cumulative case counts.
The incompatibility of this choice with the assumptions of the statistical error model has been pointed out previously \citep{Grad2012,Towers2014,Ma2014}.
In particular, the validity of the statistical estimation procedure hinges on the independence of sequential measurement errors, which is clearly violated when observations are accumulated through time (see \ref{sec:supp-results}).
To explore the impact of this violation on inferences and projections, we performed a simulation study in which we generated data using a stochastic model, then fit the corresponding deterministic model to both raw and cumulative incidence curves.
We generated 500 sets of simulated data at each of three different levels of measurement noise.
For each data set, we estimated model parameters, including transmission potential (as quantified by the basic reproduction number, $R_0$) and observation error overdispersion (as quantified by the negative binomial overdispersion parameter, $k$).
Full details of the data generation and fitting procedures are given in \ref{sec:simstudy}.
The resulting parameter estimates are shown in Fig.~\ref{fig:tm}.

Recognizing that quantification of uncertainty is prerequisite to reliable forecasting, we computed parameter estimate confidence intervals, and investigated their accuracy.
Fig.~\ref{fig:tm}A shows that, in estimating $R_0$, one finds considerable error but little evidence for bias, whether raw or cumulative incidence data are used.
Although in general one expects that violation of model assumptions will introduce some degree of bias, in this case since both the raw and cumulative incidence curves generically grow exponentially at a rate determined by $R_0$, estimates of this parameter are fairly accurate, \emph{on average}, when data are drawn, as here, from the early phase of an outbreak.
Fig.~\ref{fig:tm}B is the corresponding plot of estimated overdispersion of measurement noise.
Using the raw incidence data, one recovers the true observation variability.
When fitted to cumulative data, however, the estimates display extreme bias:
far less measurement noise is needed to explain the relatively smooth cumulative incidence.
The data superficially appear to be in very good agreement with the model.

To quantify the uncertainty in the parameter estimates, we examined the confidence intervals.
The nominal 99\% profile-likelihood confidence interval widths for $R_0$ are shown in Fig.~\ref{fig:tm}C.
When the model is fit to the simulated data, increasing levels of measurement error lead to increased variance in the estimates of $R_0$.
However, the confidence interval widths are far smaller when the cumulative data are used, superficially suggesting a higher degree of precision.
This apparent precision is an illusion however, as Fig.~\ref{fig:tm}D shows.
This figure plots the achieved coverage (probability that the true parameter value lies within the estimated confidence interval) as a function of the magnitude of measurement error and the choice of data fitted.
Given that the nominal confidence level here is 99\%, it is disturbing that the true coverage achieved is closer to 25\% when cumulative data are used.

When a deterministic model is fit to cumulative incidence data, the net result is a potentially quite over-optimistic estimate of precision, for three reasons.
First, failure to account for the non-independence of successive measurement errors leads to an under-estimate of parameter uncertainty (Fig.~\ref{fig:tm}C).
Second, as seen in Fig.~\ref{fig:tm}B, the variance of measurement noise will be substantially under-estimated.
Finally, because the model ignores environmental and demographic stochasticity, treating the unfolding outbreak as a deterministic process, forecast uncertainty will grow unrealistically slowly with the forecast horizon.
We elaborate on the last point in the Discussion.

\section{Stochastic models fit to raw incidence data: feasible and transparent}

The incorporation of demographic and/or environmental stochastic processes into models allows, on the one hand, better fits to the trends and variability in data and, on the other, improved ability to diagnose lack of model fit \citep{He2010}.
We formulated a stochastic version of the SEIR (susceptible-exposed-infectious-recovered) model as a partially observed Markov process and fit it to actual data from an early phase of the 2013--2015 West Africa EBVD outbreak.
We estimated parameters by maximum likelihood, using sequential Monte Carlo to compute the likelihood and iterated filtering to maximize it over unknown parameters \citep{Ionides2015}.
See \ref{sec:if} for details.

\begin{figure}
\begin{center}
\begin{knitrout}\scriptsize
\definecolor{shadecolor}{rgb}{0.969, 0.969, 0.969}\color{fgcolor}

{\centering \includegraphics[width=\onecolfig]{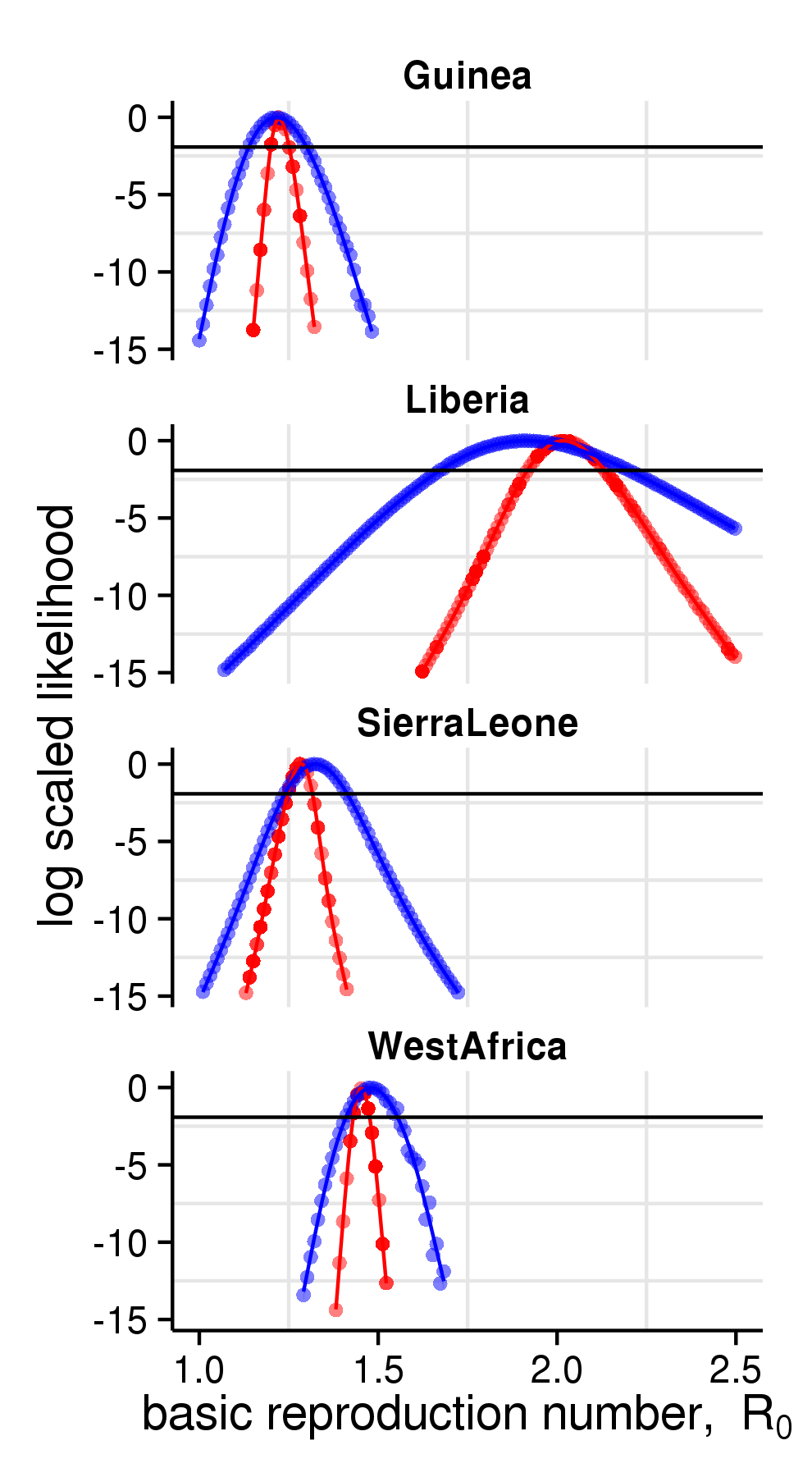} 

}

\end{knitrout}
\end{center}
\caption{
  Likelihood profiles for $R_0$ based on the stochastic model fit to raw data (blue) vs.~the deterministic model fit to cumulative incidence data (red).
  Each point represents the maximized log likelihood at each fixed value of $R_0$ relative to overall maximum.
  The maximum of each curve is achieved at the maximum-likelihood estimate (MLE) of $R_0$;
  the curvature is proportional to estimated precision.
  The horizontal line indicates the critical value of the likelihood ratio at the 95\% confidence level.
  While the (improper) use of cumulative data produces relatively small differences in the MLE for $R_0$, it does produce the illusion of high precision.
\label{fig:profiles}}
\end{figure}

Fig.~\ref{fig:profiles} shows likelihood profiles over $R_0$ for country-level data from Guinea, Liberia, and Sierra Leone.
We also wanted to explore the potential for biases associated with spatial aggregation of the data.
Hence, we fit our models to regional data, encompassing all reported cases from the three West African countries just mentioned.
In line with the lessons of Fig.~\ref{fig:tm}C, estimated confidence intervals are narrower when the cumulative reports are used.
The ``true'' parameters are, of course, unknown, but, as in the earlier example, this higher precision is probably illusory.
The somewhat, but not dramatically, larger confidence intervals that come with adherence to the independent-errors assumption (i.e., with the use of raw incidence data) lead to a quite substantial increase in forecast uncertainty, as we shall see.
Finally, the ease with which the stochastic model was fit and likelihood profiles computed testifies to the fact that, in the case of outbreaks of emerging infectious diseases, it is not particularly difficult or time-consuming to work with stochastic models.

\begin{figure}
\begin{center}
\begin{knitrout}\scriptsize
\definecolor{shadecolor}{rgb}{0.969, 0.969, 0.969}\color{fgcolor}

{\centering \includegraphics[width=\linewidth]{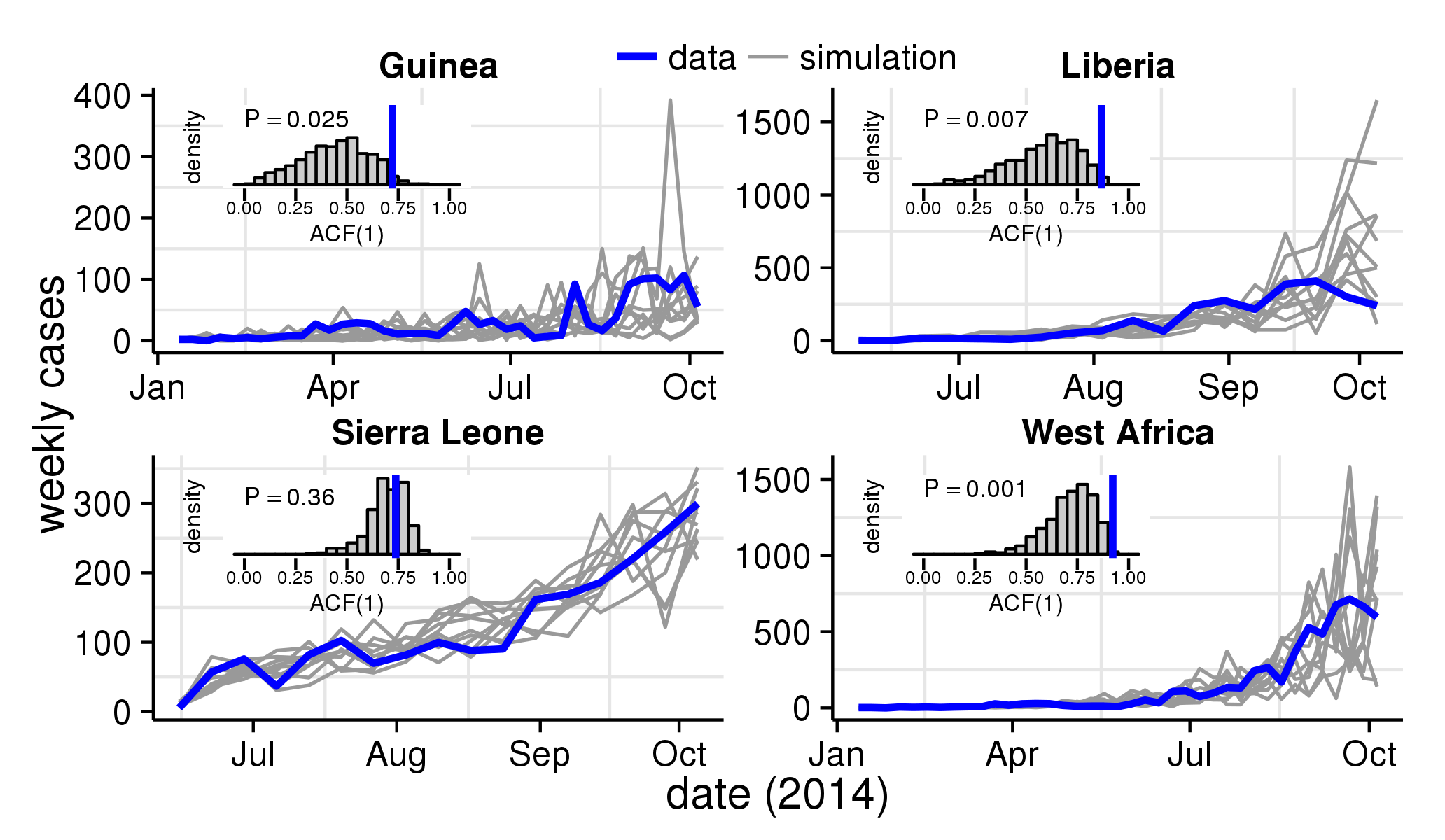} 

}

\end{knitrout}
\end{center}
\caption{Model diagnostics.
  The time series plots show the data (blue) superimposed on 10 typical simulations from the fitted model (grey).
  While the overall trend is captured by the model, the simulations display more high-frequency (week-to-week) variability than does the data.
  The insets confirm this, showing the autocorrelation function at lag 1 week (ACF(1)) in the data (blue) superimposed on the distribution of ACF(1) in 500 simulations (grey).
  For Guinea, Liberia, and the aggregated regional data (``West Africa''), the ACF(1) of the data lies in the extreme right tail of the distribution, as quantified by the one-sided $P$-values shown.
  \label{fig:diagnostics}}
\end{figure}

We took advantage of the stochastic model formulation to diagnose the fidelity of model to the data.
To do so, we simulated 10 realizations of the fitted model;
the results are plotted in Fig.~\ref{fig:diagnostics}.
While the overall trends appear similar, the model simulations display greater variability at high frequencies than do the data.
To quantify this impression, we computed the correlation between cases at weeks $t$ and $t-1$ (i.e., the autocorrelation function at lag 1~wk, ACF(1)) for both model simulations and data.
For Guinea, Liberia, and the region as a whole (``West Africa''), the observed ACF(1) lies in the extreme right tail of the model-simulated distribution, confirming our suspicion.
For Sierra Leone, the disagreement between fitted model and data is not as great, at least as measured by this criterion.
These diagnostics caution against the interpretation of the outbreaks in Guinea and Liberia as simple instances of SEIR dynamics and call for a degree of skepticism in inferences and forecasts based on this model.
On the other hand, the Sierra Leone epidemic does appear, by this single metric, to better conform to the SEIR assumptions when the data are aggregated to the country level.

\begin{figure}
\begin{center}
\begin{knitrout}\scriptsize
\definecolor{shadecolor}{rgb}{0.969, 0.969, 0.969}\color{fgcolor}

{\centering \includegraphics[width=\linewidth]{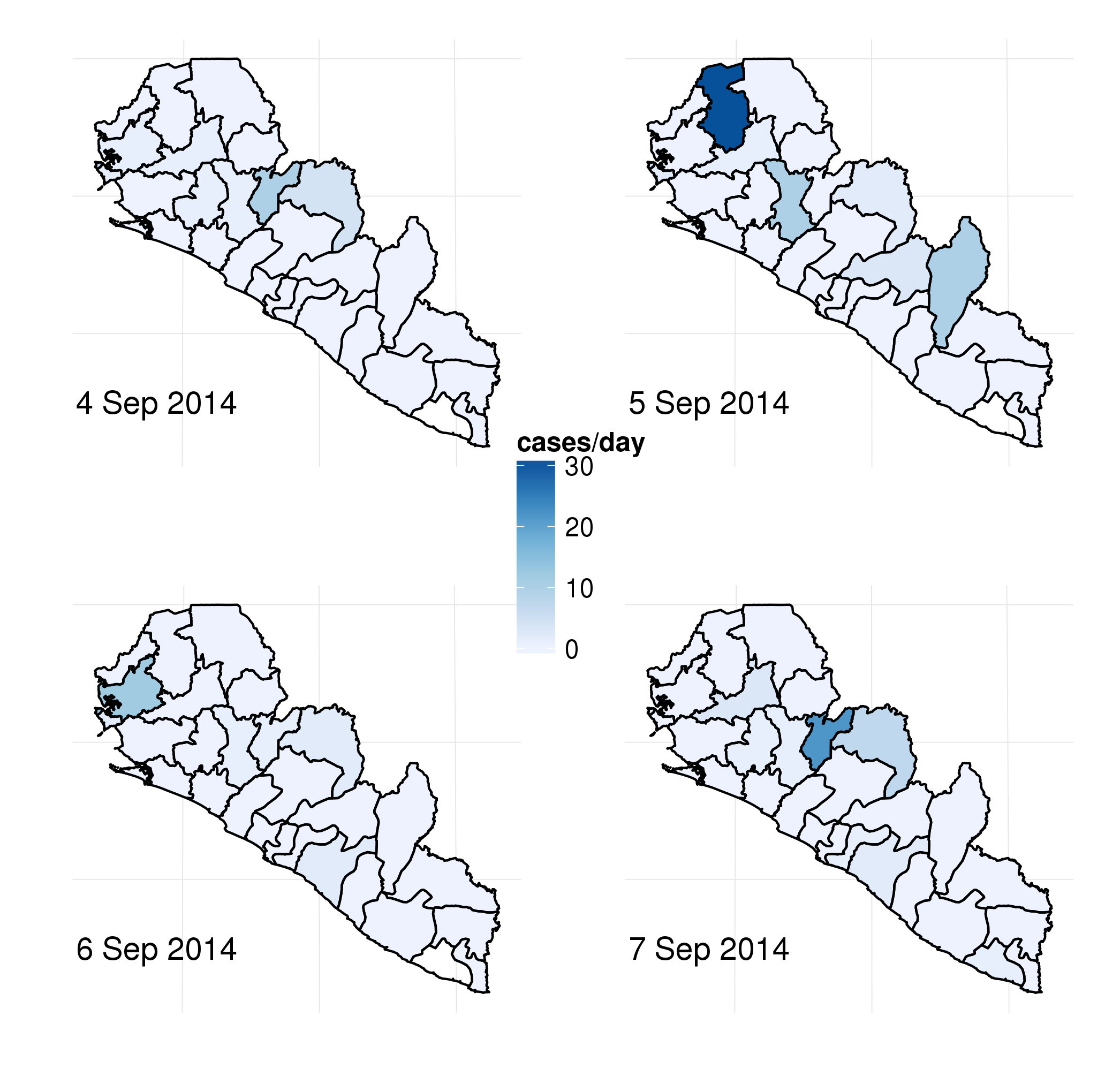} 

}

\end{knitrout}
\end{center}
\caption{
  \label{fig:maps}
  Four consecutive days of Ebola incidence in the republics of Liberia and Sierra Leone.
  In the outbreak's early stages, the spatio-temporal dynamics are highly erratic, contrary to the predictions of the well-mixed model.
}
\end{figure}

Fig.~\ref{fig:maps} suggests why the present Ebola outbreak might not be adequately described by the well-mixed dynamics of the SEIR model.
The erratically fluctuating mosaic of localized hotspots suggests spatial heterogeneity in transmission, at odds with the model's assumption of mass action.
As an aside, this heterogeneity hints at control measures beyond the purview of the SEIR model.
While the latter might provide more or less sound guidance with respect to eventual overall magnitude of the outbreak and associated demands for hospital beds, treatment centers, future vaccine coverage, etc., the former points to the potential efficacy of movement restrictions and spatial coordination of control measures.

\section{Discussion}

\begin{figure}
\begin{knitrout}\scriptsize
\definecolor{shadecolor}{rgb}{0.969, 0.969, 0.969}\color{fgcolor}

{\centering \includegraphics[width=\linewidth]{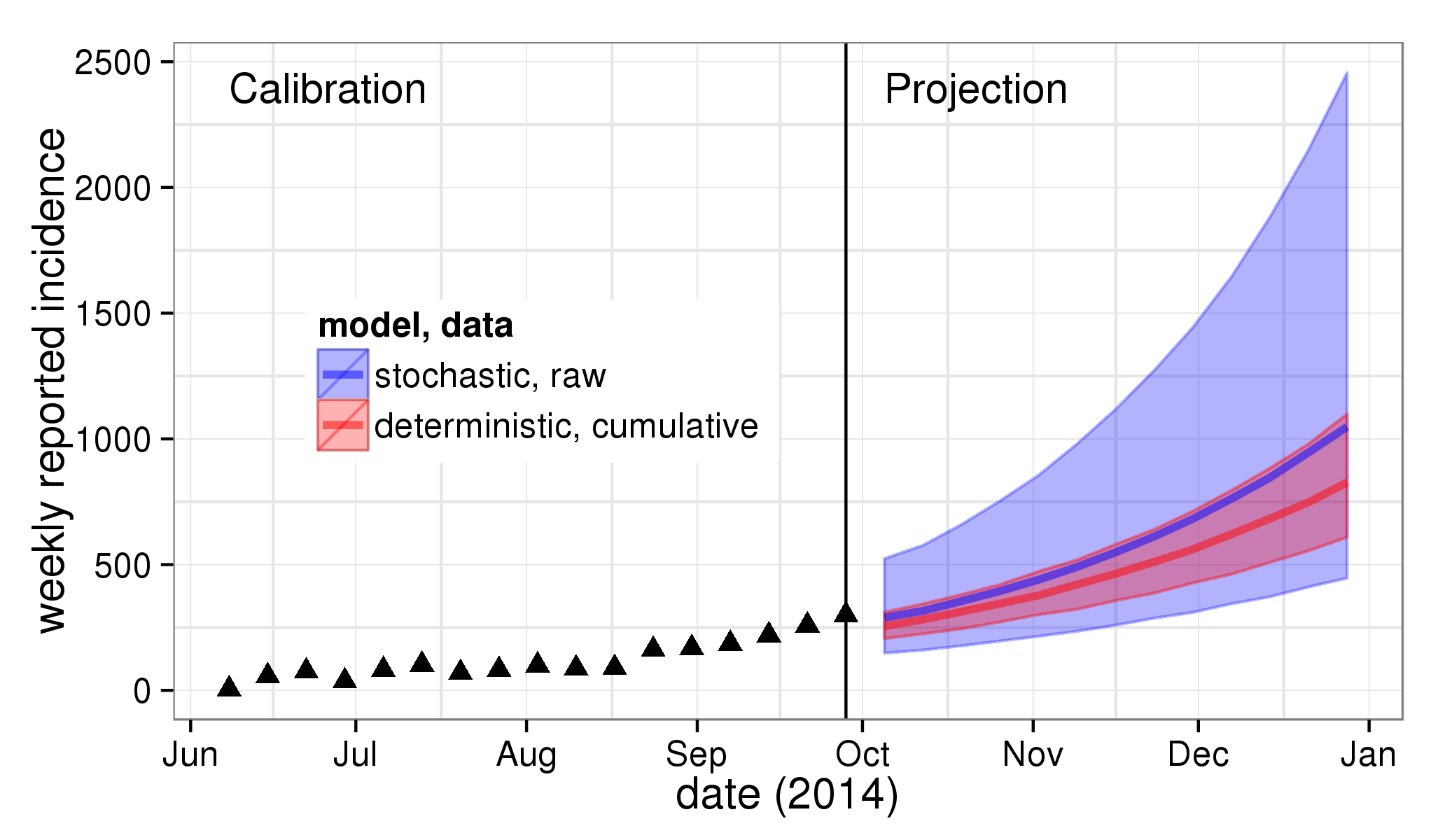} 

}

\end{knitrout}
\caption{
\label{fig:projection}
Forecast uncertainty for the Sierra Leone EBVD outbreak as a function of the model used and the data to which the model was fit.
The red ribbon shows the median and 95\% envelope of model simulations for the deterministic SEIR model fit to cumulative case reports;
the blue ribbon shows the corresponding forecast envelope for the stochastic model fit to raw incidence data.
The data used in model fitting are shown using black triangles.
}
\end{figure}

To summarize, we have here shown that the frequently adopted approach of fitting deterministic models to cumulative incidence data can lead to bias and pronounced under-estimation of the uncertainty associated with model parameters.
Not surprisingly, forecasts based on such approaches are similarly plagued by difficult-to-diagnose over-confidence as well as bias.
We illustrated this using the SEIR model---in its deterministic and stochastic incarnations---fit to data from the current West Africa EBVD outbreak.
Emphatically, we do not here assert that the SEIR model adequately captures those features of the epidemic needed to make accurate forecasts.
Indeed, when more severe diagnostic tests are applied (Fig.~\ref{fig:probes-plot}), it seems less plausible that the Sierra Leone data appear are a sample from the model distribution.
Moreover, we have side-stepped important issues of identifiability of key parameters such as route-specific transmissibility, asymptomatic ratio, and effective infectious period.
Rather, we have purposefully oversimplified, both to better reflect modeling choices often made in the early days of an outbreak and to better focus on issues of statistical practice in the context of quantities of immediate and obvious public health importance, particularly the basic reproduction number and predicted outbreak trajectory.
Fig.~\ref{fig:projection} shows projected incidence of EBVD in Sierra Leone under both the deterministic model fit to cumulative incidence data (in red) and the stochastic model fit to raw incidence data (in blue).
The shaded ribbons indicate forecast uncertainty.
In the deterministic case, the latter is due to the combined effects of estimation error and measurement noise.
As we showed above, the first contribution is unrealistically low because serial autocorrelation among measurement errors have not been properly accounted for.
The second contribution is also under-estimated because of the smoothing effect of data accumulation.
Finally, because the model ignores all process noise, it unrealistically lacks dynamic growth of forecast uncertainty.
By contrast, the stochastic model fitted to the raw incidence data show much greater levels of uncertainty.
Because measurement errors have been properly accounted for, confidence intervals more accurately reflect true uncertainty in model parameters.
Because the model accounts for process noise, uncertainty expands with the forecast horizon.
Finally, we recall once again that, because the process noise terms can to some degree compensate for model misspecification, it was possible to diagnose the latter, thus obtaining some additional qualitative appreciation of the uncertainty due to this factor.

The increasingly high expectations placed on models as tools for public policy put an ever higher premium on the reliability of model predictions and therefore on the need for accurate quantification of the associated uncertainty.
The relentless tradeoff between timeliness and reliability has with technological advance shifted steadily in favor of more complex and realistic models.
Because stochastic models with greater realism, flexibility, and transparency can be routinely and straightforwardly fit to outbreak data, there is less and less scope for older, less reliable, and more opaque methods.
In particular, the practices of fitting deterministic models and fitting models to cumulative case report data are prejudicial to accuracy and can no longer be justified on pragmatic grounds.
We propose the following principles to guide modeling responses to current and future infectious disease outbreaks:
\begin{enumerate}
\item Models should be fit to raw, disaggregated data whenever possible and never to temporally accumulated data.
\item When model assumptions, such as independence of errors, must be violated, careful checks for the effects of such violations should be performed.
\item Forecasts based on deterministic models, being by nature incapable of accurately communicating uncertainty, should be avoided.
\item Stochastic models should be preferred to deterministic models in most circumstances because they afford improved accounting for real variability and increased opportunity for quantifying uncertainty.
  \emph{Post hoc} comparison of simulated and actual data is a powerful and general procedure that can be used to distinguish model misspecification from real stochasticity.
\end{enumerate}

In closing, we are troubled that screening for lack of model fit is not a completely standard part of modeling protocol.
At best, this represents a missed opportunity, as discrepancies between the data and off-the-shelf models may suggest effective control measures.
At worst, this can lead to severely biased estimates and, worryingly, overly confident conclusions.
Fortunately, effective techniques exist by which such errors can be diagnosed and avoided, even in circumstances demanding great expedition.

\section{Methods}

\subsection{Data}

Weekly case reports in Guinea, Liberia, and Sierra Leone were digitized from the WHO situation report dated from 1~October~2014 \footnote{\url{http://www.who.int/csr/disease/ebola/situation-reports/en/}} (Fig.~\ref{fig:diagnostics}).
To compare our predictions to those of previous reports~\citep{Gomes2014}, we also aggregated those data to form a regional epidemic curve for ``West Africa''.
In Guinea, this outbreak was taken to have started in the week ending 5~January~2014 and in Sierra Leone in that ending 8~June~2014.
In Liberia, the outbreak was notified to WHO on 31~March~2014\footnote{\url{http://www.afro.who.int/en/clusters-a-programmes/dpc/epidemic-a-pandemic-alert-and-response/outbreak-news/4072-ebola-virus-disease-liberia.html}}, but few cases were reported until June;
therefore, the week ending 1~June was deemed the start of the Liberian outbreak for simulation purposes.
The data in Fig.~\ref{fig:maps} was downloaded from the repository maintained by C.~M.~Rivers\footnote{\url{https://github.com/cmrivers/ebola}} and ultimately derived from reports by the health ministries of the republics of Guinea, Sierra Leone, and Liberia.

\subsection{Model formulation}

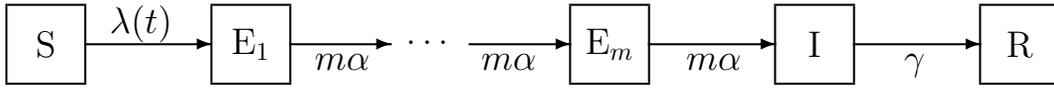
\begin{figure}[h]
  \begin{center}
    \resizebox{0.9\textwidth}{!}{
      \Large
      \setlength{\unitlength}{5pt}
	\begin{picture}(82,8)(0,0)
	  \thicklines
	  \put(0,0){\framebox(6,6){S}}
	  \put(6,3){\vector(1,0){10}}
	  \put(8,3.9){$\lambda(t)$}
	  \put(16,0){\framebox(6,6){E$_1$}}
	  \put(22,3){\vector(1,0){8}}
	  \put(24,1){$m\alpha$}
          \put(30,3){\makebox[30pt][c]{\dots}}
	  \put(36,3){\vector(1,0){8}}
	  \put(37,1){$m\alpha$}
	  \put(44,0){\framebox(6,6){E$_m$}}
	  \put(50,3){\vector(1,0){10}}
	  \put(53,1){$m\alpha$}
	  \put(60,0){\framebox(6,6){I}}
	  \put(66,3){\vector(1,0){10}}
	  \put(70,1){$\gamma$}
	  \put(76,0){\framebox(6,6){R}}
	\end{picture}
    }
  \end{center}
\caption{
  Schematic diagram of the transmission models used.
  $\lambda(t)={R_0\,\gamma\,I(t)}/{N}$ is the force of infection (i.e., the per-susceptible rate of infection).
  See the text for explanation.
  \label{fig:seir}
}
\end{figure}

The models used were variants on the basic SEIR model model (Fig.~\ref{fig:seir}), using the method of stages to allow for a more realistic (Erlang) distribution of the incubation period \citep{Lloyd2001,Wearing2005}.
The equations of the deterministic variant are:
\begin{equation*}
\begin{aligned}
\frac{dS}{dt} &= -\frac{R_{0}\gamma SI}{N}\\
\frac{dE_{1}}{dt} &= \frac{R_{0}\gamma SI}{N}-m\alpha E_{1}\\
\frac{dE_{i}}{dt} &= m\alpha(E_{i-1}-E_{i}), \qquad i=2,\dots,m\\
\frac{dI}{dt} &= m\alpha E_{m}-\gamma I\\
\frac{dR}{dt} &= \gamma I
\end{aligned}
\end{equation*}
Here, $R_{0}$ represents the basic reproduction number;
$1/\alpha$, the average incubation period;
$m$, the shape parameter for the incubation period distribution;
$1/\gamma$, the average infectious period; and
$N$, the population size, assumed constant (Table~\ref{tab:model-parameters}).

The stochastic variant was implemented as a continuous-time Markov process approximated via a multinomial modification of the $\tau$-leap algorithm \citep{He2010} with a fixed time step $\Delta t=10^{-2}$~wk.

To complete the model specification, we model the observation process.
Let ${\Delta}N_{E\rightarrow I}(t_1,t_2)$ denote the total number of transitions from latent to infectious class (E$_m$ to I) occurring between times $t_1$ and $t_2$.
Between times $t-\Delta t$ and $t$, where $\Delta t$ represents the reporting period, we write $H_{t}=\Delta N_{E\rightarrow I}(t-\Delta t,t)$ for the complete number of new infections during that time period.
When we are fitting to cumulative case counts, we change the definition accordingly to $H_{t}=\Delta N_{E\rightarrow I}(0,t)$.
When using either type of data, we modeled the corresponding case report,
$C_{t}$, as a negative binomial:
$C_{t}\sim\textrm{NegBin}(\rho H_{t},1/k)$.
Thus $\mathbb{E}[C_{t}|H_t]=\rho H_{t}$ and
$\mathrm{Var}[C_{t}|H_t]=\rho H_{t}+k\rho^{2}H_{t}^{2}$,
where $\rho$ is the reporting probability and $k$ the reporting overdispersion.

Descriptions of the methods used in the simulation study and in the model-based inferences drawn from actual data are given in the Supplementary Materials.

\section*{Acknowledgements}

We thank John Drake, Andrew Park, Robert Reiner, Jonathan Dushoff, and the two anonymous reviewers for their thoughtful comments.
PR and AAK are supported by the Research and Policy in Infectious Disease Dynamics program of the Science and Technology Directorate, Department of Homeland Security, the Fogarty International Center, National Institutes of Health
and by MIDAS, National Institute of General Medical Sciences U54-GM111274 and U01-GM110744.

\bibliography{ebola}

\appendix
\setcounter{secnumdepth}{1}
\setcounter{table}{0}
\setcounter{figure}{0}
\numberwithin{figure}{section}
\numberwithin{table}{section}
\renewcommand{\thesection}{Appendix~\Alph{section}}
\renewcommand{\thetable}{\Alph{section}\arabic{table}}
\renewcommand{\thefigure}{\Alph{section}\arabic{figure}}
\makeatletter
\def\@seccntformat#1{\csname the#1\endcsname.\quad}
\makeatother

\clearpage
\setcounter{page}{1}
\section{Simulation study}\label{sec:simstudy}

To demonstrate the differences between fitting to raw incidence vs.~cumulative incidence data, we performed a simulation study in which we fit the deterministic model variant to both types of data at three different levels of observation overdispersion: $k\in\{0,0.2,0.5\}$.
For each overdispersion treatment, 500 simulated 39-week time series were generated from the stochastic model variant.
The basic reproduction number was set to $R_0=1.4$;
the incubation and infectious periods were fixed as in Table~\ref{tab:model-parameters};
the assumed population size was taken to be that of the Republic of Guinea.
We assumed a reporting probability of $\rho=0.2$ and that, at outbreak initiation, 10 individuals were infected.
This set of parameter values yields a sample mean simulation visually comparable to the WHO data from Guinea, which display initally slow growth in the number of cases and later acceleration.

For each simulated data set, we estimated the basic reproduction number, $R_0$,
the reporting probability, $\rho$, and
the negative binomial overdispersion parameter, $k$.
All other model parameters were fixed at their true values.
Parameter estimation was accomplished using the trajectory matching algorithm (\texttt{traj.match}) from the R package \textsf{pomp}~\citep{pomp}.
We constructed likelihood profiles over $R_0$ and, from these, obtained maximum likelihood point estimates and likelihood-ratio confidence intervals.
The full process of obtaining likelihood profiles on model parameters by trajectory matching took approximately 1.2~hr on a 40-cpu cluster.

\begin{figure}
\begin{knitrout}\scriptsize
\definecolor{shadecolor}{rgb}{0.969, 0.969, 0.969}\color{fgcolor}

{\centering \includegraphics[width=\linewidth]{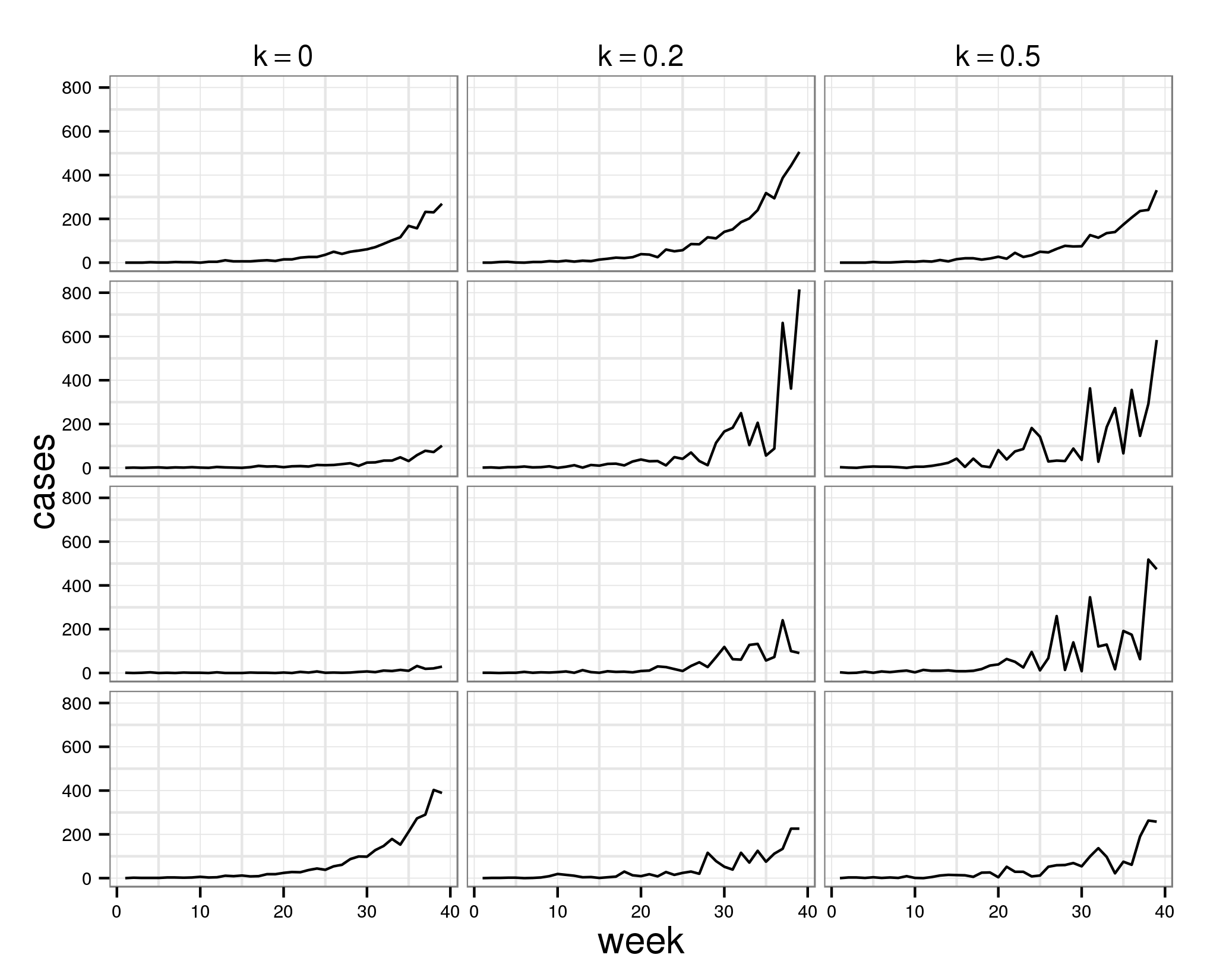} 

}

\end{knitrout}
\caption{
  Twelve randomly-selected simulated datasets from the 1500 used in the simulation study.
  Four simulations are shown for each of three values of the negative binomial overdispersion parameter, $k$.
  \label{fig:sim-study-examples}
}
\end{figure}

\begin{figure}
\begin{knitrout}\scriptsize
\definecolor{shadecolor}{rgb}{0.969, 0.969, 0.969}\color{fgcolor}

{\centering \includegraphics[width=\linewidth]{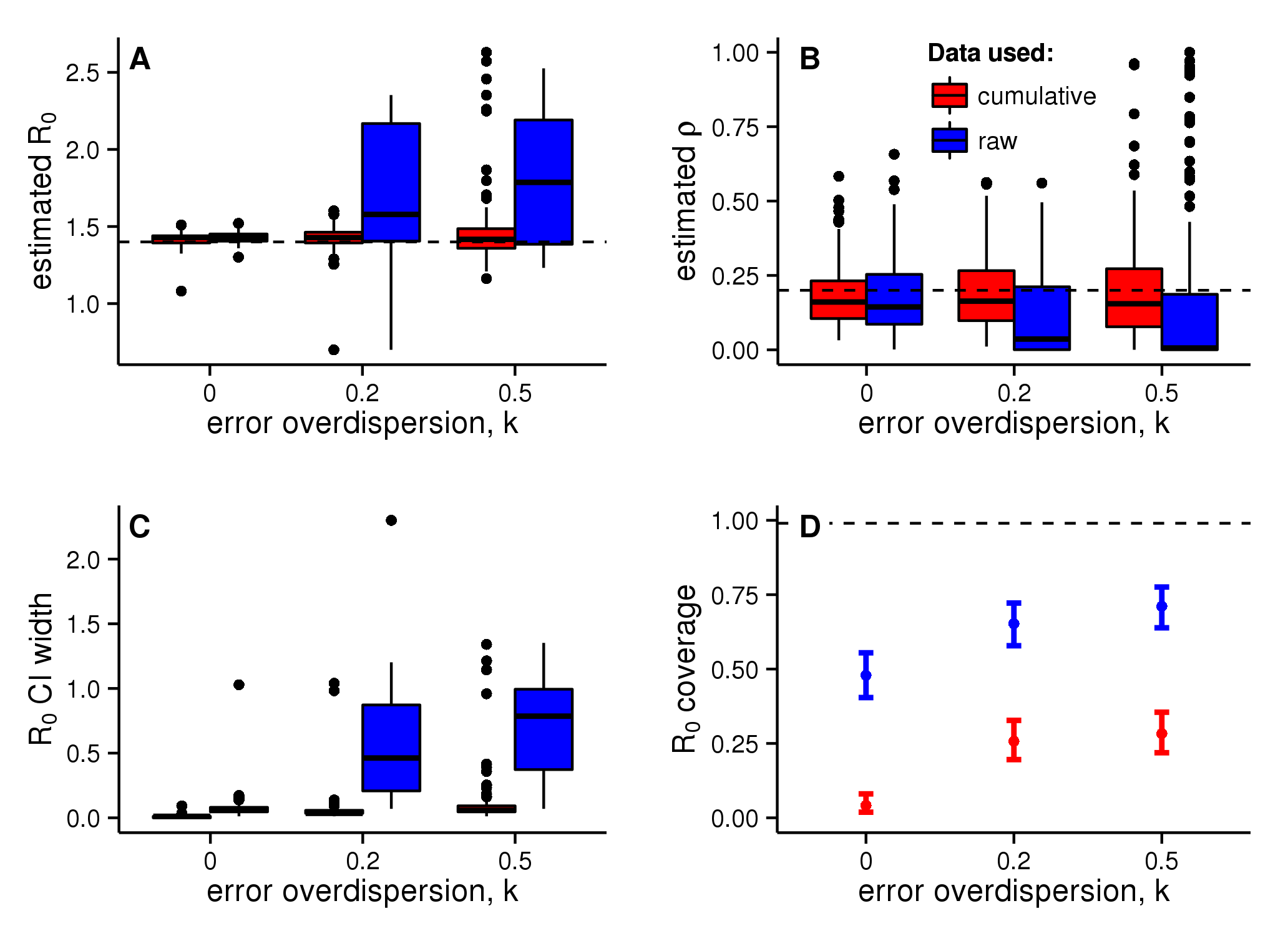} 

}

\end{knitrout}
\caption{
  Results from simulation study fitting the deterministic model to cumulative incidence data using the method of least squares.
  The model was fit to both raw (blue) and accumulated (red) simulated incidence data.
  The same 1500 simulated data sets of length 39~wk used in Fig.~\ref{fig:tm} were used here.
  (A) Estimates of $R_0$.
  True value used in generating the data is shown by the dashed line.
  (B) Estimates of reporting probability, $\rho$.
  The dashed line shows the value used to generate the data.
  (C) Widths of nominal $99\%$ profile likelihood confidence intervals (CI) for $R_0$.
  (D) Actual coverage of the CI, i.e., probability that the true value of $R_0$ lay within the CI.
  Ideally, actual coverage would agree with nominal coverage ($99\%$, dashed line).
  \label{fig:tm-ls}
}
\end{figure}

A second simulation study was performed, in which the deterministic variant of the model was fit to cumulative incidence data by ordinary least squares.
This common procedure in effect assumes that measurement errors are independent and identically normally distributed.
Results of this exercise are shown in Fig.~\ref{fig:tm-ls} in a form comparable to that of Fig.~\ref{fig:tm}.
As in the results shown in the main text, confidence interval widths are erroneously under-estimated with the result that achieved coverage is far smaller than its nominal value.

\clearpage
\section{Model-based inference}

\subsection{Trajectory matching}\label{sec:trajmatch}

Model parameters were initially estimated using trajectory matching.
As in the simulation study, we initially fitted $R_0$, $\rho$, $k$ and the initial conditions.
However, profile likelihoods over $\rho$ were flat, indicating a lack of identifiability in the reporting rate due to a trade-off between this parameter and initial conditions.
Accordingly, we fixed $\rho=0.2$.
The flatness of the likelihood profiles indicates that this assumption has no effect on the quality of fit.
All other model parameters were fixed at the known values given in Table~\ref{tab:model-parameters}.

Trajectory matching was used to compute likelihood profiles over $R_0$ and $k$.
For each point in the profile, the other parameters and initial conditions were initialized at 40 points according to a latin hypersquare (Sobol') design.
In all, the trajectory matching calculations required approximately 21~cpu~hr of computation.
Full details of the trajectory matching codes are provided in the Supplementary Material.

\begin{table}[h!]
  \caption{
    Model parameters, with their interpretations, and their assumed values (parameters estimated from incidence data are so indicated) together with the source of evidence for the assumption.
    \label{tab:model-parameters}
  }
  \begin{tabular}{|c|p{4.5cm}|p{3.5cm}|p{4cm}|}
    \hline
    \hline
    \textbf{Symbol} & \textbf{Meaning} & \textbf{Value} & \textbf{Citation}\\
    \hline
    $R_{0}$ & {\raggedright{} Basic reproduction number} & Estimated &\\
    \hline
    $1/\alpha$ & {\raggedright{} Average incubation period} & 11.4~da & {\raggedright{}\citet{WHOEbolaTeam2014}}\\
    \hline
    $m$ & {\raggedright{} Incubation period shape parameter} & 3 & {\raggedright{}\citet{WHOEbolaTeam2014}}\\
    \hline
    $1/\gamma$ & {\raggedright{} Average infectious period} & 7~da &{\raggedright{}\citet{WHOEbolaTeam2014}}\\
    \hline
    $\rho$ & {\raggedright{} Reporting probability} & 0.2 & Assumption \\
    \hline
    $k$ & {\raggedright{} Reporting overdispersion} & Estimated &\\
    \hline
    $N$ & {\raggedright{} Population size} & {\raggedright{} Guinea: 10.6M\\ Liberia: 4.1M\\ Sierra Leone: 6.2M} &\\
    \hline
    \hline
  \end{tabular}
\end{table}

\subsection{Iterated filtering}\label{sec:if}

Model parameters were estimated using the Iterated Filtering algorithm (IF2)~\citep{Ionides2015}, implemented as \texttt{mif} in the R package \textsf{pomp}~\citep{pomp}.
For each country and each type of data, the parameter estimates along the trajectory-matching profiles were used to initialize the IF2 runs.
From each initial point, we performed 60 IF2 iterations using $2\times10^{3}$ particles, hyperbolic cooling, and a random walk standard deviation (on the log scale) of 0.02 for all parameters and 1 for initial conditions.
For the parameters estimated in each IF run, the log-likelihood was computed as the log of the mean likelihoods of 10 replicate filters, each with $5\times10^{3}$ particles.
Approximate confidence intervals were then computed using the profile log-likelihood \citep{Raue2009}.
All details of these computations are provided in the Supplementary Material.
Computing each of the profile likelihoods in Fig.~\ref{fig:profiles} using iterated filtering took approximately 34~cpu~hr of computation;
all profile computations were accomplished in roughly 3.6~hr on a 100-cpu cluster.

\subsection{Results}\label{sec:supp-results}

\begin{figure}
\begin{knitrout}\scriptsize
\definecolor{shadecolor}{rgb}{0.969, 0.969, 0.969}\color{fgcolor}

{\centering \includegraphics[width=\linewidth]{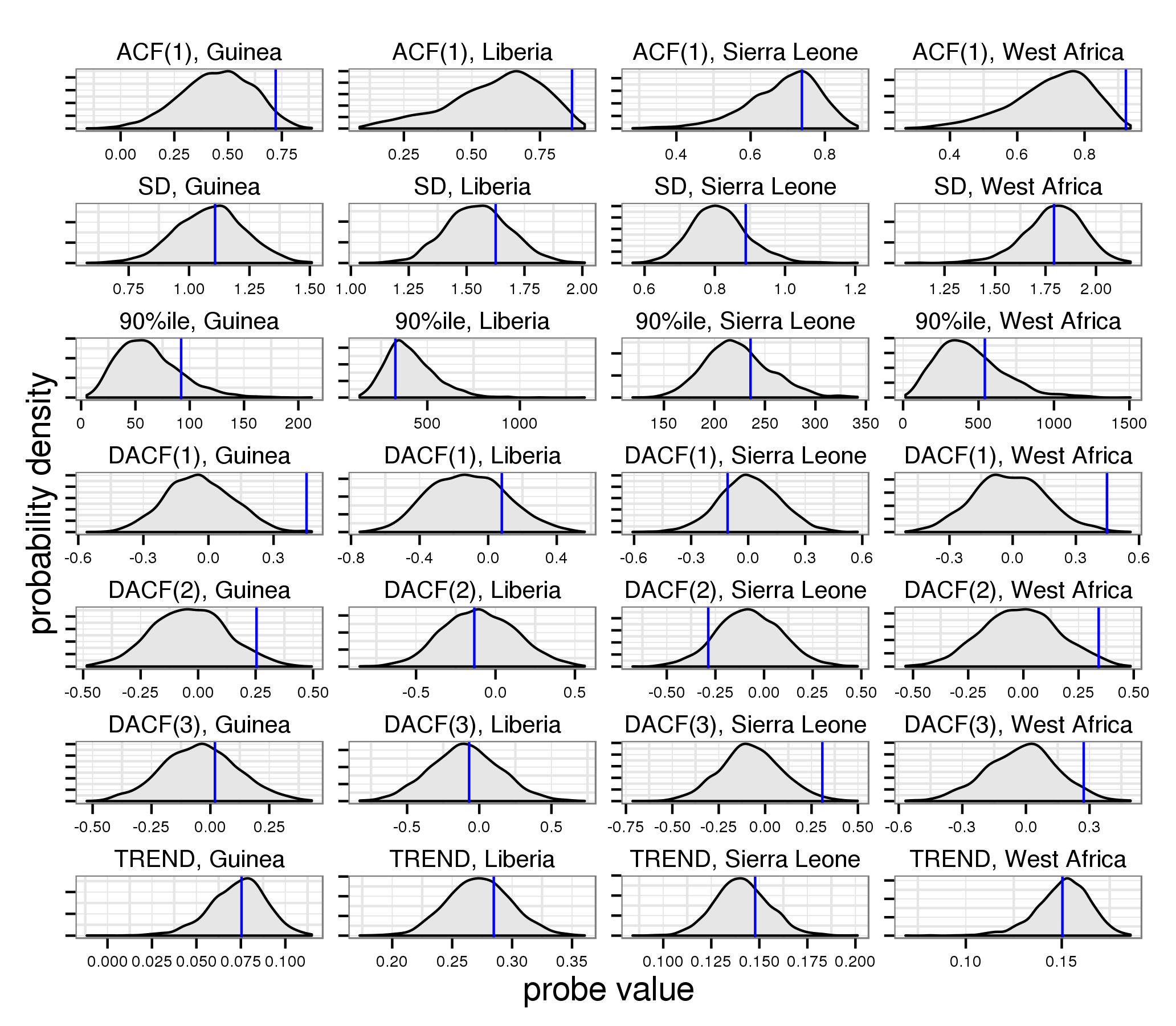} 

}

\end{knitrout}
\caption{
  Additional summary statistics, or probes, computed on both stochastic model simulations and the data.
  In each panel, the probability density of the probes on the simulated data are shown in grey;
  the blue line indicates the value of the probe on the data.
  Probes include autocorrelation at lag 1 (ACF), standard deviation (SD) on log-transformed data, 90th percentile, the autocorrelation at lags 1, 2, and 3~wk after removing an exponential trend (DACF), and the exponential growth rate as obtained by log-linear regression (TREND).
  \label{fig:probes-plot}
}
\end{figure}

Table~\ref{tab:mles} shows the maximum likelihood parameter estimates (MLE).
Fig.~\ref{fig:probes-plot} shows a comparison of various summary statistics (``probes'') computed both on the data and on model simulations.

To tease apart the consequences of failing to account for stochasticity from those of improperly fitting to cumulative data, we fit both deterministic and stochastic models each to both actual incidence and accumulated incidence data.
It is important to recognize that the exercise of fitting the stochastic model to accumulated data is not something one would ever actually do.
Indeed, at the outset incompatibility of model assumptions with the data becomes evident.
To see this, let $H_t$ be the true incidence (i.e., actual number of new infections) in reporting interval $t$ and $C_t$ be the number of reported cases in that interval.
Because of measurement error, $C_t=H_t+\varepsilon_t$, where $\varepsilon_t$ is the error.
Let $h_t=\sum_{s=1}^t\!H_t$ and $c_t=\sum_{s=1}^t\!C_t$ be the accumulated true and reported incidence, respectively.
Because $c_t=\sum_{s=1}^t\!(H_t+\varepsilon_t)$, the errors $c_t-h_t$ are not independent, which is the fundamental problem associated with fitting to cumulative incidence data, irrespective of whether the model for $H_t$ is deterministic or stochastic.
If one attempts to fit a stochastic model to $c_t$ by modeling $c_t=h_t+\xi_t$, where $\xi_t$ are measurement errors, one is confronted with the fact that, even though the accumulated data, $c_t$, and simulations of $h_t$ are guaranteed to increase with time, simulations of $c_t$ under this model will not in general be monotonically increasing.

\begin{figure}[h!]
\begin{knitrout}\scriptsize
\definecolor{shadecolor}{rgb}{0.969, 0.969, 0.969}\color{fgcolor}

{\centering \includegraphics[width=\linewidth]{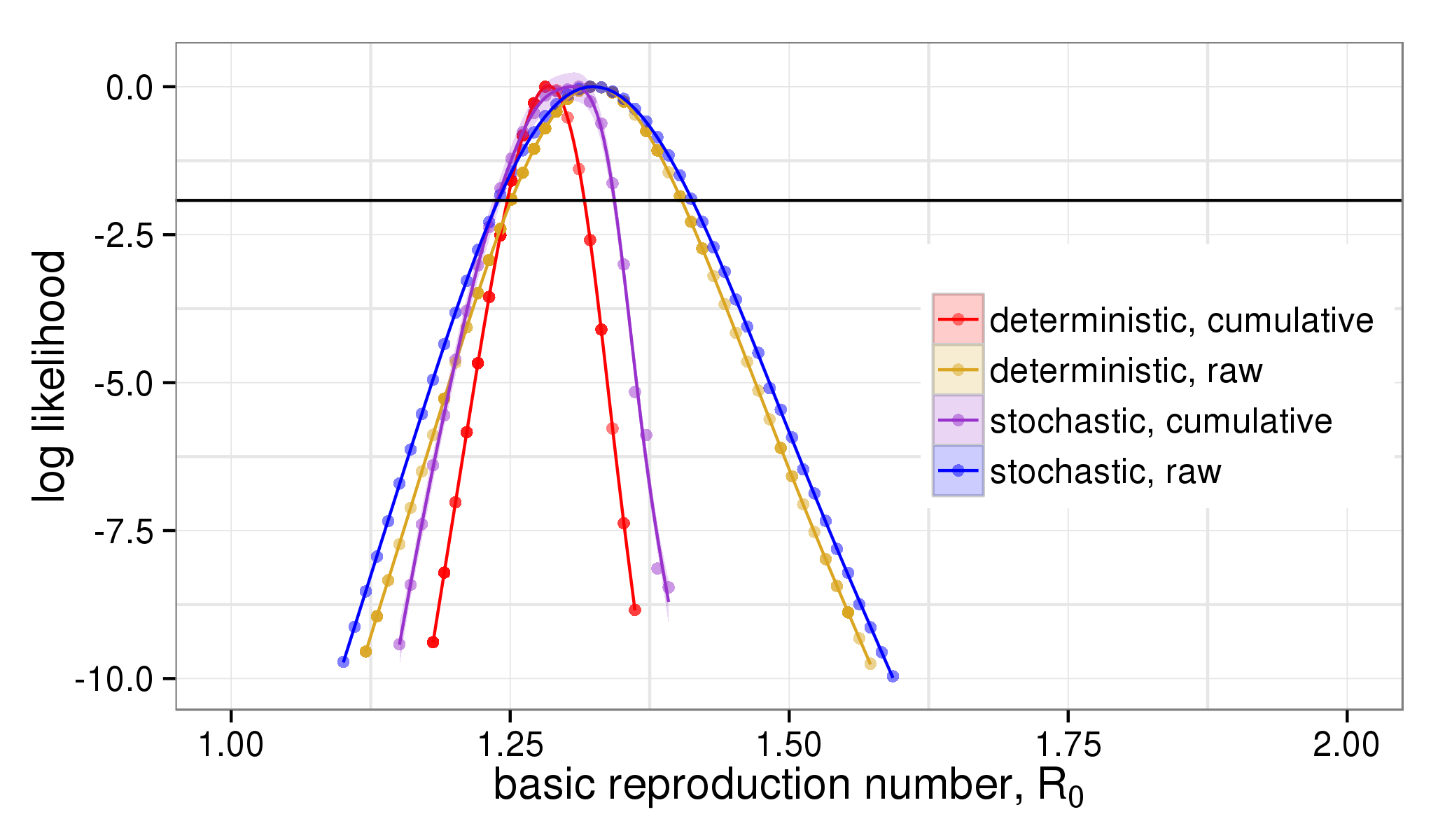} 

}

\end{knitrout}
\caption{
  $R_0$ likelihood profiles for four model-data combinations for the SEIR model fit to the Sierra Leone outbreak.
  \label{fig:profiles-four}
}
\end{figure}

\begin{figure}
\begin{knitrout}\scriptsize
\definecolor{shadecolor}{rgb}{0.969, 0.969, 0.969}\color{fgcolor}

{\centering \includegraphics[width=\linewidth]{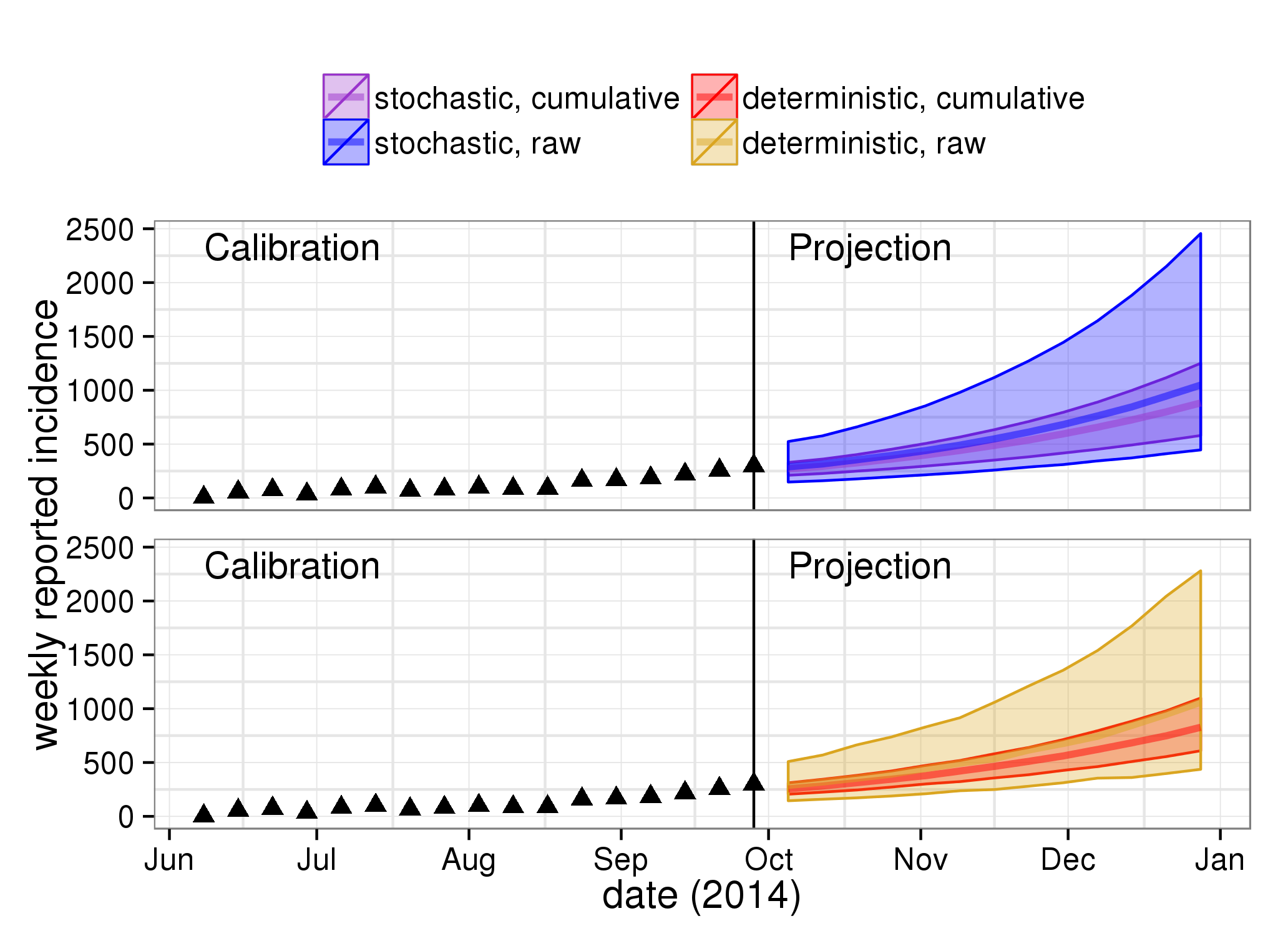} 

}

\end{knitrout}
\caption{
  \label{fig:forecasts-four-obs}
  Forecast uncertainty for the Sierra Leone EBVD outbreak as a function of the model used and the data to which the model was fit.
  The ribbons show the median and 95\% envelope of model simulations for the various models fit to raw and cumulative incidence data from the Sierra Leone outbreak.
  The data used in model fitting are shown using black triangles.
}
\end{figure}

Nevertheless, one naturally wonders about the relative importance of the choice to use a deterministic or stochastic model vs.\ using raw or accumulated incidence data.
Although the answer will certainly depend on both model and data, and therefore vary from situation to situation, we present the comparison in the present case to partially satisfy this natural curiosity.
For the SEIR model fit to the Sierra Leone outbreak data, Fig.~\ref{fig:profiles-four} shows likelihood profiles for the four model-data combinations and Fig.~\ref{fig:forecasts-four-obs} shows the corresponding forecasts.

\begin{table}[h!]
\caption{
  Parameter estimates for the stochastic and deterministic models on the raw data.
  MLE point estimates with nominal $95\%$ confidence intervals are shown.
  \label{tab:mles}
}
\begin{center}
\begin{tabular}{|l|l l|l l|}
\hline
\hline
Parameter & \multicolumn{2}{| c |}{$R_0$} & \multicolumn{2}{| c |}{$k$} \\
\hline
\multicolumn{5}{| c |}{\textbf{Stochastic model, raw data}}\\
\hline
{Guinea} & 1.2 & (1.1--1.3) & 0.37 & (0.22--0.62) \\
\hline
{Liberia} & 1.9 & (1.7--2.2) & 0.24 & (0.12--0.52) \\
\hline
{Sierra Leone} & 1.3 & (1.2--1.4) & 0.038 & (0.02--0.1) \\
\hline
{West Africa} & 1.5 & (1.4--1.6) & 0.17 & (0.091--0.3) \\
\hline
\multicolumn{5}{| c |}{\textbf{Deterministic model, raw data}}\\
\hline
{Guinea} & 1.2 & (1.2--1.3) & 0.37 & (0.23--0.63) \\
\hline
{Liberia} & 1.9 & (1.7--2.2) & 0.23 & (0.12--0.49) \\
\hline
{Sierra Leone} & 1.3 & (1.3--1.4) & 0.04 & (0.02--0.1) \\
\hline
{West Africa} & 1.5 & (1.4--1.5) & 0.18 & (0.11--0.32) \\
\hline
\multicolumn{5}{| c |}{\textbf{Stochastic model, cumulative data}}\\
\hline
{Guinea} & 1.2 & (1.2--1.3) & 0.01 & (0.01--0.014) \\
\hline
{Liberia} & 2 & (1.9--2.1) & 0.02 & (0.01--0.051) \\
\hline
{Sierra Leone} & 1.3 & (1.2--1.3) & 0.0005 & (0--0.0022) \\
\hline
{West Africa} & 1.5 & (1.45--1.51) & 0.01 & (0.01--0.02) \\
\hline
\multicolumn{5}{| c |}{\textbf{Deterministic model, cumulative data}}\\
\hline
{Guinea} & 1.2 & (1.2--1.24) & 0.024 & (0.02--0.051) \\
\hline
{Liberia} & 2 & (1.9--2.1) & 0.02 & (0.01--0.04) \\
\hline
{Sierra Leone} & 1.3 & (1.25--1.31) & 0.0011 & (0--0.002) \\
\hline
{West Africa} & 1.5 & (1.4--1.5) & 0.02 & (0.02--0.03) \\
\hline
\hline
\end{tabular}
\end{center}
\end{table}

\clearpage
\section{Data and codes}

This appendix describes and displays the data and codes needed to fully reproduce all the results of the paper.
All referenced files are provided on \url{datadryad.org}: 
\href{http://dx.doi.org/10.5061/dryad.r5f30}{DOI:10.5061/dryad.r5f30}.

\subsection{Preliminaries}

\subsubsection{R packages}

The following installs the necessary packages, if they are not already installed.
\begin{knitrout}\scriptsize
\definecolor{shadecolor}{rgb}{0.969, 0.969, 0.969}\color{fgcolor}\begin{kframe}
\begin{alltt}
\hlstd{pkgs} \hlkwb{<-} \hlkwd{c}\hlstd{(}\hlstr{"plyr"}\hlstd{,}\hlstr{"pomp"}\hlstd{,}\hlstr{"reshape2"}\hlstd{,}\hlstr{"magrittr"}\hlstd{,}\hlstr{"ggplot2"}\hlstd{,}\hlstr{"scales"}\hlstd{,}
          \hlstr{"foreach"}\hlstd{,}\hlstr{"doMC"}\hlstd{,}\hlstr{"doMPI"}\hlstd{,}\hlstr{"iterators"}\hlstd{)}
\end{alltt}
\end{kframe}
\end{knitrout}
\begin{knitrout}\scriptsize
\definecolor{shadecolor}{rgb}{0.969, 0.969, 0.969}\color{fgcolor}\begin{kframe}
\begin{alltt}
\hlstd{ipkgs} \hlkwb{<-} \hlkwd{rownames}\hlstd{(}\hlkwd{installed.packages}\hlstd{())}
\hlstd{npkgs} \hlkwb{<-} \hlkwd{setdiff}\hlstd{(pkgs,ipkgs)}
\hlkwa{if} \hlstd{(}\hlkwd{length}\hlstd{(npkgs)}\hlopt{>}\hlnum{0}\hlstd{)} \hlkwd{install.packages}\hlstd{(npkgs)}
\hlkwa{if} \hlstd{(}\hlkwd{packageVersion}\hlstd{(}\hlstr{"pomp"}\hlstd{)}\hlopt{<}\hlstr{"0.62-5"}\hlstd{) \{}
  \hlkwd{install.packages}\hlstd{(}\hlstr{"pomp"}\hlstd{,} \hlkwc{repos}\hlstd{=}\hlstr{"http://R-Forge.R-project.org"}\hlstd{)}
  \hlstd{\}}
\end{alltt}
\end{kframe}
\end{knitrout}
The following are the version numbers of these packages used in performing the study.
\begin{knitrout}\scriptsize
\definecolor{shadecolor}{rgb}{0.969, 0.969, 0.969}\color{fgcolor}\begin{kframe}
\begin{alltt}
\hlstd{R.version.string}
\hlkwd{sapply}\hlstd{(pkgs,}\hlkwa{function}\hlstd{(}\hlkwc{x}\hlstd{)}\hlkwd{as.character}\hlstd{(}\hlkwd{packageVersion}\hlstd{(x)))}
\end{alltt}
\begin{verbatim}
## [1] "R version 3.1.3 (2015-03-09)"
##      plyr      pomp  reshape2  magrittr   ggplot2    scales   foreach 
##   "1.8.1"  "0.63.5"   "1.4.1"     "1.5"   "1.0.1"   "0.2.4"   "1.4.2" 
##      doMC     doMPI iterators 
##   "1.3.3"   "0.2.1"   "1.0.7"
\end{verbatim}
\end{kframe}
\end{knitrout}

\subsubsection{MPI}

Most of the codes described below are designed to be run on a cluster using OpenMPI.
It will be assumed that an MPI hostfile, \code{hosts}, exists, which specifes the hosts to be used.
An example \code{hosts} file has contents
\begin{verbatim}
localhost slots=1 max-slots=1
node1 slots=0 max-slots=10
node2 slots=0 max-slots=10
node3 slots=0 max-slots=10
node4 slots=0 max-slots=10
\end{verbatim}
Here, \code{node1}, \code{node2}, \code{node3}, and \code{node4} are the compute nodes in this small cluster.
Each has at least 10 cores.

\subsection{Model implementation}

The file \code{ebola.R}, shown below, will be sourced by the each of the codes below.
It defines a function \code{ebolaModel} that constructs a \code{pomp} object holding the Ebola data and the model codes.

Contents of the file \code{ebola.R}:
\begin{knitrout}\scriptsize
\definecolor{shadecolor}{rgb}{0.969, 0.969, 0.969}\color{fgcolor}\begin{kframe}
\begin{alltt}
\hlkwd{require}\hlstd{(pomp)}
\hlkwd{require}\hlstd{(plyr)}
\hlkwd{require}\hlstd{(reshape2)}
\hlkwd{require}\hlstd{(magrittr)}

\hlkwd{stopifnot}\hlstd{(}\hlkwd{packageVersion}\hlstd{(}\hlstr{"pomp"}\hlstd{)}\hlopt{>=}\hlstr{"0.62-4"}\hlstd{)}

\hlstd{WHO.situation.report.Oct.1} \hlkwb{<-} \hlstr{'
week,Guinea,Liberia,SierraLeone
1,2.244,,
2,2.244,,
3,0.073,,
4,5.717,,
5,3.954,,
6,5.444,,
7,3.274,,
8,5.762,,
9,7.615,,
10,7.615,,
11,27.392,,
12,17.387,,
13,27.115,,
14,29.29,,
15,27.84,,
16,16.345,,
17,10.917,,
18,11.959,,
19,11.959,,
20,8.657,,
21,26.537,,
22,47.764,3.517,
23,26.582,1.043,5.494
24,32.967,18,57.048
25,18.707,16.34,76.022
26,24.322,13.742,36.768
27,4.719,10.155,81.929
28,7.081,24.856,102.632
29,8.527,53.294,69.823
30,92.227,70.146,81.783
31,26.423,139.269,99.775
32,16.549,65.66,88.17
33,36.819,240.645,90.489
34,92.08,274.826,161.54
35,101.03,215.56,168.966
36,102.113,388.553,186.144
37,83.016,410.299,220.442
38,106.674,300.989,258.693
39,55.522,240.237,299.546
'}

\hlcom{## Population sizes in Guinea, Liberia, and Sierra Leone (census 2014)}
\hlstd{populations} \hlkwb{<-} \hlkwd{c}\hlstd{(}\hlkwc{Guinea}\hlstd{=}\hlnum{10628972}\hlstd{,}\hlkwc{Liberia}\hlstd{=}\hlnum{4092310}\hlstd{,}\hlkwc{SierraLeone}\hlstd{=}\hlnum{6190280}\hlstd{)}
\hlstd{populations[}\hlstr{"WestAfrica"}\hlstd{]} \hlkwb{<-} \hlkwd{sum}\hlstd{(populations)}

\hlkwd{read.csv}\hlstd{(}\hlkwc{text}\hlstd{=WHO.situation.report.Oct.1,}\hlkwc{stringsAsFactors}\hlstd{=}\hlnum{FALSE}\hlstd{)} \hlopt{%>%}
  \hlkwd{mutate}\hlstd{(}\hlkwc{date}\hlstd{=}\hlkwd{seq}\hlstd{(}\hlkwc{from}\hlstd{=}\hlkwd{as.Date}\hlstd{(}\hlstr{"2014-01-05"}\hlstd{),}\hlkwc{length}\hlstd{=}\hlkwd{length}\hlstd{(week),}\hlkwc{by}\hlstd{=}\hlstr{'week'}\hlstd{))} \hlopt{%>%}
  \hlkwd{melt}\hlstd{(}\hlkwc{id}\hlstd{=}\hlkwd{c}\hlstd{(}\hlstr{"week"}\hlstd{,}\hlstr{"date"}\hlstd{),}\hlkwc{variable.name}\hlstd{=}\hlstr{"country"}\hlstd{,}\hlkwc{value.name}\hlstd{=}\hlstr{"cases"}\hlstd{)} \hlopt{%>%}
  \hlkwd{mutate}\hlstd{(}\hlkwc{deaths}\hlstd{=}\hlnum{NA}\hlstd{)} \hlkwb{->} \hlstd{dat}

\hlcom{## Parameter transformations}

\hlstd{paruntrans} \hlkwb{<-} \hlkwd{Csnippet}\hlstd{(}\hlstr{'
  double *IC = &S_0;
  double *TIC = &TS_0;
  TR0 = log(R0);
  Trho = logit(rho);
  Tk = log(k);
  to_log_barycentric(TIC,IC,4);
'}\hlstd{)}

\hlstd{partrans} \hlkwb{<-} \hlkwd{Csnippet}\hlstd{(}\hlstr{'
  double *IC = &S_0;
  double *TIC = &TS_0;
  TR0 = exp(R0);
  Trho = expit(rho);
  Tk = exp(k);
  from_log_barycentric(TIC,IC,4);
'}\hlstd{)}

\hlcom{##  Measurement model: hierarchical model for cases}
\hlcom{## p(C_t | H_t): Negative binomial with mean rho*H_t and variance rho*H_t*(1+k*rho*H_t)}
\hlstd{dObs} \hlkwb{<-} \hlkwd{Csnippet}\hlstd{(}\hlstr{'
  double f;
  if (k > 0.0)
    f = dnbinom_mu(nearbyint(cases),1.0/k,rho*N_EI,1);
  else
    f = dpois(nearbyint(cases),rho*N_EI,1);
  lik = (give_log) ? f : exp(f);
'}\hlstd{)}

\hlstd{rObs} \hlkwb{<-} \hlkwd{Csnippet}\hlstd{(}\hlstr{'
  if (k > 0) \{
    cases = rnbinom_mu(1.0/k,rho*N_EI);
    deaths = rnbinom_mu(1.0/k,rho*cfr*N_IR);
  \} else \{
    cases = rpois(rho*N_EI);
    deaths = rpois(rho*cfr*N_IR);
  \}'}\hlstd{)}

\hlcom{### measurement model for ordinary least-squares}
\hlstd{dObsLS} \hlkwb{<-} \hlkwd{Csnippet}\hlstd{(}\hlstr{'
  double f;
  f = dnorm(cases,rho*N_EI,k,1);
  lik = (give_log) ? f : exp(f);
'}\hlstd{)}

\hlstd{rObsLS} \hlkwb{<-} \hlkwd{Csnippet}\hlstd{(}\hlstr{'
  cases = rnorm(rho*N_EI,k);
  deaths = NA_REAL;
'}\hlstd{)}

\hlcom{## Process model simulator}
\hlstd{rSim} \hlkwb{<-} \hlkwd{Csnippet}\hlstd{(}\hlstr{'
  double lambda, beta;
  double *E = &E1;
  beta = R0 * gamma; // Transmission rate
  lambda = beta * I / N; // Force of infection
  int i;

  // Transitions
  // From class S
  double transS = rbinom(S, 1.0 - exp(- lambda * dt)); // No of infections
  // From class E
  double transE[nstageE]; // No of transitions between classes E
  for(i = 0; i < nstageE; i++)\{
    transE[i] = rbinom(E[i], 1.0 - exp(- nstageE * alpha * dt));
  \}
  // From class I
  double transI = rbinom(I, 1.0 - exp(- gamma * dt)); // No of transitions I->R

  // Balance the equations
  S -= transS;
  E[0] += transS - transE[0];
  for(i=1; i < nstageE; i++) \{
    E[i] += transE[i-1] - transE[i];
  \}
  I += transE[nstageE - 1] - transI;
  R += transI;
  N_EI += transE[nstageE - 1]; // No of transitions from E to I
  N_IR += transI; // No of transitions from I to R
'}\hlstd{)}

\hlcom{## Deterministic skeleton (an ODE), used in trajectory matching}
\hlstd{skel} \hlkwb{<-} \hlkwd{Csnippet}\hlstd{(}\hlstr{'
  double lambda, beta;
  double *E = &E1;
  double *DE = &DE1;
  beta = R0 * gamma; // Transmission rate
  lambda = beta * I / N; // Force of infection
  int i;

  // Balance the equations
  DS = - lambda * S;
  DE[0] = lambda * S - nstageE * alpha * E[0];
  for (i=1; i < nstageE; i++)
    DE[i] = nstageE * alpha * (E[i-1]-E[i]);
  DI = nstageE * alpha * E[nstageE-1] - gamma * I;
  DR = gamma * I;
  DN_EI = nstageE * alpha * E[nstageE-1];
  DN_IR = gamma * I;
'}\hlstd{)}

\hlstd{ebolaModel} \hlkwb{<-} \hlkwa{function} \hlstd{(}\hlkwc{country}\hlstd{=}\hlkwd{c}\hlstd{(}\hlstr{"Guinea"}\hlstd{,} \hlstr{"SierraLeone"}\hlstd{,} \hlstr{"Liberia"}\hlstd{,} \hlstr{"WestAfrica"}\hlstd{),}
                        \hlkwc{data} \hlstd{=} \hlkwa{NULL}\hlstd{,}
                        \hlkwc{timestep} \hlstd{=} \hlnum{0.01}\hlstd{,} \hlkwc{nstageE} \hlstd{=} \hlnum{3L}\hlstd{,}
                        \hlkwc{type} \hlstd{=} \hlkwd{c}\hlstd{(}\hlstr{"raw"}\hlstd{,}\hlstr{"cum"}\hlstd{),} \hlkwc{na.rm} \hlstd{=} \hlnum{FALSE}\hlstd{,} \hlkwc{least.sq} \hlstd{=} \hlnum{FALSE}\hlstd{) \{}

  \hlstd{type} \hlkwb{<-} \hlkwd{match.arg}\hlstd{(type)}
  \hlstd{ctry} \hlkwb{<-} \hlkwd{match.arg}\hlstd{(country)}
  \hlstd{pop} \hlkwb{<-} \hlkwd{unname}\hlstd{(populations[ctry])}

  \hlcom{## Incubation period is supposed to be Gamma distributed with shape parameter 3}
  \hlcom{## and mean 11.4 days.  The discrete-time formula is used to calculate the}
  \hlcom{## corresponding alpha (cf He et al., Interface 2010).}
  \hlcom{## Case-fatality ratio is fixed at 0.7 (cf WHO Ebola response team, NEJM 2014)}
  \hlstd{incubation_period} \hlkwb{<-} \hlnum{11.4}\hlopt{/}\hlnum{7}
  \hlstd{infectious_period} \hlkwb{<-} \hlnum{7}\hlopt{/}\hlnum{7}
  \hlstd{index_case} \hlkwb{<-} \hlnum{10}\hlopt{/}\hlstd{pop}
  \hlstd{dt} \hlkwb{<-} \hlstd{timestep}
  \hlstd{nstageE} \hlkwb{<-} \hlkwd{as.integer}\hlstd{(nstageE)}

  \hlstd{globs} \hlkwb{<-} \hlkwd{paste0}\hlstd{(}\hlstr{"static int nstageE = "}\hlstd{,nstageE,}\hlstr{";"}\hlstd{);}

  \hlstd{theta} \hlkwb{<-} \hlkwd{c}\hlstd{(}\hlkwc{N}\hlstd{=pop,}\hlkwc{R0}\hlstd{=}\hlnum{1.4}\hlstd{,}
             \hlkwc{alpha}\hlstd{=}\hlopt{-}\hlnum{1}\hlopt{/}\hlstd{(nstageE}\hlopt{*}\hlstd{dt)}\hlopt{*}\hlkwd{log}\hlstd{(}\hlnum{1}\hlopt{-}\hlstd{nstageE}\hlopt{*}\hlstd{dt}\hlopt{/}\hlstd{incubation_period),}
             \hlkwc{gamma}\hlstd{=}\hlopt{-}\hlkwd{log}\hlstd{(}\hlnum{1}\hlopt{-}\hlstd{dt}\hlopt{/}\hlstd{infectious_period)}\hlopt{/}\hlstd{dt,}
             \hlkwc{rho}\hlstd{=}\hlnum{0.2}\hlstd{,}\hlkwc{cfr}\hlstd{=}\hlnum{0.7}\hlstd{,}
             \hlkwc{k}\hlstd{=}\hlnum{0}\hlstd{,}
             \hlkwc{S_0}\hlstd{=}\hlnum{1}\hlopt{-}\hlstd{index_case,}\hlkwc{E_0}\hlstd{=index_case}\hlopt{/}\hlnum{2}\hlopt{-}\hlnum{5e-9}\hlstd{,}
             \hlkwc{I_0}\hlstd{=index_case}\hlopt{/}\hlnum{2}\hlopt{-}\hlnum{5e-9}\hlstd{,}\hlkwc{R_0}\hlstd{=}\hlnum{1e-8}\hlstd{)}

  \hlkwa{if} \hlstd{(}\hlkwd{is.null}\hlstd{(data)) \{}
    \hlkwa{if} \hlstd{(ctry}\hlopt{==}\hlstr{"WestAfrica"}\hlstd{) \{}
      \hlstd{dat} \hlkwb{<-} \hlkwd{ddply}\hlstd{(dat,}\hlopt{~}\hlstd{week,summarize,}
                   \hlkwc{cases}\hlstd{=}\hlkwd{sum}\hlstd{(cases,}\hlkwc{na.rm}\hlstd{=}\hlnum{TRUE}\hlstd{),}
                   \hlkwc{deaths}\hlstd{=}\hlkwd{sum}\hlstd{(deaths,}\hlkwc{na.rm}\hlstd{=}\hlnum{TRUE}\hlstd{))}
      \hlstd{\}} \hlkwa{else} \hlstd{\{}
        \hlstd{dat} \hlkwb{<-} \hlkwd{subset}\hlstd{(dat,country}\hlopt{==}\hlstd{ctry,}\hlkwc{select}\hlstd{=}\hlopt{-}\hlstd{country)}
        \hlstd{\}}
    \hlstd{\}} \hlkwa{else} \hlstd{\{}
      \hlstd{dat} \hlkwb{<-} \hlstd{data}
      \hlstd{\}}

  \hlkwa{if} \hlstd{(na.rm) \{}
    \hlstd{dat} \hlkwb{<-} \hlkwd{mutate}\hlstd{(}\hlkwd{subset}\hlstd{(dat,}\hlopt{!}\hlkwd{is.na}\hlstd{(cases)),}\hlkwc{week}\hlstd{=week}\hlopt{-}\hlkwd{min}\hlstd{(week)}\hlopt{+}\hlnum{1}\hlstd{)}
    \hlstd{\}}
  \hlkwa{if} \hlstd{(type}\hlopt{==}\hlstr{"cum"}\hlstd{) \{}
    \hlstd{dat} \hlkwb{<-} \hlkwd{mutate}\hlstd{(dat,}\hlkwc{cases}\hlstd{=}\hlkwd{cumsum}\hlstd{(cases),}\hlkwc{deaths}\hlstd{=}\hlkwd{cumsum}\hlstd{(deaths))}
    \hlstd{\}}

  \hlcom{## Create the pomp object}
  \hlkwd{pomp}\hlstd{(}
    \hlkwc{data}\hlstd{=dat[}\hlkwd{c}\hlstd{(}\hlstr{"week"}\hlstd{,}\hlstr{"cases"}\hlstd{,}\hlstr{"deaths"}\hlstd{)],}
    \hlkwc{times}\hlstd{=}\hlstr{"week"}\hlstd{,}
    \hlkwc{t0}\hlstd{=}\hlnum{0}\hlstd{,}
    \hlkwc{params}\hlstd{=theta,}
    \hlkwc{globals}\hlstd{=globs,}
    \hlkwc{obsnames}\hlstd{=}\hlkwd{c}\hlstd{(}\hlstr{"cases"}\hlstd{,}\hlstr{"deaths"}\hlstd{),}
    \hlkwc{statenames}\hlstd{=}\hlkwd{c}\hlstd{(}\hlstr{"S"}\hlstd{,}\hlstr{"E1"}\hlstd{,}\hlstr{"I"}\hlstd{,}\hlstr{"R"}\hlstd{,}\hlstr{"N_EI"}\hlstd{,}\hlstr{"N_IR"}\hlstd{),}
    \hlkwc{zeronames}\hlstd{=}\hlkwa{if} \hlstd{(type}\hlopt{==}\hlstr{"raw"}\hlstd{)} \hlkwd{c}\hlstd{(}\hlstr{"N_EI"}\hlstd{,}\hlstr{"N_IR"}\hlstd{)} \hlkwa{else} \hlkwd{character}\hlstd{(}\hlnum{0}\hlstd{),}
    \hlkwc{paramnames}\hlstd{=}\hlkwd{c}\hlstd{(}\hlstr{"N"}\hlstd{,}\hlstr{"R0"}\hlstd{,}\hlstr{"alpha"}\hlstd{,}\hlstr{"gamma"}\hlstd{,}\hlstr{"rho"}\hlstd{,}\hlstr{"k"}\hlstd{,}\hlstr{"cfr"}\hlstd{,}
                 \hlstr{"S_0"}\hlstd{,}\hlstr{"E_0"}\hlstd{,}\hlstr{"I_0"}\hlstd{,}\hlstr{"R_0"}\hlstd{),}
    \hlkwc{nstageE}\hlstd{=nstageE,}
    \hlkwc{dmeasure}\hlstd{=}\hlkwa{if} \hlstd{(least.sq) dObsLS} \hlkwa{else} \hlstd{dObs,}
    \hlkwc{rmeasure}\hlstd{=}\hlkwa{if} \hlstd{(least.sq) rObsLS} \hlkwa{else} \hlstd{rObs,}
    \hlkwc{rprocess}\hlstd{=}\hlkwd{discrete.time.sim}\hlstd{(}\hlkwc{step.fun}\hlstd{=rSim,}\hlkwc{delta.t}\hlstd{=timestep),}
    \hlkwc{skeleton}\hlstd{=skel,}
    \hlkwc{skeleton.type}\hlstd{=}\hlstr{"vectorfield"}\hlstd{,}
    \hlkwc{parameter.transform}\hlstd{=partrans,}
    \hlkwc{parameter.inv.transform}\hlstd{=paruntrans,}
    \hlkwc{initializer}\hlstd{=}\hlkwa{function} \hlstd{(}\hlkwc{params}\hlstd{,} \hlkwc{t0}\hlstd{,} \hlkwc{nstageE}\hlstd{,} \hlkwc{...}\hlstd{) \{}
      \hlstd{all.state.names} \hlkwb{<-} \hlkwd{c}\hlstd{(}\hlstr{"S"}\hlstd{,}\hlkwd{paste0}\hlstd{(}\hlstr{"E"}\hlstd{,}\hlnum{1}\hlopt{:}\hlstd{nstageE),}\hlstr{"I"}\hlstd{,}\hlstr{"R"}\hlstd{,}\hlstr{"N_EI"}\hlstd{,}\hlstr{"N_IR"}\hlstd{)}
      \hlstd{comp.names} \hlkwb{<-} \hlkwd{c}\hlstd{(}\hlstr{"S"}\hlstd{,}\hlkwd{paste0}\hlstd{(}\hlstr{"E"}\hlstd{,}\hlnum{1}\hlopt{:}\hlstd{nstageE),}\hlstr{"I"}\hlstd{,}\hlstr{"R"}\hlstd{)}
      \hlstd{x0} \hlkwb{<-} \hlkwd{setNames}\hlstd{(}\hlkwd{numeric}\hlstd{(}\hlkwd{length}\hlstd{(all.state.names)),all.state.names)}
      \hlstd{frac} \hlkwb{<-} \hlkwd{c}\hlstd{(params[}\hlstr{"S_0"}\hlstd{],}\hlkwd{rep}\hlstd{(params[}\hlstr{"E_0"}\hlstd{]}\hlopt{/}\hlstd{nstageE,nstageE),params[}\hlstr{"I_0"}\hlstd{],params[}\hlstr{"R_0"}\hlstd{])}
      \hlstd{x0[comp.names]} \hlkwb{<-} \hlkwd{round}\hlstd{(params[}\hlstr{"N"}\hlstd{]}\hlopt{*}\hlstd{frac}\hlopt{/}\hlkwd{sum}\hlstd{(frac))}
      \hlstd{x0}
      \hlstd{\}}
    \hlstd{)} \hlkwb{->} \hlstd{po}
  \hlstd{\}}

\hlkwd{c}\hlstd{(}\hlstr{"ebolaModel"}\hlstd{)}
\end{alltt}
\end{kframe}
\end{knitrout}

\subsection{Simulation study}

The codes in file, \code{simstudy.R}, shown below, perform the simulation study computations.
Note that it assumes that file \code{ebola.R} (contents displayed above) is present in the working directory.
On a cluster with \code{R} and \code{OpenMPI} installed, in a directory containing \code{simstudy.R}, \code{ebola.R}, and the hostfile \code{hosts}, execute the computations with a command such as:
\begin{verbatim}
mpirun -hostfile hosts -np 41 Rscript --vanilla simstudy.R
\end{verbatim}

Contents of file \code{simstudy.R}:
\begin{knitrout}\scriptsize
\definecolor{shadecolor}{rgb}{0.969, 0.969, 0.969}\color{fgcolor}\begin{kframe}
\begin{alltt}
\hlkwd{require}\hlstd{(pomp)}
\hlkwd{require}\hlstd{(plyr)}
\hlkwd{require}\hlstd{(reshape2)}
\hlkwd{require}\hlstd{(magrittr)}
\hlkwd{options}\hlstd{(}\hlkwc{stringsAsFactors}\hlstd{=}\hlnum{FALSE}\hlstd{)}

\hlkwd{require}\hlstd{(foreach)}
\hlkwd{require}\hlstd{(doMPI)}
\hlkwd{require}\hlstd{(iterators)}

\hlkwd{source}\hlstd{(}\hlstr{"ebola.R"}\hlstd{)}
\hlstd{noexport} \hlkwb{<-} \hlkwd{c}\hlstd{(}\hlstr{"ebolaModel"}\hlstd{)}

\hlstd{cl} \hlkwb{<-} \hlkwd{startMPIcluster}\hlstd{()}
\hlkwd{registerDoMPI}\hlstd{(cl)}

\hlstd{bake} \hlkwb{<-} \hlkwa{function} \hlstd{(}\hlkwc{file}\hlstd{,} \hlkwc{expr}\hlstd{) \{}
  \hlkwa{if} \hlstd{(}\hlkwd{file.exists}\hlstd{(file)) \{}
    \hlkwd{readRDS}\hlstd{(file)}
    \hlstd{\}} \hlkwa{else} \hlstd{\{}
      \hlstd{val} \hlkwb{<-} \hlkwd{eval}\hlstd{(expr)}
      \hlkwd{saveRDS}\hlstd{(val,}\hlkwc{file}\hlstd{=file)}
      \hlstd{val}
      \hlstd{\}}
  \hlstd{\}}

\hlcom{## trajectory matching simulation study}

\hlstd{tic} \hlkwb{<-} \hlkwd{Sys.time}\hlstd{()}

\hlstd{po} \hlkwb{<-} \hlkwd{ebolaModel}\hlstd{(}\hlkwc{country}\hlstd{=}\hlstr{"Guinea"}\hlstd{,}\hlkwc{timestep}\hlstd{=}\hlnum{0.1}\hlstd{)}
\hlstd{params} \hlkwb{<-} \hlkwd{parmat}\hlstd{(}\hlkwd{coef}\hlstd{(po),}\hlnum{3}\hlstd{)}
\hlstd{params[}\hlstr{"k"}\hlstd{,]} \hlkwb{<-} \hlkwd{c}\hlstd{(}\hlnum{0}\hlstd{,}\hlnum{0.2}\hlstd{,}\hlnum{0.5}\hlstd{)}
\hlstd{paramnames} \hlkwb{<-} \hlkwd{names}\hlstd{(}\hlkwd{coef}\hlstd{(po))}

\hlstd{nsims} \hlkwb{<-} \hlnum{500}

\hlkwd{bake}\hlstd{(}\hlkwc{file}\hlstd{=}\hlstr{"sims.rds"}\hlstd{,\{}
  \hlkwd{simulate}\hlstd{(po,}\hlkwc{params}\hlstd{=params,}\hlkwc{nsim}\hlstd{=nsims,}\hlkwc{seed}\hlstd{=}\hlnum{208335746L}\hlstd{,}
           \hlkwc{as.data.frame}\hlstd{=}\hlnum{TRUE}\hlstd{,}\hlkwc{obs}\hlstd{=}\hlnum{TRUE}\hlstd{)} \hlopt{%>%}
    \hlkwd{rename}\hlstd{(}\hlkwd{c}\hlstd{(}\hlkwc{time}\hlstd{=}\hlstr{"week"}\hlstd{))} \hlopt{%>%}
    \hlkwd{mutate}\hlstd{(}\hlkwc{k}\hlstd{=params[}\hlstr{"k"}\hlstd{,((}\hlkwd{as.integer}\hlstd{(sim)}\hlopt{-}\hlnum{1}\hlstd{)}\hlopt{%%}\hlkwd{ncol}\hlstd{(params))}\hlopt{+}\hlnum{1}\hlstd{])}
  \hlstd{\})} \hlkwb{->} \hlstd{simdat}

\hlkwd{pompUnload}\hlstd{(po)}

\hlkwd{bake}\hlstd{(}\hlkwc{file}\hlstd{=}\hlstr{"tm-sim-profiles-R0.rds"}\hlstd{,\{}
  \hlkwd{foreach}\hlstd{(}\hlkwc{simul}\hlstd{=}\hlnum{1}\hlopt{:}\hlstd{nsims,}
          \hlkwc{.combine}\hlstd{=rbind,}\hlkwc{.inorder}\hlstd{=}\hlnum{FALSE}\hlstd{,}
          \hlkwc{.noexport}\hlstd{=noexport,}
          \hlkwc{.options.mpi}\hlstd{=}\hlkwd{list}\hlstd{(}\hlkwc{chunkSize}\hlstd{=}\hlnum{10}\hlstd{,}\hlkwc{seed}\hlstd{=}\hlnum{1598260027L}\hlstd{,}\hlkwc{info}\hlstd{=}\hlnum{TRUE}\hlstd{))} \hlopt{%:%}
    \hlkwd{foreach}\hlstd{(}\hlkwc{type}\hlstd{=}\hlkwd{c}\hlstd{(}\hlstr{"raw"}\hlstd{,}\hlstr{"cum"}\hlstd{),}\hlkwc{.combine}\hlstd{=rbind,}\hlkwc{.inorder}\hlstd{=}\hlnum{FALSE}\hlstd{)} \hlopt{%dopar%}
    \hlstd{\{}
      \hlstd{dat} \hlkwb{<-} \hlkwd{subset}\hlstd{(simdat,sim}\hlopt{==}\hlstd{simul,}\hlkwc{select}\hlstd{=}\hlkwd{c}\hlstd{(week,cases,deaths))}
      \hlstd{tm} \hlkwb{<-} \hlkwd{ebolaModel}\hlstd{(}\hlkwc{country}\hlstd{=}\hlstr{"Guinea"}\hlstd{,}\hlkwc{data}\hlstd{=dat,}\hlkwc{type}\hlstd{=}\hlkwd{as.character}\hlstd{(type))}
      \hlstd{st} \hlkwb{<-} \hlstd{params[,(simul}\hlopt{-}\hlnum{1}\hlstd{)}\hlopt{%%}\hlnum{3}\hlopt{+}\hlnum{1}\hlstd{]}
      \hlstd{true.k} \hlkwb{<-} \hlkwd{unname}\hlstd{(st[}\hlstr{"k"}\hlstd{])}
      \hlstd{true.R0} \hlkwb{<-} \hlkwd{unname}\hlstd{(st[}\hlstr{"R0"}\hlstd{])}
      \hlstd{true.rho} \hlkwb{<-} \hlkwd{unname}\hlstd{(st[}\hlstr{"rho"}\hlstd{])}
      \hlstd{st[}\hlstr{"k"}\hlstd{]} \hlkwb{<-} \hlstd{st[}\hlstr{"k"}\hlstd{]}\hlopt{+}\hlnum{1e-6}
      \hlstd{tm} \hlkwb{<-} \hlkwd{traj.match}\hlstd{(tm,}\hlkwc{start}\hlstd{=st,}\hlkwc{est}\hlstd{=}\hlkwd{c}\hlstd{(}\hlstr{"R0"}\hlstd{,}\hlstr{"k"}\hlstd{,}\hlstr{"rho"}\hlstd{),}\hlkwc{transform}\hlstd{=}\hlnum{TRUE}\hlstd{)}
      \hlkwa{if} \hlstd{(}\hlkwd{coef}\hlstd{(tm,}\hlstr{"rho"}\hlstd{)}\hlopt{==}\hlnum{1}\hlstd{)} \hlkwd{coef}\hlstd{(tm,}\hlstr{"rho"}\hlstd{)} \hlkwb{<-} \hlnum{0.999}
      \hlkwa{if} \hlstd{(}\hlkwd{coef}\hlstd{(tm,}\hlstr{"rho"}\hlstd{)}\hlopt{==}\hlnum{0}\hlstd{)} \hlkwd{coef}\hlstd{(tm,}\hlstr{"rho"}\hlstd{)} \hlkwb{<-} \hlnum{0.001}
      \hlstd{tm} \hlkwb{<-} \hlkwd{traj.match}\hlstd{(tm,}\hlkwc{est}\hlstd{=}\hlkwd{c}\hlstd{(}\hlstr{"R0"}\hlstd{,}\hlstr{"k"}\hlstd{,}\hlstr{"rho"}\hlstd{),}\hlkwc{method}\hlstd{=}\hlstr{'subplex'}\hlstd{,}\hlkwc{transform}\hlstd{=}\hlnum{TRUE}\hlstd{)}
      \hlkwd{pompUnload}\hlstd{(tm)}
      \hlkwd{data.frame}\hlstd{(}\hlkwc{sim}\hlstd{=simul,}\hlkwc{type}\hlstd{=}\hlkwd{as.character}\hlstd{(type),}
                 \hlkwc{true.k}\hlstd{=true.k,}\hlkwc{true.R0}\hlstd{=true.R0,}\hlkwc{true.rho}\hlstd{=true.rho,}
                 \hlkwd{as.list}\hlstd{(}\hlkwd{coef}\hlstd{(tm)),}\hlkwc{loglik}\hlstd{=}\hlkwd{logLik}\hlstd{(tm),}
                 \hlkwc{conv}\hlstd{=tm}\hlopt{$}\hlstd{convergence)}
      \hlstd{\}} \hlkwb{->} \hlstd{fits}

  \hlkwd{foreach} \hlstd{(}\hlkwc{fit}\hlstd{=}\hlkwd{iter}\hlstd{(fits,}\hlkwc{by}\hlstd{=}\hlstr{"row"}\hlstd{),}
           \hlkwc{.noexport}\hlstd{=noexport,}
           \hlkwc{.combine}\hlstd{=rbind,}\hlkwc{.inorder}\hlstd{=}\hlnum{FALSE}\hlstd{)} \hlopt{%:%}
    \hlkwd{foreach} \hlstd{(}\hlkwc{r0}\hlstd{=}\hlkwd{seq}\hlstd{(}\hlkwc{from}\hlstd{=}\hlnum{0.7}\hlstd{,}\hlkwc{to}\hlstd{=}\hlnum{3}\hlstd{,}\hlkwc{length}\hlstd{=}\hlnum{200}\hlstd{),}
             \hlkwc{.combine}\hlstd{=rbind,}\hlkwc{.inorder}\hlstd{=}\hlnum{FALSE}\hlstd{,}
             \hlkwc{.options.mpi}\hlstd{=}\hlkwd{list}\hlstd{(}\hlkwc{chunkSize}\hlstd{=}\hlnum{200}\hlstd{,}\hlkwc{seed}\hlstd{=}\hlnum{1598260027L}\hlstd{,}\hlkwc{info}\hlstd{=}\hlnum{TRUE}\hlstd{))} \hlopt{%dopar%}
    \hlstd{\{}
      \hlstd{dat} \hlkwb{<-} \hlkwd{subset}\hlstd{(simdat,sim}\hlopt{==}\hlstd{fit}\hlopt{$}\hlstd{sim,}\hlkwc{select}\hlstd{=}\hlkwd{c}\hlstd{(week,cases,deaths))}
      \hlstd{tm} \hlkwb{<-} \hlkwd{ebolaModel}\hlstd{(}\hlkwc{country}\hlstd{=}\hlstr{"Guinea"}\hlstd{,}\hlkwc{data}\hlstd{=dat,}\hlkwc{type}\hlstd{=}\hlkwd{as.character}\hlstd{(fit}\hlopt{$}\hlstd{type))}
      \hlkwd{coef}\hlstd{(tm)} \hlkwb{<-} \hlkwd{unlist}\hlstd{(fit[paramnames])}
      \hlkwd{coef}\hlstd{(tm,}\hlstr{"R0"}\hlstd{)} \hlkwb{<-} \hlstd{r0}
      \hlkwa{if} \hlstd{(}\hlkwd{coef}\hlstd{(tm,}\hlstr{"rho"}\hlstd{)}\hlopt{==}\hlnum{1}\hlstd{)} \hlkwd{coef}\hlstd{(tm,}\hlstr{"rho"}\hlstd{)} \hlkwb{<-} \hlnum{0.999}
      \hlkwa{if} \hlstd{(}\hlkwd{coef}\hlstd{(tm,}\hlstr{"rho"}\hlstd{)}\hlopt{==}\hlnum{0}\hlstd{)} \hlkwd{coef}\hlstd{(tm,}\hlstr{"rho"}\hlstd{)} \hlkwb{<-} \hlnum{0.001}
      \hlstd{tm} \hlkwb{<-} \hlkwd{traj.match}\hlstd{(tm,}\hlkwc{est}\hlstd{=}\hlkwd{c}\hlstd{(}\hlstr{"k"}\hlstd{,}\hlstr{"rho"}\hlstd{),}\hlkwc{transform}\hlstd{=}\hlnum{TRUE}\hlstd{)}
      \hlkwa{if} \hlstd{(}\hlkwd{coef}\hlstd{(tm,}\hlstr{"rho"}\hlstd{)}\hlopt{==}\hlnum{1}\hlstd{)} \hlkwd{coef}\hlstd{(tm,}\hlstr{"rho"}\hlstd{)} \hlkwb{<-} \hlnum{0.999}
      \hlkwa{if} \hlstd{(}\hlkwd{coef}\hlstd{(tm,}\hlstr{"rho"}\hlstd{)}\hlopt{==}\hlnum{0}\hlstd{)} \hlkwd{coef}\hlstd{(tm,}\hlstr{"rho"}\hlstd{)} \hlkwb{<-} \hlnum{0.001}
      \hlstd{tm} \hlkwb{<-} \hlkwd{traj.match}\hlstd{(tm,}\hlkwc{est}\hlstd{=}\hlkwd{c}\hlstd{(}\hlstr{"k"}\hlstd{,}\hlstr{"rho"}\hlstd{),}\hlkwc{transform}\hlstd{=}\hlnum{TRUE}\hlstd{,}\hlkwc{method}\hlstd{=}\hlstr{'subplex'}\hlstd{)}
      \hlkwd{pompUnload}\hlstd{(tm)}
      \hlkwd{data.frame}\hlstd{(}\hlkwc{sim}\hlstd{=fit}\hlopt{$}\hlstd{sim,}\hlkwc{type}\hlstd{=fit}\hlopt{$}\hlstd{type,}
                 \hlkwc{true.k}\hlstd{=fit}\hlopt{$}\hlstd{true.k,}\hlkwc{true.R0}\hlstd{=fit}\hlopt{$}\hlstd{true.R0,}\hlkwc{true.rho}\hlstd{=fit}\hlopt{$}\hlstd{true.rho,}
                 \hlkwd{as.list}\hlstd{(}\hlkwd{coef}\hlstd{(tm)),}\hlkwc{loglik}\hlstd{=}\hlkwd{logLik}\hlstd{(tm),}
                 \hlkwc{conv}\hlstd{=tm}\hlopt{$}\hlstd{convergence)}
      \hlstd{\}}
  \hlstd{\})} \hlkwb{->} \hlstd{profiles}

\hlkwd{bake}\hlstd{(}\hlkwc{file}\hlstd{=}\hlstr{"tm-sim-fits.rds"}\hlstd{,\{}

  \hlkwd{ddply}\hlstd{(profiles,}\hlopt{~}\hlstd{type}\hlopt{+}\hlstd{sim}\hlopt{+}\hlstd{true.k,subset,loglik}\hlopt{==}\hlkwd{max}\hlstd{(loglik))} \hlkwb{->} \hlstd{starts}

  \hlkwd{foreach}\hlstd{(}\hlkwc{fit}\hlstd{=}\hlkwd{iter}\hlstd{(starts,}\hlkwc{by}\hlstd{=}\hlstr{"row"}\hlstd{),}\hlkwc{.combine}\hlstd{=rbind,}\hlkwc{.inorder}\hlstd{=}\hlnum{FALSE}\hlstd{,}
          \hlkwc{.noexport}\hlstd{=noexport,}
          \hlkwc{.options.mpi}\hlstd{=}\hlkwd{list}\hlstd{(}\hlkwc{chunkSize}\hlstd{=}\hlnum{30}\hlstd{,}\hlkwc{seed}\hlstd{=}\hlnum{1598260027L}\hlstd{,}\hlkwc{info}\hlstd{=}\hlnum{TRUE}\hlstd{))} \hlopt{%dopar%}
    \hlstd{\{}
      \hlstd{dat} \hlkwb{<-} \hlkwd{subset}\hlstd{(simdat,sim}\hlopt{==}\hlstd{fit}\hlopt{$}\hlstd{sim,}\hlkwc{select}\hlstd{=}\hlkwd{c}\hlstd{(week,cases,deaths))}
      \hlstd{tm} \hlkwb{<-} \hlkwd{ebolaModel}\hlstd{(}\hlkwc{country}\hlstd{=}\hlstr{"Guinea"}\hlstd{,}\hlkwc{data}\hlstd{=dat,}\hlkwc{type}\hlstd{=}\hlkwd{as.character}\hlstd{(fit}\hlopt{$}\hlstd{type))}
      \hlkwd{coef}\hlstd{(tm)} \hlkwb{<-} \hlkwd{unlist}\hlstd{(fit[paramnames])}
      \hlkwa{if} \hlstd{(}\hlkwd{coef}\hlstd{(tm,}\hlstr{"rho"}\hlstd{)}\hlopt{==}\hlnum{1}\hlstd{)} \hlkwd{coef}\hlstd{(tm,}\hlstr{"rho"}\hlstd{)} \hlkwb{<-} \hlnum{0.999}
      \hlkwa{if} \hlstd{(}\hlkwd{coef}\hlstd{(tm,}\hlstr{"rho"}\hlstd{)}\hlopt{==}\hlnum{0}\hlstd{)} \hlkwd{coef}\hlstd{(tm,}\hlstr{"rho"}\hlstd{)} \hlkwb{<-} \hlnum{0.001}
      \hlstd{tm} \hlkwb{<-} \hlkwd{traj.match}\hlstd{(tm,}\hlkwc{est}\hlstd{=}\hlkwd{c}\hlstd{(}\hlstr{"k"}\hlstd{,}\hlstr{"rho"}\hlstd{),}\hlkwc{transform}\hlstd{=}\hlnum{TRUE}\hlstd{)}
      \hlkwa{if} \hlstd{(}\hlkwd{coef}\hlstd{(tm,}\hlstr{"rho"}\hlstd{)}\hlopt{==}\hlnum{1}\hlstd{)} \hlkwd{coef}\hlstd{(tm,}\hlstr{"rho"}\hlstd{)} \hlkwb{<-} \hlnum{0.999}
      \hlkwa{if} \hlstd{(}\hlkwd{coef}\hlstd{(tm,}\hlstr{"rho"}\hlstd{)}\hlopt{==}\hlnum{0}\hlstd{)} \hlkwd{coef}\hlstd{(tm,}\hlstr{"rho"}\hlstd{)} \hlkwb{<-} \hlnum{0.001}
      \hlstd{tm} \hlkwb{<-} \hlkwd{traj.match}\hlstd{(tm,}\hlkwc{est}\hlstd{=}\hlkwd{c}\hlstd{(}\hlstr{"k"}\hlstd{,}\hlstr{"rho"}\hlstd{),}\hlkwc{transform}\hlstd{=}\hlnum{TRUE}\hlstd{,}\hlkwc{method}\hlstd{=}\hlstr{'subplex'}\hlstd{)}
      \hlkwd{pompUnload}\hlstd{(tm)}
      \hlkwd{data.frame}\hlstd{(}\hlkwc{sim}\hlstd{=fit}\hlopt{$}\hlstd{sim,}\hlkwc{type}\hlstd{=fit}\hlopt{$}\hlstd{type,}
                 \hlkwc{true.k}\hlstd{=fit}\hlopt{$}\hlstd{true.k,}\hlkwc{true.R0}\hlstd{=fit}\hlopt{$}\hlstd{true.R0,}\hlkwc{true.rho}\hlstd{=fit}\hlopt{$}\hlstd{true.rho,}
                 \hlkwd{as.list}\hlstd{(}\hlkwd{coef}\hlstd{(tm)),}\hlkwc{loglik}\hlstd{=}\hlkwd{logLik}\hlstd{(tm),}
                 \hlkwc{conv}\hlstd{=tm}\hlopt{$}\hlstd{convergence)}
      \hlstd{\}}
  \hlstd{\})} \hlkwb{->} \hlstd{fits}

\hlstd{toc} \hlkwb{<-} \hlkwd{Sys.time}\hlstd{()}
\hlkwd{print}\hlstd{(toc}\hlopt{-}\hlstd{tic)}

\hlcom{## trajectory matching with least squares simulation study}

\hlstd{tic} \hlkwb{<-} \hlkwd{Sys.time}\hlstd{()}

\hlkwd{bake}\hlstd{(}\hlkwc{file}\hlstd{=}\hlstr{"ls-sim-profiles-R0.rds"}\hlstd{,\{}

  \hlkwd{foreach}\hlstd{(}\hlkwc{simul}\hlstd{=}\hlnum{1}\hlopt{:}\hlstd{nsims,}
          \hlkwc{.combine}\hlstd{=rbind,}\hlkwc{.inorder}\hlstd{=}\hlnum{FALSE}\hlstd{,}
          \hlkwc{.noexport}\hlstd{=noexport,}
          \hlkwc{.options.mpi}\hlstd{=}\hlkwd{list}\hlstd{(}\hlkwc{chunkSize}\hlstd{=}\hlnum{10}\hlstd{,}\hlkwc{seed}\hlstd{=}\hlnum{1598260027L}\hlstd{,}\hlkwc{info}\hlstd{=}\hlnum{TRUE}\hlstd{))} \hlopt{%:%}
    \hlkwd{foreach}\hlstd{(}\hlkwc{type}\hlstd{=}\hlkwd{c}\hlstd{(}\hlstr{"raw"}\hlstd{,}\hlstr{"cum"}\hlstd{),}\hlkwc{.combine}\hlstd{=rbind,}\hlkwc{.inorder}\hlstd{=}\hlnum{FALSE}\hlstd{)} \hlopt{%dopar%}
    \hlstd{\{}
      \hlstd{dat} \hlkwb{<-} \hlkwd{subset}\hlstd{(simdat,sim}\hlopt{==}\hlstd{simul,}\hlkwc{select}\hlstd{=}\hlkwd{c}\hlstd{(week,cases,deaths))}
      \hlstd{tm} \hlkwb{<-} \hlkwd{ebolaModel}\hlstd{(}\hlkwc{country}\hlstd{=}\hlstr{"Guinea"}\hlstd{,}\hlkwc{data}\hlstd{=dat,}\hlkwc{type}\hlstd{=}\hlkwd{as.character}\hlstd{(type),}
                       \hlkwc{least.sq}\hlstd{=}\hlnum{TRUE}\hlstd{)}
      \hlstd{st} \hlkwb{<-} \hlstd{params[,(simul}\hlopt{-}\hlnum{1}\hlstd{)}\hlopt{%%}\hlnum{3}\hlopt{+}\hlnum{1}\hlstd{]}
      \hlstd{true.k} \hlkwb{<-} \hlkwd{unname}\hlstd{(st[}\hlstr{"k"}\hlstd{])}
      \hlstd{true.R0} \hlkwb{<-} \hlkwd{unname}\hlstd{(st[}\hlstr{"R0"}\hlstd{])}
      \hlstd{true.rho} \hlkwb{<-} \hlkwd{unname}\hlstd{(st[}\hlstr{"rho"}\hlstd{])}
      \hlstd{st[}\hlstr{"k"}\hlstd{]} \hlkwb{<-} \hlnum{10}
      \hlstd{tm} \hlkwb{<-} \hlkwd{traj.match}\hlstd{(tm,}\hlkwc{start}\hlstd{=st,}\hlkwc{est}\hlstd{=}\hlkwd{c}\hlstd{(}\hlstr{"R0"}\hlstd{,}\hlstr{"k"}\hlstd{,}\hlstr{"rho"}\hlstd{),}\hlkwc{transform}\hlstd{=}\hlnum{TRUE}\hlstd{)}
      \hlkwa{if} \hlstd{(}\hlkwd{coef}\hlstd{(tm,}\hlstr{"rho"}\hlstd{)}\hlopt{==}\hlnum{1}\hlstd{)} \hlkwd{coef}\hlstd{(tm,}\hlstr{"rho"}\hlstd{)} \hlkwb{<-} \hlnum{0.999}
      \hlkwa{if} \hlstd{(}\hlkwd{coef}\hlstd{(tm,}\hlstr{"rho"}\hlstd{)}\hlopt{==}\hlnum{0}\hlstd{)} \hlkwd{coef}\hlstd{(tm,}\hlstr{"rho"}\hlstd{)} \hlkwb{<-} \hlnum{0.001}
      \hlstd{tm} \hlkwb{<-} \hlkwd{traj.match}\hlstd{(tm,}\hlkwc{est}\hlstd{=}\hlkwd{c}\hlstd{(}\hlstr{"R0"}\hlstd{,}\hlstr{"k"}\hlstd{,}\hlstr{"rho"}\hlstd{),}\hlkwc{method}\hlstd{=}\hlstr{'subplex'}\hlstd{,}\hlkwc{transform}\hlstd{=}\hlnum{TRUE}\hlstd{)}
      \hlkwd{pompUnload}\hlstd{(tm)}
      \hlkwd{data.frame}\hlstd{(}\hlkwc{sim}\hlstd{=simul,}\hlkwc{type}\hlstd{=}\hlkwd{as.character}\hlstd{(type),}
                 \hlkwc{true.k}\hlstd{=true.k,}\hlkwc{true.R0}\hlstd{=true.R0,}\hlkwc{true.rho}\hlstd{=true.rho,}
                 \hlkwd{as.list}\hlstd{(}\hlkwd{coef}\hlstd{(tm)),}\hlkwc{loglik}\hlstd{=}\hlkwd{logLik}\hlstd{(tm),}
                 \hlkwc{conv}\hlstd{=tm}\hlopt{$}\hlstd{convergence)}
      \hlstd{\}} \hlkwb{->} \hlstd{fits}

  \hlkwd{foreach} \hlstd{(}\hlkwc{fit}\hlstd{=}\hlkwd{iter}\hlstd{(fits,}\hlkwc{by}\hlstd{=}\hlstr{"row"}\hlstd{),}
           \hlkwc{.noexport}\hlstd{=noexport,}
           \hlkwc{.combine}\hlstd{=rbind,}\hlkwc{.inorder}\hlstd{=}\hlnum{FALSE}\hlstd{)} \hlopt{%:%}
    \hlkwd{foreach} \hlstd{(}\hlkwc{r0}\hlstd{=}\hlkwd{seq}\hlstd{(}\hlkwc{from}\hlstd{=}\hlnum{0.7}\hlstd{,}\hlkwc{to}\hlstd{=}\hlnum{3}\hlstd{,}\hlkwc{length}\hlstd{=}\hlnum{200}\hlstd{),}
             \hlkwc{.combine}\hlstd{=rbind,}\hlkwc{.inorder}\hlstd{=}\hlnum{FALSE}\hlstd{,}
             \hlkwc{.options.mpi}\hlstd{=}\hlkwd{list}\hlstd{(}\hlkwc{chunkSize}\hlstd{=}\hlnum{200}\hlstd{,}\hlkwc{seed}\hlstd{=}\hlnum{1598260027L}\hlstd{,}\hlkwc{info}\hlstd{=}\hlnum{TRUE}\hlstd{))} \hlopt{%dopar%}
    \hlstd{\{}
      \hlstd{dat} \hlkwb{<-} \hlkwd{subset}\hlstd{(simdat,sim}\hlopt{==}\hlstd{fit}\hlopt{$}\hlstd{sim,}\hlkwc{select}\hlstd{=}\hlkwd{c}\hlstd{(week,cases,deaths))}
      \hlstd{tm} \hlkwb{<-} \hlkwd{ebolaModel}\hlstd{(}\hlkwc{country}\hlstd{=}\hlstr{"Guinea"}\hlstd{,}\hlkwc{data}\hlstd{=dat,}\hlkwc{type}\hlstd{=}\hlkwd{as.character}\hlstd{(fit}\hlopt{$}\hlstd{type),}
                       \hlkwc{least.sq}\hlstd{=}\hlnum{TRUE}\hlstd{)}
      \hlkwd{coef}\hlstd{(tm)} \hlkwb{<-} \hlkwd{unlist}\hlstd{(fit[paramnames])}
      \hlkwd{coef}\hlstd{(tm,}\hlstr{"R0"}\hlstd{)} \hlkwb{<-} \hlstd{r0}
      \hlkwa{if} \hlstd{(}\hlkwd{coef}\hlstd{(tm,}\hlstr{"rho"}\hlstd{)}\hlopt{==}\hlnum{1}\hlstd{)} \hlkwd{coef}\hlstd{(tm,}\hlstr{"rho"}\hlstd{)} \hlkwb{<-} \hlnum{0.999}
      \hlkwa{if} \hlstd{(}\hlkwd{coef}\hlstd{(tm,}\hlstr{"rho"}\hlstd{)}\hlopt{==}\hlnum{0}\hlstd{)} \hlkwd{coef}\hlstd{(tm,}\hlstr{"rho"}\hlstd{)} \hlkwb{<-} \hlnum{0.001}
      \hlstd{tm} \hlkwb{<-} \hlkwd{traj.match}\hlstd{(tm,}\hlkwc{est}\hlstd{=}\hlkwd{c}\hlstd{(}\hlstr{"k"}\hlstd{,}\hlstr{"rho"}\hlstd{),}\hlkwc{transform}\hlstd{=}\hlnum{TRUE}\hlstd{)}
      \hlkwa{if} \hlstd{(}\hlkwd{coef}\hlstd{(tm,}\hlstr{"rho"}\hlstd{)}\hlopt{==}\hlnum{1}\hlstd{)} \hlkwd{coef}\hlstd{(tm,}\hlstr{"rho"}\hlstd{)} \hlkwb{<-} \hlnum{0.999}
      \hlkwa{if} \hlstd{(}\hlkwd{coef}\hlstd{(tm,}\hlstr{"rho"}\hlstd{)}\hlopt{==}\hlnum{0}\hlstd{)} \hlkwd{coef}\hlstd{(tm,}\hlstr{"rho"}\hlstd{)} \hlkwb{<-} \hlnum{0.001}
      \hlstd{tm} \hlkwb{<-} \hlkwd{traj.match}\hlstd{(tm,}\hlkwc{est}\hlstd{=}\hlkwd{c}\hlstd{(}\hlstr{"k"}\hlstd{,}\hlstr{"rho"}\hlstd{),}\hlkwc{transform}\hlstd{=}\hlnum{TRUE}\hlstd{,}\hlkwc{method}\hlstd{=}\hlstr{'subplex'}\hlstd{)}
      \hlkwd{pompUnload}\hlstd{(tm)}
      \hlkwd{data.frame}\hlstd{(}\hlkwc{sim}\hlstd{=fit}\hlopt{$}\hlstd{sim,}\hlkwc{type}\hlstd{=fit}\hlopt{$}\hlstd{type,}
                 \hlkwc{true.k}\hlstd{=fit}\hlopt{$}\hlstd{true.k,}\hlkwc{true.R0}\hlstd{=fit}\hlopt{$}\hlstd{true.R0,}\hlkwc{true.rho}\hlstd{=fit}\hlopt{$}\hlstd{true.rho,}
                 \hlkwd{as.list}\hlstd{(}\hlkwd{coef}\hlstd{(tm)),}\hlkwc{loglik}\hlstd{=}\hlkwd{logLik}\hlstd{(tm),}
                 \hlkwc{conv}\hlstd{=tm}\hlopt{$}\hlstd{convergence)}
      \hlstd{\}}
  \hlstd{\})} \hlkwb{->} \hlstd{profiles}

\hlkwd{bake}\hlstd{(}\hlkwc{file}\hlstd{=}\hlstr{"ls-sim-fits.rds"}\hlstd{,\{}

  \hlkwd{ddply}\hlstd{(profiles,}\hlopt{~}\hlstd{type}\hlopt{+}\hlstd{sim}\hlopt{+}\hlstd{true.k,subset,loglik}\hlopt{==}\hlkwd{max}\hlstd{(loglik))} \hlkwb{->} \hlstd{starts}

  \hlkwd{foreach}\hlstd{(}\hlkwc{fit}\hlstd{=}\hlkwd{iter}\hlstd{(starts,}\hlkwc{by}\hlstd{=}\hlstr{"row"}\hlstd{),}\hlkwc{.combine}\hlstd{=rbind,}\hlkwc{.inorder}\hlstd{=}\hlnum{FALSE}\hlstd{,}
          \hlkwc{.noexport}\hlstd{=noexport,}
          \hlkwc{.options.mpi}\hlstd{=}\hlkwd{list}\hlstd{(}\hlkwc{chunkSize}\hlstd{=}\hlnum{30}\hlstd{,}\hlkwc{seed}\hlstd{=}\hlnum{1598260027L}\hlstd{,}\hlkwc{info}\hlstd{=}\hlnum{TRUE}\hlstd{))} \hlopt{%dopar%}
    \hlstd{\{}
      \hlstd{dat} \hlkwb{<-} \hlkwd{subset}\hlstd{(simdat,sim}\hlopt{==}\hlstd{fit}\hlopt{$}\hlstd{sim,}\hlkwc{select}\hlstd{=}\hlkwd{c}\hlstd{(week,cases,deaths))}
      \hlstd{tm} \hlkwb{<-} \hlkwd{ebolaModel}\hlstd{(}\hlkwc{country}\hlstd{=}\hlstr{"Guinea"}\hlstd{,}\hlkwc{data}\hlstd{=dat,}\hlkwc{type}\hlstd{=}\hlkwd{as.character}\hlstd{(fit}\hlopt{$}\hlstd{type),}
                       \hlkwc{least.sq}\hlstd{=}\hlnum{TRUE}\hlstd{)}
      \hlkwd{coef}\hlstd{(tm)} \hlkwb{<-} \hlkwd{unlist}\hlstd{(fit[paramnames])}
      \hlkwa{if} \hlstd{(}\hlkwd{coef}\hlstd{(tm,}\hlstr{"rho"}\hlstd{)}\hlopt{==}\hlnum{1}\hlstd{)} \hlkwd{coef}\hlstd{(tm,}\hlstr{"rho"}\hlstd{)} \hlkwb{<-} \hlnum{0.999}
      \hlkwa{if} \hlstd{(}\hlkwd{coef}\hlstd{(tm,}\hlstr{"rho"}\hlstd{)}\hlopt{==}\hlnum{0}\hlstd{)} \hlkwd{coef}\hlstd{(tm,}\hlstr{"rho"}\hlstd{)} \hlkwb{<-} \hlnum{0.001}
      \hlstd{tm} \hlkwb{<-} \hlkwd{traj.match}\hlstd{(tm,}\hlkwc{est}\hlstd{=}\hlkwd{c}\hlstd{(}\hlstr{"k"}\hlstd{,}\hlstr{"rho"}\hlstd{),}\hlkwc{transform}\hlstd{=}\hlnum{TRUE}\hlstd{)}
      \hlkwa{if} \hlstd{(}\hlkwd{coef}\hlstd{(tm,}\hlstr{"rho"}\hlstd{)}\hlopt{==}\hlnum{1}\hlstd{)} \hlkwd{coef}\hlstd{(tm,}\hlstr{"rho"}\hlstd{)} \hlkwb{<-} \hlnum{0.999}
      \hlkwa{if} \hlstd{(}\hlkwd{coef}\hlstd{(tm,}\hlstr{"rho"}\hlstd{)}\hlopt{==}\hlnum{0}\hlstd{)} \hlkwd{coef}\hlstd{(tm,}\hlstr{"rho"}\hlstd{)} \hlkwb{<-} \hlnum{0.001}
      \hlstd{tm} \hlkwb{<-} \hlkwd{traj.match}\hlstd{(tm,}\hlkwc{est}\hlstd{=}\hlkwd{c}\hlstd{(}\hlstr{"k"}\hlstd{,}\hlstr{"rho"}\hlstd{),}\hlkwc{transform}\hlstd{=}\hlnum{TRUE}\hlstd{,}\hlkwc{method}\hlstd{=}\hlstr{'subplex'}\hlstd{)}
      \hlkwd{pompUnload}\hlstd{(tm)}
      \hlkwd{data.frame}\hlstd{(}\hlkwc{sim}\hlstd{=fit}\hlopt{$}\hlstd{sim,}\hlkwc{type}\hlstd{=fit}\hlopt{$}\hlstd{type,}
                 \hlkwc{true.k}\hlstd{=fit}\hlopt{$}\hlstd{true.k,}\hlkwc{true.R0}\hlstd{=fit}\hlopt{$}\hlstd{true.R0,}\hlkwc{true.rho}\hlstd{=fit}\hlopt{$}\hlstd{true.rho,}
                 \hlkwd{as.list}\hlstd{(}\hlkwd{coef}\hlstd{(tm)),}\hlkwc{loglik}\hlstd{=}\hlkwd{logLik}\hlstd{(tm),}
                 \hlkwc{conv}\hlstd{=tm}\hlopt{$}\hlstd{convergence)}
      \hlstd{\}}
  \hlstd{\})} \hlkwb{->} \hlstd{fits}

\hlstd{toc} \hlkwb{<-} \hlkwd{Sys.time}\hlstd{()}
\hlkwd{print}\hlstd{(toc}\hlopt{-}\hlstd{tic)}

\hlkwd{closeCluster}\hlstd{(cl)}
\hlkwd{mpi.quit}\hlstd{()}
\end{alltt}
\end{kframe}
\end{knitrout}

\subsection{Parameter estimation}

The codes in file \code{profiles.R} use trajectory matching and iterated filtering to estimate model parameters for all four countries and both types of data.
Again, this is designed to be run on an MPI cluster.
In a directory with the files \code{hosts} and \code{ebola.R} (see above), execute these computations via a command like
\begin{verbatim}
mpirun -hostfile hosts -np 101 Rscript --vanilla profiles.R
\end{verbatim}

Contents of file \code{profiles.R}:
\begin{knitrout}\scriptsize
\definecolor{shadecolor}{rgb}{0.969, 0.969, 0.969}\color{fgcolor}\begin{kframe}
\begin{alltt}
\hlkwd{require}\hlstd{(pomp)}
\hlkwd{require}\hlstd{(plyr)}
\hlkwd{require}\hlstd{(reshape2)}
\hlkwd{require}\hlstd{(magrittr)}
\hlkwd{options}\hlstd{(}\hlkwc{stringsAsFactors}\hlstd{=}\hlnum{FALSE}\hlstd{)}

\hlkwd{require}\hlstd{(foreach)}
\hlkwd{require}\hlstd{(doMPI)}
\hlkwd{require}\hlstd{(iterators)}

\hlkwd{source}\hlstd{(}\hlstr{"ebola.R"}\hlstd{)}

\hlkwd{foreach} \hlstd{(}\hlkwc{country}\hlstd{=}\hlkwd{c}\hlstd{(}\hlstr{"SierraLeone"}\hlstd{,}\hlstr{"Liberia"}\hlstd{,}\hlstr{"Guinea"}\hlstd{,}\hlstr{"WestAfrica"}\hlstd{),}
         \hlkwc{.inorder}\hlstd{=}\hlnum{TRUE}\hlstd{,}\hlkwc{.combine}\hlstd{=c)} \hlopt{%:%}
  \hlkwd{foreach} \hlstd{(}\hlkwc{type}\hlstd{=}\hlkwd{c}\hlstd{(}\hlstr{"raw"}\hlstd{,}\hlstr{"cum"}\hlstd{),}\hlkwc{.inorder}\hlstd{=}\hlnum{TRUE}\hlstd{,}\hlkwc{.combine}\hlstd{=c)} \hlopt{%do%}
  \hlstd{\{}
    \hlkwd{ebolaModel}\hlstd{(}\hlkwc{country}\hlstd{=country,}\hlkwc{type}\hlstd{=type,}\hlkwc{na.rm}\hlstd{=}\hlnum{TRUE}\hlstd{)}
    \hlstd{\}} \hlkwb{->} \hlstd{models}
\hlkwd{dim}\hlstd{(models)} \hlkwb{<-} \hlkwd{c}\hlstd{(}\hlnum{2}\hlstd{,}\hlnum{4}\hlstd{)}
\hlkwd{dimnames}\hlstd{(models)} \hlkwb{<-} \hlkwd{list}\hlstd{(}\hlkwd{c}\hlstd{(}\hlstr{"raw"}\hlstd{,}\hlstr{"cum"}\hlstd{),}
                         \hlkwd{c}\hlstd{(}\hlstr{"SierraLeone"}\hlstd{,}\hlstr{"Liberia"}\hlstd{,}\hlstr{"Guinea"}\hlstd{,}\hlstr{"WestAfrica"}\hlstd{))}

\hlstd{noexport} \hlkwb{<-} \hlkwd{c}\hlstd{(}\hlstr{"models"}\hlstd{)}

\hlstd{cl} \hlkwb{<-} \hlkwd{startMPIcluster}\hlstd{()}
\hlkwd{registerDoMPI}\hlstd{(cl)}

\hlstd{bake} \hlkwb{<-} \hlkwa{function} \hlstd{(}\hlkwc{file}\hlstd{,} \hlkwc{expr}\hlstd{) \{}
  \hlkwa{if} \hlstd{(}\hlkwd{file.exists}\hlstd{(file)) \{}
    \hlkwd{readRDS}\hlstd{(file)}
    \hlstd{\}} \hlkwa{else} \hlstd{\{}
      \hlstd{val} \hlkwb{<-} \hlkwd{eval}\hlstd{(expr)}
      \hlkwd{saveRDS}\hlstd{(val,}\hlkwc{file}\hlstd{=file)}
      \hlstd{val}
      \hlstd{\}}
  \hlstd{\}}

\hlcom{## trajectory matching: R0 profile}

\hlkwd{bake}\hlstd{(}\hlkwc{file}\hlstd{=}\hlstr{"tm-fits-R0.rds"}\hlstd{,\{}

  \hlstd{starts} \hlkwb{<-} \hlkwd{profileDesign}\hlstd{(}\hlkwc{R0}\hlstd{=}\hlkwd{seq}\hlstd{(}\hlnum{1}\hlstd{,}\hlnum{3}\hlstd{,}\hlkwc{length}\hlstd{=}\hlnum{200}\hlstd{),}
                          \hlkwc{upper}\hlstd{=}\hlkwd{c}\hlstd{(}\hlkwc{k}\hlstd{=}\hlnum{1}\hlstd{),}
                          \hlkwc{lower}\hlstd{=}\hlkwd{c}\hlstd{(}\hlkwc{k}\hlstd{=}\hlnum{1e-8}\hlstd{),}
                          \hlkwc{nprof}\hlstd{=}\hlnum{40}\hlstd{)}

  \hlkwd{foreach} \hlstd{(}\hlkwc{start}\hlstd{=}\hlkwd{iter}\hlstd{(starts,}\hlkwc{by}\hlstd{=}\hlstr{'row'}\hlstd{),}
           \hlkwc{.combine}\hlstd{=rbind,}\hlkwc{.inorder}\hlstd{=}\hlnum{FALSE}\hlstd{,}
           \hlkwc{.noexport}\hlstd{=noexport,}
           \hlkwc{.options.mpi}\hlstd{=}\hlkwd{list}\hlstd{(}\hlkwc{chunkSize}\hlstd{=}\hlnum{100}\hlstd{,}\hlkwc{seed}\hlstd{=}\hlnum{2016138277L}\hlstd{,}\hlkwc{info}\hlstd{=}\hlnum{TRUE}\hlstd{)}
           \hlstd{)} \hlopt{%:%}
    \hlkwd{foreach} \hlstd{(}\hlkwc{type}\hlstd{=}\hlkwd{c}\hlstd{(}\hlstr{"raw"}\hlstd{,}\hlstr{"cum"}\hlstd{),}\hlkwc{.combine}\hlstd{=rbind,}\hlkwc{.inorder}\hlstd{=}\hlnum{FALSE}\hlstd{)} \hlopt{%:%}
    \hlkwd{foreach} \hlstd{(}\hlkwc{country}\hlstd{=}\hlkwd{c}\hlstd{(}\hlstr{"SierraLeone"}\hlstd{,}\hlstr{"Liberia"}\hlstd{,}\hlstr{"Guinea"}\hlstd{,}\hlstr{"WestAfrica"}\hlstd{),}
             \hlkwc{.combine}\hlstd{=rbind,}\hlkwc{.inorder}\hlstd{=}\hlnum{FALSE}\hlstd{)} \hlopt{%dopar%}
    \hlstd{\{}
      \hlstd{tm} \hlkwb{<-} \hlstd{models[type,country][[}\hlnum{1}\hlstd{]]}
      \hlstd{tic} \hlkwb{<-} \hlkwd{Sys.time}\hlstd{()}
      \hlkwd{coef}\hlstd{(tm,}\hlkwd{names}\hlstd{(start))} \hlkwb{<-} \hlkwd{unname}\hlstd{(}\hlkwd{unlist}\hlstd{(start))}
      \hlkwd{coef}\hlstd{(tm,}\hlstr{"rho"}\hlstd{)} \hlkwb{<-} \hlnum{0.2}
      \hlstd{tm} \hlkwb{<-} \hlkwd{traj.match}\hlstd{(tm,}\hlkwc{est}\hlstd{=}\hlkwd{c}\hlstd{(}\hlstr{"k"}\hlstd{,}\hlstr{"E_0"}\hlstd{,}\hlstr{"I_0"}\hlstd{),}\hlkwc{transform}\hlstd{=}\hlnum{TRUE}\hlstd{)}
      \hlkwa{if} \hlstd{(}\hlkwd{coef}\hlstd{(tm,}\hlstr{"k"}\hlstd{)}\hlopt{==}\hlnum{0}\hlstd{)} \hlkwd{coef}\hlstd{(tm,}\hlstr{"k"}\hlstd{)} \hlkwb{<-} \hlnum{1e-8}
      \hlkwa{if} \hlstd{(}\hlkwd{coef}\hlstd{(tm,}\hlstr{"E_0"}\hlstd{)}\hlopt{==}\hlnum{0}\hlstd{)} \hlkwd{coef}\hlstd{(tm,}\hlstr{"E_0"}\hlstd{)} \hlkwb{<-} \hlnum{1e-12}
      \hlkwa{if} \hlstd{(}\hlkwd{coef}\hlstd{(tm,}\hlstr{"I_0"}\hlstd{)}\hlopt{==}\hlnum{0}\hlstd{)} \hlkwd{coef}\hlstd{(tm,}\hlstr{"I_0"}\hlstd{)} \hlkwb{<-} \hlnum{1e-12}
      \hlstd{tm} \hlkwb{<-} \hlkwd{traj.match}\hlstd{(tm,}\hlkwc{method}\hlstd{=}\hlstr{'subplex'}\hlstd{,}\hlkwc{control}\hlstd{=}\hlkwd{list}\hlstd{(}\hlkwc{maxit}\hlstd{=}\hlnum{1e5}\hlstd{))}
      \hlstd{toc} \hlkwb{<-} \hlkwd{Sys.time}\hlstd{()}
      \hlstd{etime} \hlkwb{<-} \hlstd{toc}\hlopt{-}\hlstd{tic}
      \hlkwd{units}\hlstd{(etime)} \hlkwb{<-} \hlstr{"hours"}
      \hlkwd{data.frame}\hlstd{(}\hlkwc{country}\hlstd{=country,}\hlkwc{type}\hlstd{=type,}\hlkwd{as.list}\hlstd{(}\hlkwd{coef}\hlstd{(tm)),}
                 \hlkwc{loglik}\hlstd{=}\hlkwd{logLik}\hlstd{(tm),}\hlkwc{conv}\hlstd{=tm}\hlopt{$}\hlstd{convergence,}
                 \hlkwc{etime}\hlstd{=}\hlkwd{as.numeric}\hlstd{(etime))}
      \hlstd{\}} \hlopt{%>%} \hlkwd{mutate}\hlstd{(}\hlkwc{sum}\hlstd{=S_0}\hlopt{+}\hlstd{E_0}\hlopt{+}\hlstd{I_0}\hlopt{+}\hlstd{R_0,}
                   \hlkwc{S_0}\hlstd{=}\hlkwd{round}\hlstd{(N}\hlopt{*}\hlstd{S_0}\hlopt{/}\hlstd{sum),}
                   \hlkwc{E_0}\hlstd{=}\hlkwd{round}\hlstd{(N}\hlopt{*}\hlstd{E_0}\hlopt{/}\hlstd{sum),}
                   \hlkwc{I_0}\hlstd{=}\hlkwd{round}\hlstd{(N}\hlopt{*}\hlstd{I_0}\hlopt{/}\hlstd{sum),}
                   \hlkwc{R_0}\hlstd{=}\hlkwd{round}\hlstd{(N}\hlopt{*}\hlstd{R_0}\hlopt{/}\hlstd{sum))} \hlopt{%>%}
    \hlkwd{subset}\hlstd{(conv} \hlopt{%in%} \hlkwd{c}\hlstd{(}\hlnum{0}\hlstd{,}\hlnum{1}\hlstd{),}\hlkwc{select}\hlstd{=}\hlopt{-}\hlstd{sum)} \hlopt{%>%}
    \hlkwd{unique}\hlstd{()}

  \hlstd{\})} \hlkwb{->} \hlstd{profR0}

\hlcom{## trajectory matching: k profile}

\hlkwd{bake}\hlstd{(}\hlkwc{file}\hlstd{=}\hlstr{"tm-fits-k.rds"}\hlstd{,\{}

  \hlstd{starts} \hlkwb{<-} \hlkwd{profileDesign}\hlstd{(}\hlkwc{k}\hlstd{=}\hlkwd{seq}\hlstd{(}\hlnum{0}\hlstd{,}\hlnum{1}\hlstd{,}\hlkwc{length}\hlstd{=}\hlnum{100}\hlstd{),}
                          \hlkwc{upper}\hlstd{=}\hlkwd{c}\hlstd{(}\hlkwc{R0}\hlstd{=}\hlnum{1}\hlstd{),}
                          \hlkwc{lower}\hlstd{=}\hlkwd{c}\hlstd{(}\hlkwc{R0}\hlstd{=}\hlnum{3}\hlstd{),}
                          \hlkwc{nprof}\hlstd{=}\hlnum{40}\hlstd{)}

  \hlkwd{foreach} \hlstd{(}\hlkwc{start}\hlstd{=}\hlkwd{iter}\hlstd{(starts,}\hlkwc{by}\hlstd{=}\hlstr{'row'}\hlstd{),}
           \hlkwc{.combine}\hlstd{=rbind,}\hlkwc{.inorder}\hlstd{=}\hlnum{FALSE}\hlstd{,}
           \hlkwc{.noexport}\hlstd{=noexport,}
           \hlkwc{.options.mpi}\hlstd{=}\hlkwd{list}\hlstd{(}\hlkwc{chunkSize}\hlstd{=}\hlnum{100}\hlstd{,}\hlkwc{seed}\hlstd{=}\hlnum{2016138277L}\hlstd{,}\hlkwc{info}\hlstd{=}\hlnum{TRUE}\hlstd{)}
           \hlstd{)} \hlopt{%:%}
    \hlkwd{foreach} \hlstd{(}\hlkwc{type}\hlstd{=}\hlkwd{c}\hlstd{(}\hlstr{"raw"}\hlstd{,}\hlstr{"cum"}\hlstd{),}\hlkwc{.combine}\hlstd{=rbind,}\hlkwc{.inorder}\hlstd{=}\hlnum{FALSE}\hlstd{)} \hlopt{%:%}
    \hlkwd{foreach} \hlstd{(}\hlkwc{country}\hlstd{=}\hlkwd{c}\hlstd{(}\hlstr{"SierraLeone"}\hlstd{,}\hlstr{"Liberia"}\hlstd{,}\hlstr{"Guinea"}\hlstd{,}\hlstr{"WestAfrica"}\hlstd{),}
             \hlkwc{.combine}\hlstd{=rbind,}\hlkwc{.inorder}\hlstd{=}\hlnum{FALSE}\hlstd{)} \hlopt{%dopar%}
    \hlstd{\{}
      \hlstd{tm} \hlkwb{<-} \hlstd{models[type,country][[}\hlnum{1}\hlstd{]]}
      \hlstd{tic} \hlkwb{<-} \hlkwd{Sys.time}\hlstd{()}
      \hlkwd{coef}\hlstd{(tm,}\hlkwd{names}\hlstd{(start))} \hlkwb{<-} \hlkwd{unname}\hlstd{(}\hlkwd{unlist}\hlstd{(start))}
      \hlkwd{coef}\hlstd{(tm,}\hlstr{"rho"}\hlstd{)} \hlkwb{<-} \hlnum{0.2}
      \hlstd{tm} \hlkwb{<-} \hlkwd{traj.match}\hlstd{(tm,}\hlkwc{est}\hlstd{=}\hlkwd{c}\hlstd{(}\hlstr{"R0"}\hlstd{,}\hlstr{"E_0"}\hlstd{,}\hlstr{"I_0"}\hlstd{),}\hlkwc{transform}\hlstd{=}\hlnum{TRUE}\hlstd{)}
      \hlkwa{if} \hlstd{(}\hlkwd{coef}\hlstd{(tm,}\hlstr{"E_0"}\hlstd{)}\hlopt{==}\hlnum{0}\hlstd{)} \hlkwd{coef}\hlstd{(tm,}\hlstr{"E_0"}\hlstd{)} \hlkwb{<-} \hlnum{1e-12}
      \hlkwa{if} \hlstd{(}\hlkwd{coef}\hlstd{(tm,}\hlstr{"I_0"}\hlstd{)}\hlopt{==}\hlnum{0}\hlstd{)} \hlkwd{coef}\hlstd{(tm,}\hlstr{"I_0"}\hlstd{)} \hlkwb{<-} \hlnum{1e-12}
      \hlstd{tm} \hlkwb{<-} \hlkwd{traj.match}\hlstd{(tm,}\hlkwc{method}\hlstd{=}\hlstr{'subplex'}\hlstd{,}\hlkwc{control}\hlstd{=}\hlkwd{list}\hlstd{(}\hlkwc{maxit}\hlstd{=}\hlnum{1e5}\hlstd{))}
      \hlstd{toc} \hlkwb{<-} \hlkwd{Sys.time}\hlstd{()}
      \hlstd{etime} \hlkwb{<-} \hlstd{toc}\hlopt{-}\hlstd{tic}
      \hlkwd{units}\hlstd{(etime)} \hlkwb{<-} \hlstr{"hours"}
      \hlkwd{data.frame}\hlstd{(}\hlkwc{country}\hlstd{=country,}\hlkwc{type}\hlstd{=type,}\hlkwd{as.list}\hlstd{(}\hlkwd{coef}\hlstd{(tm)),}
                 \hlkwc{loglik}\hlstd{=}\hlkwd{logLik}\hlstd{(tm),}\hlkwc{conv}\hlstd{=tm}\hlopt{$}\hlstd{convergence,}
                 \hlkwc{etime}\hlstd{=}\hlkwd{as.numeric}\hlstd{(etime))}
      \hlstd{\}} \hlopt{%>%} \hlkwd{mutate}\hlstd{(}\hlkwc{sum}\hlstd{=S_0}\hlopt{+}\hlstd{E_0}\hlopt{+}\hlstd{I_0}\hlopt{+}\hlstd{R_0,}
                   \hlkwc{S_0}\hlstd{=}\hlkwd{round}\hlstd{(N}\hlopt{*}\hlstd{S_0}\hlopt{/}\hlstd{sum),}
                   \hlkwc{E_0}\hlstd{=}\hlkwd{round}\hlstd{(N}\hlopt{*}\hlstd{E_0}\hlopt{/}\hlstd{sum),}
                   \hlkwc{I_0}\hlstd{=}\hlkwd{round}\hlstd{(N}\hlopt{*}\hlstd{I_0}\hlopt{/}\hlstd{sum),}
                   \hlkwc{R_0}\hlstd{=}\hlkwd{round}\hlstd{(N}\hlopt{*}\hlstd{R_0}\hlopt{/}\hlstd{sum))} \hlopt{%>%}
    \hlkwd{subset}\hlstd{(conv} \hlopt{%in%} \hlkwd{c}\hlstd{(}\hlnum{0}\hlstd{,}\hlnum{1}\hlstd{),}\hlkwc{select}\hlstd{=}\hlopt{-}\hlstd{sum)} \hlopt{%>%}
    \hlkwd{unique}\hlstd{()}

  \hlstd{\})} \hlkwb{->} \hlstd{profk}

\hlcom{## All trajectory matching computations}
\hlkwd{ldply}\hlstd{(}\hlkwd{list}\hlstd{(}\hlkwc{R0}\hlstd{=profR0,}\hlkwc{k}\hlstd{=profk),}\hlkwc{.id}\hlstd{=}\hlstr{'profile'}\hlstd{)} \hlkwb{->} \hlstd{profTM}

\hlcom{## Iterated filtering, R0 profile}

\hlkwd{bake}\hlstd{(}\hlkwc{file}\hlstd{=}\hlstr{"if-fits-R0_a.rds"}\hlstd{,\{}

  \hlstd{profTM} \hlopt{%>%} \hlkwd{subset}\hlstd{(profile}\hlopt{==}\hlstr{"R0"}\hlstd{)} \hlopt{%>%}
    \hlkwd{ddply}\hlstd{(}\hlopt{~}\hlstd{country}\hlopt{+}\hlstd{type,subset,}
          \hlkwd{is.finite}\hlstd{(loglik)}\hlopt{&}\hlstd{loglik}\hlopt{>}\hlkwd{max}\hlstd{(loglik)}\hlopt{-}\hlnum{20}\hlstd{)} \hlopt{%>%}
    \hlkwd{ddply}\hlstd{(}\hlopt{~}\hlstd{country}\hlopt{+}\hlstd{type}\hlopt{+}\hlstd{R0,subset,}
          \hlstd{loglik}\hlopt{==}\hlkwd{max}\hlstd{(loglik),}
          \hlkwc{select}\hlstd{=}\hlopt{-}\hlkwd{c}\hlstd{(loglik,etime,conv,profile))} \hlkwb{->} \hlstd{pars}

  \hlkwd{foreach} \hlstd{(}\hlkwc{start}\hlstd{=}\hlkwd{iter}\hlstd{(pars,}\hlkwc{by}\hlstd{=}\hlstr{'row'}\hlstd{),}
           \hlkwc{.combine}\hlstd{=rbind,}\hlkwc{.inorder}\hlstd{=}\hlnum{FALSE}\hlstd{,}
           \hlkwc{.options.mpi}\hlstd{=}\hlkwd{list}\hlstd{(}\hlkwc{chunkSize}\hlstd{=}\hlnum{1}\hlstd{,}\hlkwc{seed}\hlstd{=}\hlnum{1264624821L}\hlstd{),}
           \hlkwc{.noexport}\hlstd{=noexport)} \hlopt{%dopar%}
    \hlstd{\{}
      \hlstd{tic} \hlkwb{<-} \hlkwd{Sys.time}\hlstd{()}

      \hlstd{country} \hlkwb{<-} \hlkwd{as.character}\hlstd{(start}\hlopt{$}\hlstd{country)}
      \hlstd{type} \hlkwb{<-} \hlkwd{as.character}\hlstd{(start}\hlopt{$}\hlstd{type)}
      \hlstd{st} \hlkwb{<-} \hlkwd{unlist}\hlstd{(}\hlkwd{subset}\hlstd{(start,}\hlkwc{select}\hlstd{=}\hlopt{-}\hlkwd{c}\hlstd{(country,type)))}

      \hlstd{po} \hlkwb{<-} \hlstd{models[type,country][[}\hlnum{1}\hlstd{]]}
      \hlkwd{coef}\hlstd{(po,}\hlkwd{names}\hlstd{(st))} \hlkwb{<-} \hlstd{st}
      \hlkwa{if} \hlstd{(}\hlkwd{coef}\hlstd{(po,}\hlstr{"E_0"}\hlstd{)}\hlopt{==}\hlnum{0}\hlstd{)} \hlkwd{coef}\hlstd{(po,}\hlstr{"E_0"}\hlstd{)} \hlkwb{<-} \hlnum{1e-5}
      \hlkwa{if} \hlstd{(}\hlkwd{coef}\hlstd{(po,}\hlstr{"I_0"}\hlstd{)}\hlopt{==}\hlnum{0}\hlstd{)} \hlkwd{coef}\hlstd{(po,}\hlstr{"I_0"}\hlstd{)} \hlkwb{<-} \hlnum{1e-5}

      \hlstd{mf} \hlkwb{<-} \hlkwd{mif}\hlstd{(po,} \hlkwc{Nmif}\hlstd{=}\hlnum{10}\hlstd{,}
                \hlkwc{rw.sd} \hlstd{=} \hlkwd{c}\hlstd{(}\hlkwc{k}\hlstd{=}\hlnum{0.02}\hlstd{,}\hlkwc{E_0}\hlstd{=}\hlnum{1}\hlstd{,}\hlkwc{I_0}\hlstd{=}\hlnum{1}\hlstd{),}
                \hlkwc{ivps} \hlstd{=} \hlkwd{c}\hlstd{(}\hlstr{"E_0"}\hlstd{,}\hlstr{"I_0"}\hlstd{),}
                \hlkwc{Np} \hlstd{=} \hlnum{2000}\hlstd{,}
                \hlkwc{var.factor} \hlstd{=} \hlnum{2}\hlstd{,}
                \hlkwc{method} \hlstd{=} \hlstr{"mif2"}\hlstd{,}
                \hlkwc{cooling.type} \hlstd{=} \hlstr{"hyperbolic"}\hlstd{,}
                \hlkwc{cooling.fraction} \hlstd{=} \hlnum{0.5}\hlstd{,}
                \hlkwc{transform} \hlstd{=} \hlnum{TRUE}\hlstd{,}
                \hlkwc{verbose} \hlstd{=} \hlnum{FALSE}\hlstd{)}
      \hlstd{mf} \hlkwb{<-} \hlkwd{continue}\hlstd{(mf,} \hlkwc{Nmif} \hlstd{=} \hlnum{50}\hlstd{,} \hlkwc{cooling.fraction} \hlstd{=} \hlnum{0.1}\hlstd{)}

      \hlcom{## Runs 10 particle filters to assess Monte Carlo error in likelihood}
      \hlstd{pf} \hlkwb{<-} \hlkwd{replicate}\hlstd{(}\hlnum{10}\hlstd{,}\hlkwd{pfilter}\hlstd{(mf,}\hlkwc{Np}\hlstd{=}\hlnum{5000}\hlstd{,}\hlkwc{max.fail}\hlstd{=}\hlnum{Inf}\hlstd{))}
      \hlstd{ll} \hlkwb{<-} \hlkwd{sapply}\hlstd{(pf,logLik)}
      \hlstd{ll} \hlkwb{<-} \hlkwd{logmeanexp}\hlstd{(ll,} \hlkwc{se} \hlstd{=} \hlnum{TRUE}\hlstd{)}
      \hlstd{nfail} \hlkwb{<-} \hlkwd{sapply}\hlstd{(pf,getElement,}\hlstr{"nfail"}\hlstd{)}
      \hlstd{toc} \hlkwb{<-} \hlkwd{Sys.time}\hlstd{()}
      \hlstd{etime} \hlkwb{<-} \hlstd{toc}\hlopt{-}\hlstd{tic}
      \hlkwd{units}\hlstd{(etime)} \hlkwb{<-} \hlstr{"hours"}

      \hlkwd{data.frame}\hlstd{(}\hlkwc{country}\hlstd{=country,}\hlkwc{type}\hlstd{=type,}\hlkwd{as.list}\hlstd{(}\hlkwd{coef}\hlstd{(mf)),}
                 \hlkwc{loglik} \hlstd{= ll[}\hlnum{1}\hlstd{],}
                 \hlkwc{loglik.se} \hlstd{= ll[}\hlnum{2}\hlstd{],}
                 \hlkwc{nfail.min} \hlstd{=} \hlkwd{min}\hlstd{(nfail),}
                 \hlkwc{nfail.max} \hlstd{=} \hlkwd{max}\hlstd{(nfail),}
                 \hlkwc{etime} \hlstd{=} \hlkwd{as.numeric}\hlstd{(etime))}
      \hlstd{\}}
  \hlstd{\})} \hlkwb{->} \hlstd{profR0}

\hlcom{## Filter once more on maxima}

\hlkwd{bake}\hlstd{(}\hlkwc{file}\hlstd{=}\hlstr{"if-fits-R0.rds"}\hlstd{,\{}

  \hlstd{profR0} \hlopt{%>%} \hlkwd{subset}\hlstd{(}\hlkwd{is.finite}\hlstd{(loglik)}\hlopt{&}\hlstd{nfail.max}\hlopt{==}\hlnum{0}\hlstd{)} \hlopt{%>%}
    \hlkwd{ddply}\hlstd{(}\hlopt{~}\hlstd{country}\hlopt{+}\hlstd{type}\hlopt{+}\hlstd{R0,subset,}\hlkwd{rank}\hlstd{(}\hlopt{-}\hlstd{loglik)}\hlopt{<=}\hlnum{5}\hlstd{)} \hlopt{%>%}
    \hlkwd{subset}\hlstd{(}\hlkwc{select}\hlstd{=}\hlopt{-}\hlkwd{c}\hlstd{(loglik,loglik.se,nfail.max,nfail.min,etime))} \hlkwb{->} \hlstd{pars}

  \hlkwd{foreach} \hlstd{(}\hlkwc{start}\hlstd{=}\hlkwd{iter}\hlstd{(pars,}\hlkwc{by}\hlstd{=}\hlstr{'row'}\hlstd{),}
           \hlkwc{.combine}\hlstd{=rbind,}\hlkwc{.inorder}\hlstd{=}\hlnum{FALSE}\hlstd{,}
           \hlkwc{.options.mpi}\hlstd{=}\hlkwd{list}\hlstd{(}\hlkwc{chunkSize}\hlstd{=}\hlnum{1}\hlstd{,}\hlkwc{seed}\hlstd{=}\hlnum{1264624821L}\hlstd{),}
           \hlkwc{.noexport}\hlstd{=noexport)} \hlopt{%dopar%}
    \hlstd{\{}
      \hlstd{tic} \hlkwb{<-} \hlkwd{Sys.time}\hlstd{()}

      \hlstd{country} \hlkwb{<-} \hlkwd{as.character}\hlstd{(start}\hlopt{$}\hlstd{country)}
      \hlstd{type} \hlkwb{<-} \hlkwd{as.character}\hlstd{(start}\hlopt{$}\hlstd{type)}
      \hlstd{st} \hlkwb{<-} \hlkwd{unlist}\hlstd{(}\hlkwd{subset}\hlstd{(start,}\hlkwc{select}\hlstd{=}\hlopt{-}\hlkwd{c}\hlstd{(country,type)))}

      \hlstd{po} \hlkwb{<-} \hlstd{models[type,country][[}\hlnum{1}\hlstd{]]}
      \hlkwd{coef}\hlstd{(po,}\hlkwd{names}\hlstd{(st))} \hlkwb{<-} \hlkwd{unname}\hlstd{(st)}

      \hlcom{## Runs 10 particle filters to assess Monte Carlo error in likelihood}
      \hlstd{pf} \hlkwb{<-} \hlkwd{try}\hlstd{(}\hlkwd{replicate}\hlstd{(}\hlnum{10}\hlstd{,}\hlkwd{pfilter}\hlstd{(po,}\hlkwc{Np}\hlstd{=}\hlnum{5000}\hlstd{,}\hlkwc{max.fail}\hlstd{=}\hlnum{Inf}\hlstd{)))}

      \hlstd{toc} \hlkwb{<-} \hlkwd{Sys.time}\hlstd{()}
      \hlstd{etime} \hlkwb{<-} \hlstd{toc}\hlopt{-}\hlstd{tic}
      \hlkwd{units}\hlstd{(etime)} \hlkwb{<-} \hlstr{"hours"}

      \hlstd{ll} \hlkwb{<-} \hlkwd{sapply}\hlstd{(pf,logLik)}
      \hlstd{ll} \hlkwb{<-} \hlkwd{logmeanexp}\hlstd{(ll,} \hlkwc{se} \hlstd{=} \hlnum{TRUE}\hlstd{)}
      \hlstd{nfail} \hlkwb{<-} \hlkwd{sapply}\hlstd{(pf,getElement,}\hlstr{"nfail"}\hlstd{)}

      \hlkwd{data.frame}\hlstd{(}\hlkwc{country}\hlstd{=country,}\hlkwc{type}\hlstd{=type,}\hlkwd{as.list}\hlstd{(}\hlkwd{coef}\hlstd{(po)),}
                 \hlkwc{loglik} \hlstd{= ll[}\hlnum{1}\hlstd{],}
                 \hlkwc{loglik.se} \hlstd{= ll[}\hlnum{2}\hlstd{],}
                 \hlkwc{nfail.min} \hlstd{=} \hlkwd{min}\hlstd{(nfail),}
                 \hlkwc{nfail.max} \hlstd{=} \hlkwd{max}\hlstd{(nfail),}
                 \hlkwc{etime} \hlstd{=} \hlkwd{as.numeric}\hlstd{(etime))}
      \hlstd{\}}
  \hlstd{\})} \hlkwb{->} \hlstd{profR0}

\hlcom{## Iterated filtering, k profile}

\hlkwd{bake}\hlstd{(}\hlkwc{file}\hlstd{=}\hlstr{"if-fits-k_a.rds"}\hlstd{,\{}

  \hlstd{profTM} \hlopt{%>%} \hlkwd{subset}\hlstd{(profile}\hlopt{==}\hlstr{"k"}\hlstd{)} \hlopt{%>%}
    \hlkwd{ddply}\hlstd{(}\hlopt{~}\hlstd{country}\hlopt{+}\hlstd{type,subset,}
          \hlkwd{is.finite}\hlstd{(loglik)}\hlopt{&}\hlstd{loglik}\hlopt{>}\hlkwd{max}\hlstd{(loglik)}\hlopt{-}\hlnum{20}\hlstd{)} \hlopt{%>%}
    \hlkwd{ddply}\hlstd{(}\hlopt{~}\hlstd{country}\hlopt{+}\hlstd{type}\hlopt{+}\hlstd{k,subset,}
          \hlstd{loglik}\hlopt{==}\hlkwd{max}\hlstd{(loglik),}
          \hlkwc{select}\hlstd{=}\hlopt{-}\hlkwd{c}\hlstd{(loglik,etime,conv,profile))} \hlkwb{->} \hlstd{pars}

  \hlkwd{foreach} \hlstd{(}\hlkwc{start}\hlstd{=}\hlkwd{iter}\hlstd{(pars,}\hlkwc{by}\hlstd{=}\hlstr{'row'}\hlstd{),}
           \hlkwc{.combine}\hlstd{=rbind,}\hlkwc{.inorder}\hlstd{=}\hlnum{FALSE}\hlstd{,}
           \hlkwc{.options.mpi}\hlstd{=}\hlkwd{list}\hlstd{(}\hlkwc{chunkSize}\hlstd{=}\hlnum{1}\hlstd{,}\hlkwc{seed}\hlstd{=}\hlnum{1264624821L}\hlstd{),}
           \hlkwc{.noexport}\hlstd{=noexport)} \hlopt{%dopar%}
    \hlstd{\{}
      \hlstd{tic} \hlkwb{<-} \hlkwd{Sys.time}\hlstd{()}

      \hlstd{country} \hlkwb{<-} \hlkwd{as.character}\hlstd{(start}\hlopt{$}\hlstd{country)}
      \hlstd{type} \hlkwb{<-} \hlkwd{as.character}\hlstd{(start}\hlopt{$}\hlstd{type)}
      \hlstd{st} \hlkwb{<-} \hlkwd{unlist}\hlstd{(}\hlkwd{subset}\hlstd{(start,}\hlkwc{select}\hlstd{=}\hlopt{-}\hlkwd{c}\hlstd{(country,type)))}

      \hlstd{po} \hlkwb{<-} \hlstd{models[type,country][[}\hlnum{1}\hlstd{]]}
      \hlkwd{coef}\hlstd{(po,}\hlkwd{names}\hlstd{(st))} \hlkwb{<-} \hlstd{st}
      \hlkwa{if} \hlstd{(}\hlkwd{coef}\hlstd{(po,}\hlstr{"E_0"}\hlstd{)}\hlopt{==}\hlnum{0}\hlstd{)} \hlkwd{coef}\hlstd{(po,}\hlstr{"E_0"}\hlstd{)} \hlkwb{<-} \hlnum{1e-5}
      \hlkwa{if} \hlstd{(}\hlkwd{coef}\hlstd{(po,}\hlstr{"I_0"}\hlstd{)}\hlopt{==}\hlnum{0}\hlstd{)} \hlkwd{coef}\hlstd{(po,}\hlstr{"I_0"}\hlstd{)} \hlkwb{<-} \hlnum{1e-5}

      \hlstd{mf} \hlkwb{<-} \hlkwd{mif}\hlstd{(po,} \hlkwc{Nmif}\hlstd{=}\hlnum{10}\hlstd{,}
                \hlkwc{rw.sd} \hlstd{=} \hlkwd{c}\hlstd{(}\hlkwc{R0}\hlstd{=}\hlnum{0.02}\hlstd{,}\hlkwc{E_0}\hlstd{=}\hlnum{1}\hlstd{,}\hlkwc{I_0}\hlstd{=}\hlnum{1}\hlstd{),}
                \hlkwc{ivps} \hlstd{=} \hlkwd{c}\hlstd{(}\hlstr{"E_0"}\hlstd{,}\hlstr{"I_0"}\hlstd{),}
                \hlkwc{Np} \hlstd{=} \hlnum{2000}\hlstd{,}
                \hlkwc{var.factor} \hlstd{=} \hlnum{2}\hlstd{,}
                \hlkwc{method} \hlstd{=} \hlstr{"mif2"}\hlstd{,}
                \hlkwc{cooling.type} \hlstd{=} \hlstr{"hyperbolic"}\hlstd{,}
                \hlkwc{cooling.fraction} \hlstd{=} \hlnum{0.5}\hlstd{,}
                \hlkwc{transform} \hlstd{=} \hlnum{TRUE}\hlstd{,}
                \hlkwc{verbose} \hlstd{=} \hlnum{FALSE}\hlstd{)}
      \hlstd{mf} \hlkwb{<-} \hlkwd{continue}\hlstd{(mf,} \hlkwc{Nmif} \hlstd{=} \hlnum{50}\hlstd{,} \hlkwc{cooling.fraction} \hlstd{=} \hlnum{0.1}\hlstd{)}

      \hlcom{## Runs 10 particle filters to assess Monte Carlo error in likelihood}
      \hlstd{pf} \hlkwb{<-} \hlkwd{replicate}\hlstd{(}\hlnum{10}\hlstd{,}\hlkwd{pfilter}\hlstd{(mf,}\hlkwc{Np}\hlstd{=}\hlnum{5000}\hlstd{,}\hlkwc{max.fail}\hlstd{=}\hlnum{Inf}\hlstd{))}
      \hlstd{ll} \hlkwb{<-} \hlkwd{sapply}\hlstd{(pf,logLik)}
      \hlstd{ll} \hlkwb{<-} \hlkwd{logmeanexp}\hlstd{(ll,} \hlkwc{se} \hlstd{=} \hlnum{TRUE}\hlstd{)}
      \hlstd{nfail} \hlkwb{<-} \hlkwd{sapply}\hlstd{(pf,getElement,}\hlstr{"nfail"}\hlstd{)}
      \hlstd{toc} \hlkwb{<-} \hlkwd{Sys.time}\hlstd{()}
      \hlstd{etime} \hlkwb{<-} \hlstd{toc}\hlopt{-}\hlstd{tic}
      \hlkwd{units}\hlstd{(etime)} \hlkwb{<-} \hlstr{"hours"}

      \hlkwd{data.frame}\hlstd{(}\hlkwc{country}\hlstd{=country,}\hlkwc{type}\hlstd{=type,}\hlkwd{as.list}\hlstd{(}\hlkwd{coef}\hlstd{(mf)),}
                 \hlkwc{loglik} \hlstd{= ll[}\hlnum{1}\hlstd{],}
                 \hlkwc{loglik.se} \hlstd{= ll[}\hlnum{2}\hlstd{],}
                 \hlkwc{nfail.min} \hlstd{=} \hlkwd{min}\hlstd{(nfail),}
                 \hlkwc{nfail.max} \hlstd{=} \hlkwd{max}\hlstd{(nfail),}
                 \hlkwc{etime} \hlstd{=} \hlkwd{as.numeric}\hlstd{(etime))}
      \hlstd{\}}
  \hlstd{\})} \hlkwb{->} \hlstd{profk}

\hlcom{## Filter once more on maxima}

\hlkwd{bake}\hlstd{(}\hlkwc{file}\hlstd{=}\hlstr{"if-fits-k.rds"}\hlstd{,\{}

  \hlstd{profk} \hlopt{%>%} \hlkwd{subset}\hlstd{(}\hlkwd{is.finite}\hlstd{(loglik)}\hlopt{&}\hlstd{nfail.max}\hlopt{==}\hlnum{0}\hlstd{)} \hlopt{%>%}
    \hlkwd{ddply}\hlstd{(}\hlopt{~}\hlstd{country}\hlopt{+}\hlstd{type}\hlopt{+}\hlstd{R0,subset,}\hlkwd{rank}\hlstd{(}\hlopt{-}\hlstd{loglik)}\hlopt{<=}\hlnum{5}\hlstd{)} \hlopt{%>%}
    \hlkwd{subset}\hlstd{(}\hlkwc{select}\hlstd{=}\hlopt{-}\hlkwd{c}\hlstd{(loglik,loglik.se,nfail.max,nfail.min,etime))} \hlkwb{->} \hlstd{pars}

  \hlkwd{foreach} \hlstd{(}\hlkwc{start}\hlstd{=}\hlkwd{iter}\hlstd{(pars,}\hlkwc{by}\hlstd{=}\hlstr{'row'}\hlstd{),}
           \hlkwc{.combine}\hlstd{=rbind,}\hlkwc{.inorder}\hlstd{=}\hlnum{FALSE}\hlstd{,}
           \hlkwc{.options.mpi}\hlstd{=}\hlkwd{list}\hlstd{(}\hlkwc{chunkSize}\hlstd{=}\hlnum{1}\hlstd{,}\hlkwc{seed}\hlstd{=}\hlnum{1264624821L}\hlstd{),}
           \hlkwc{.noexport}\hlstd{=noexport)} \hlopt{%dopar%}
    \hlstd{\{}
      \hlstd{tic} \hlkwb{<-} \hlkwd{Sys.time}\hlstd{()}

      \hlstd{country} \hlkwb{<-} \hlkwd{as.character}\hlstd{(start}\hlopt{$}\hlstd{country)}
      \hlstd{type} \hlkwb{<-} \hlkwd{as.character}\hlstd{(start}\hlopt{$}\hlstd{type)}
      \hlstd{st} \hlkwb{<-} \hlkwd{unlist}\hlstd{(}\hlkwd{subset}\hlstd{(start,}\hlkwc{select}\hlstd{=}\hlopt{-}\hlkwd{c}\hlstd{(country,type)))}

      \hlstd{po} \hlkwb{<-} \hlstd{models[type,country][[}\hlnum{1}\hlstd{]]}
      \hlkwd{coef}\hlstd{(po,}\hlkwd{names}\hlstd{(st))} \hlkwb{<-} \hlkwd{unname}\hlstd{(st)}

      \hlcom{## Runs 10 particle filters to assess Monte Carlo error in likelihood}
      \hlstd{pf} \hlkwb{<-} \hlkwd{try}\hlstd{(}\hlkwd{replicate}\hlstd{(}\hlnum{10}\hlstd{,}\hlkwd{pfilter}\hlstd{(po,}\hlkwc{Np}\hlstd{=}\hlnum{5000}\hlstd{,}\hlkwc{max.fail}\hlstd{=}\hlnum{Inf}\hlstd{)))}

      \hlstd{toc} \hlkwb{<-} \hlkwd{Sys.time}\hlstd{()}
      \hlstd{etime} \hlkwb{<-} \hlstd{toc}\hlopt{-}\hlstd{tic}
      \hlkwd{units}\hlstd{(etime)} \hlkwb{<-} \hlstr{"hours"}

      \hlstd{ll} \hlkwb{<-} \hlkwd{sapply}\hlstd{(pf,logLik)}
      \hlstd{ll} \hlkwb{<-} \hlkwd{logmeanexp}\hlstd{(ll,} \hlkwc{se} \hlstd{=} \hlnum{TRUE}\hlstd{)}
      \hlstd{nfail} \hlkwb{<-} \hlkwd{sapply}\hlstd{(pf,getElement,}\hlstr{"nfail"}\hlstd{)}

      \hlkwd{data.frame}\hlstd{(}\hlkwc{country}\hlstd{=country,}\hlkwc{type}\hlstd{=type,}\hlkwd{as.list}\hlstd{(}\hlkwd{coef}\hlstd{(po)),}
                 \hlkwc{loglik} \hlstd{= ll[}\hlnum{1}\hlstd{],}
                 \hlkwc{loglik.se} \hlstd{= ll[}\hlnum{2}\hlstd{],}
                 \hlkwc{nfail.min} \hlstd{=} \hlkwd{min}\hlstd{(nfail),}
                 \hlkwc{nfail.max} \hlstd{=} \hlkwd{max}\hlstd{(nfail),}
                 \hlkwc{etime} \hlstd{=} \hlkwd{as.numeric}\hlstd{(etime))}
      \hlstd{\}}
  \hlstd{\})} \hlkwb{->} \hlstd{profk}

\hlkwd{ldply}\hlstd{(}\hlkwd{list}\hlstd{(}\hlkwc{R0}\hlstd{=profR0,}\hlkwc{k}\hlstd{=profk),}\hlkwc{.id}\hlstd{=}\hlstr{'profile'}\hlstd{)} \hlkwb{->} \hlstd{profIF}

\hlkwd{ldply}\hlstd{(}\hlkwd{list}\hlstd{(}\hlkwc{det}\hlstd{=profTM,}\hlkwc{stoch}\hlstd{=profIF),}\hlkwc{.id}\hlstd{=}\hlstr{'model'}\hlstd{)} \hlopt{%>%}
  \hlkwd{saveRDS}\hlstd{(}\hlkwc{file}\hlstd{=}\hlstr{'profiles.rds'}\hlstd{)}

\hlkwd{closeCluster}\hlstd{(cl)}
\hlkwd{mpi.quit}\hlstd{()}
\end{alltt}
\end{kframe}
\end{knitrout}

Executing this code will result in the creation of several files, of which \code{profiles.rds} is the most important, containing as it does the results of all the parameter estimation.


\subsection{Diagnostics}

The codes in file \code{diagnostics.R} compute the diagnostics displayed in the paper.
These computations are not very heavy, but can be accelerated using \code{doMC} if run on a multi-core workstation.
In a directory containing \code{profiles.rds} (see above) and \code{ebola.R}, execute these codes with a command like
\begin{verbatim}
Rscript --vanilla diagnostics.R
\end{verbatim}

Contents of file \code{diagnostics.R}:
\begin{knitrout}\scriptsize
\definecolor{shadecolor}{rgb}{0.969, 0.969, 0.969}\color{fgcolor}\begin{kframe}
\begin{alltt}
\hlkwd{library}\hlstd{(pomp)}
\hlkwd{library}\hlstd{(plyr)}
\hlkwd{library}\hlstd{(reshape2)}
\hlkwd{library}\hlstd{(magrittr)}
\hlkwd{options}\hlstd{(}\hlkwc{stringsAsFactors}\hlstd{=}\hlnum{FALSE}\hlstd{)}

\hlkwd{require}\hlstd{(foreach)}
\hlkwd{require}\hlstd{(doMC)}
\hlkwd{require}\hlstd{(iterators)}

\hlkwd{source}\hlstd{(}\hlstr{"ebola.R"}\hlstd{)}

\hlkwd{readRDS}\hlstd{(}\hlstr{"profiles.rds"}\hlstd{)} \hlopt{%>%}
  \hlkwd{ddply}\hlstd{(}\hlopt{~}\hlstd{country}\hlopt{+}\hlstd{type}\hlopt{+}\hlstd{model,subset,loglik}\hlopt{==}\hlkwd{max}\hlstd{(loglik))} \hlopt{%>%}
  \hlkwd{subset}\hlstd{(type}\hlopt{==}\hlstr{"raw"}\hlopt{&}\hlstd{model}\hlopt{==}\hlstr{"stoch"}\hlstd{)} \hlkwb{->} \hlstd{mles}

\hlstd{time1} \hlkwb{<-} \hlkwd{c}\hlstd{(}\hlkwc{Guinea}\hlstd{=}\hlstr{"2014-01-05"}\hlstd{,}\hlkwc{Liberia}\hlstd{=}\hlstr{"2014-06-01"}\hlstd{,}
           \hlkwc{SierraLeone}\hlstd{=}\hlstr{"2014-06-08"}\hlstd{,}\hlkwc{WestAfrica}\hlstd{=}\hlstr{"2014-01-05"}\hlstd{)}

\hlkwd{registerDoMC}\hlstd{(}\hlnum{4}\hlstd{)}

\hlkwd{foreach} \hlstd{(}\hlkwc{mle}\hlstd{=}\hlkwd{iter}\hlstd{(mles,}\hlkwc{by}\hlstd{=}\hlstr{'row'}\hlstd{),}\hlkwc{.combine}\hlstd{=rbind)} \hlopt{%dopar%}
  \hlstd{\{}
    \hlstd{country}\hlkwb{=}\hlkwd{as.character}\hlstd{(mle}\hlopt{$}\hlstd{country)}
    \hlstd{type}\hlkwb{=}\hlkwd{as.character}\hlstd{(mle}\hlopt{$}\hlstd{type)}
    \hlstd{M} \hlkwb{<-} \hlkwd{ebolaModel}\hlstd{(}\hlkwc{country}\hlstd{=country,}\hlkwc{type}\hlstd{=type,}
                    \hlkwc{na.rm}\hlstd{=}\hlnum{TRUE}\hlstd{,}\hlkwc{nstage}\hlstd{=}\hlnum{3}\hlstd{,}\hlkwc{timestep}\hlstd{=}\hlnum{0.01}\hlstd{)}
    \hlstd{p} \hlkwb{<-} \hlkwd{unlist}\hlstd{(}\hlkwd{subset}\hlstd{(mle,}\hlkwc{select}\hlstd{=}\hlopt{-}\hlkwd{c}\hlstd{(country,type,model,profile,}
                                     \hlstd{loglik,loglik.se,}
                                     \hlstd{nfail.min,nfail.max,conv,etime)))}
    \hlkwd{coef}\hlstd{(M,}\hlkwd{names}\hlstd{(p))} \hlkwb{<-} \hlkwd{unname}\hlstd{(}\hlkwd{unlist}\hlstd{(p))}

    \hlstd{t0} \hlkwb{<-} \hlkwd{as.Date}\hlstd{(time1[country])}

    \hlkwd{simulate}\hlstd{(M,}\hlkwc{nsim}\hlstd{=}\hlnum{10}\hlstd{,}\hlkwc{as.data.frame}\hlstd{=}\hlnum{TRUE}\hlstd{,}\hlkwc{obs}\hlstd{=}\hlnum{TRUE}\hlstd{,}\hlkwc{include.data}\hlstd{=}\hlnum{TRUE}\hlstd{,}\hlkwc{seed}\hlstd{=}\hlnum{2186L}\hlstd{)} \hlopt{%>%}
      \hlkwd{mutate}\hlstd{(}\hlkwc{date}\hlstd{=t0}\hlopt{+}\hlstd{time}\hlopt{*}\hlnum{7}\hlstd{,}\hlkwc{country}\hlstd{=country,}\hlkwc{type}\hlstd{=type)}
    \hlstd{\}} \hlopt{%>%} \hlkwd{saveRDS}\hlstd{(}\hlkwc{file}\hlstd{=}\hlstr{"diagnostics-sim.rds"}\hlstd{)}

\hlkwd{foreach} \hlstd{(}\hlkwc{mle}\hlstd{=}\hlkwd{iter}\hlstd{(mles,}\hlkwc{by}\hlstd{=}\hlstr{'row'}\hlstd{),}\hlkwc{.combine}\hlstd{=rbind)} \hlopt{%dopar%}
  \hlstd{\{}
    \hlstd{country}\hlkwb{=}\hlkwd{as.character}\hlstd{(mle}\hlopt{$}\hlstd{country)}
    \hlstd{type}\hlkwb{=}\hlkwd{as.character}\hlstd{(mle}\hlopt{$}\hlstd{type)}
    \hlstd{M} \hlkwb{<-} \hlkwd{ebolaModel}\hlstd{(}\hlkwc{country}\hlstd{=country,}\hlkwc{type}\hlstd{=type,}
                    \hlkwc{na.rm}\hlstd{=}\hlnum{TRUE}\hlstd{,}\hlkwc{nstage}\hlstd{=}\hlnum{3}\hlstd{,}\hlkwc{timestep}\hlstd{=}\hlnum{0.01}\hlstd{)}
    \hlstd{p} \hlkwb{<-} \hlkwd{unlist}\hlstd{(}\hlkwd{subset}\hlstd{(mle,}\hlkwc{select}\hlstd{=}\hlopt{-}\hlkwd{c}\hlstd{(country,type,model,profile,}
                                     \hlstd{loglik,loglik.se,}
                                     \hlstd{nfail.min,nfail.max,conv,etime)))}
    \hlkwd{coef}\hlstd{(M,}\hlkwd{names}\hlstd{(p))} \hlkwb{<-} \hlkwd{unname}\hlstd{(}\hlkwd{unlist}\hlstd{(p))}

    \hlkwd{probe}\hlstd{(M,}\hlkwc{probes}\hlstd{=}\hlkwd{list}\hlstd{(}\hlkwd{probe.acf}\hlstd{(}\hlkwc{var}\hlstd{=}\hlstr{"cases"}\hlstd{,}\hlkwc{lags}\hlstd{=}\hlnum{1}\hlstd{,}\hlkwc{type}\hlstd{=}\hlstr{"correlation"}\hlstd{)),}
          \hlkwc{nsim}\hlstd{=}\hlnum{500}\hlstd{,}\hlkwc{seed}\hlstd{=}\hlnum{1878812716L}\hlstd{)} \hlopt{%>%}
      \hlkwd{as.data.frame}\hlstd{()} \hlkwb{->} \hlstd{pb}
    \hlstd{pb} \hlopt{%>%} \hlkwd{mutate}\hlstd{(}\hlkwc{sim}\hlstd{=}\hlkwd{rownames}\hlstd{(pb),}
                  \hlkwc{data}\hlstd{=}\hlkwd{ifelse}\hlstd{(sim}\hlopt{==}\hlstr{"data"}\hlstd{,}\hlstr{"data"}\hlstd{,}\hlstr{"simulation"}\hlstd{),}
                  \hlkwc{type}\hlstd{=type,}
                  \hlkwc{country}\hlstd{=country)}
    \hlstd{\}} \hlopt{%>%} \hlkwd{saveRDS}\hlstd{(}\hlkwc{file}\hlstd{=}\hlstr{"diagnostics-probes.rds"}\hlstd{)}

\hlcom{## Additional diagnostics}

\hlcom{## Run probes for each country}
\hlcom{## Custom probe: exponential growth rate}
\hlstd{probe.trend} \hlkwb{<-} \hlkwa{function} \hlstd{(}\hlkwc{y}\hlstd{) \{}
  \hlstd{cases} \hlkwb{<-} \hlstd{y[}\hlstr{"cases"}\hlstd{,]}
  \hlstd{df} \hlkwb{<-} \hlkwd{data.frame}\hlstd{(}\hlkwc{week}\hlstd{=}\hlkwd{seq_along}\hlstd{(cases),}\hlkwc{cases}\hlstd{=cases)}
  \hlstd{fit} \hlkwb{<-} \hlkwd{lm}\hlstd{(}\hlkwd{log1p}\hlstd{(cases)}\hlopt{~}\hlstd{week,}\hlkwc{data}\hlstd{=df)}
  \hlkwd{unname}\hlstd{(}\hlkwd{coef}\hlstd{(fit)[}\hlnum{2}\hlstd{])}
  \hlstd{\}}

\hlkwd{foreach} \hlstd{(}\hlkwc{mle}\hlstd{=}\hlkwd{iter}\hlstd{(mles,}\hlkwc{by}\hlstd{=}\hlstr{'row'}\hlstd{),}\hlkwc{.combine}\hlstd{=rbind)} \hlopt{%dopar%}
  \hlstd{\{}
    \hlstd{country}\hlkwb{=}\hlkwd{as.character}\hlstd{(mle}\hlopt{$}\hlstd{country)}
    \hlstd{type}\hlkwb{=}\hlkwd{as.character}\hlstd{(mle}\hlopt{$}\hlstd{type)}
    \hlstd{M} \hlkwb{<-} \hlkwd{ebolaModel}\hlstd{(}\hlkwc{country}\hlstd{=country,}\hlkwc{type}\hlstd{=type,}
                    \hlkwc{na.rm}\hlstd{=}\hlnum{TRUE}\hlstd{,}\hlkwc{nstage}\hlstd{=}\hlnum{3}\hlstd{,}\hlkwc{timestep}\hlstd{=}\hlnum{0.01}\hlstd{)}
    \hlstd{p} \hlkwb{<-} \hlkwd{unlist}\hlstd{(}\hlkwd{subset}\hlstd{(mle,}\hlkwc{select}\hlstd{=}\hlopt{-}\hlkwd{c}\hlstd{(country,type,model,profile,}
                                     \hlstd{loglik,loglik.se,}
                                     \hlstd{nfail.min,nfail.max,conv,etime)))}
    \hlkwd{coef}\hlstd{(M,}\hlkwd{names}\hlstd{(p))} \hlkwb{<-} \hlkwd{unname}\hlstd{(}\hlkwd{unlist}\hlstd{(p))}

    \hlcom{## remove an exponential trend, give residuals on the log scale}
    \hlstd{dm} \hlkwb{<-} \hlkwd{model.matrix}\hlstd{(}\hlkwd{lm}\hlstd{(}\hlkwd{log1p}\hlstd{(cases)}\hlopt{~}\hlstd{time,}\hlkwc{data}\hlstd{=}\hlkwd{as.data.frame}\hlstd{(M)))}
    \hlstd{rm} \hlkwb{<-} \hlkwd{diag}\hlstd{(}\hlkwd{nrow}\hlstd{(dm))}\hlopt{-}\hlstd{dm}\hlopt{%*%}\hlkwd{solve}\hlstd{(}\hlkwd{crossprod}\hlstd{(dm))}\hlopt{%*%}\hlkwd{t}\hlstd{(dm)}
    \hlstd{detrend} \hlkwb{<-} \hlkwa{function} \hlstd{(}\hlkwc{x}\hlstd{)} \hlkwd{log1p}\hlstd{(x)}\hlopt{%*%}\hlstd{rm}

    \hlkwd{probe}\hlstd{(M,}\hlkwc{probes}\hlstd{=}\hlkwd{list}\hlstd{(}
      \hlkwd{probe.acf}\hlstd{(}\hlkwc{var}\hlstd{=}\hlstr{"cases"}\hlstd{,}\hlkwc{lags}\hlstd{=}\hlnum{1}\hlstd{,}\hlkwc{type}\hlstd{=}\hlstr{"correlation"}\hlstd{),}
      \hlkwc{sd}\hlstd{=}\hlkwd{probe.sd}\hlstd{(}\hlkwc{var}\hlstd{=}\hlstr{"cases"}\hlstd{,}\hlkwc{transform}\hlstd{=log1p),}
      \hlkwd{probe.quantile}\hlstd{(}\hlkwc{var}\hlstd{=}\hlstr{"cases"}\hlstd{,}\hlkwc{prob}\hlstd{=}\hlkwd{c}\hlstd{(}\hlnum{0.9}\hlstd{)),}
      \hlkwc{d}\hlstd{=}\hlkwd{probe.acf}\hlstd{(}\hlkwc{var}\hlstd{=}\hlstr{"cases"}\hlstd{,}\hlkwc{lags}\hlstd{=}\hlkwd{c}\hlstd{(}\hlnum{1}\hlstd{,}\hlnum{2}\hlstd{,}\hlnum{3}\hlstd{),}\hlkwc{type}\hlstd{=}\hlstr{"correlation"}\hlstd{,}
                  \hlkwc{transform}\hlstd{=detrend),}
      \hlkwc{trend}\hlstd{=probe.trend),}
      \hlkwc{nsim}\hlstd{=}\hlnum{2000}\hlstd{,}\hlkwc{seed}\hlstd{=}\hlnum{2186L}
      \hlstd{)} \hlopt{%>%} \hlkwd{as.data.frame}\hlstd{()} \hlkwb{->} \hlstd{pb}
    \hlstd{pb} \hlopt{%>%} \hlkwd{mutate}\hlstd{(}\hlkwc{sim}\hlstd{=}\hlkwd{rownames}\hlstd{(pb),}
                  \hlkwc{kind}\hlstd{=}\hlkwd{ifelse}\hlstd{(sim}\hlopt{==}\hlstr{"data"}\hlstd{,}\hlstr{"data"}\hlstd{,}\hlstr{"simulation"}\hlstd{),}
                  \hlkwc{type}\hlstd{=type,}
                  \hlkwc{country}\hlstd{=country)}
    \hlstd{\}} \hlopt{%>%}
  \hlkwd{melt}\hlstd{(}\hlkwc{id}\hlstd{=}\hlkwd{c}\hlstd{(}\hlstr{"country"}\hlstd{,}\hlstr{"type"}\hlstd{,}\hlstr{"kind"}\hlstd{,}\hlstr{"sim"}\hlstd{),}\hlkwc{variable.name}\hlstd{=}\hlstr{"probe"}\hlstd{)} \hlopt{%>%}
  \hlkwd{arrange}\hlstd{(country,type,probe,kind,sim)} \hlopt{%>%}
  \hlkwd{saveRDS}\hlstd{(}\hlkwc{file}\hlstd{=}\hlstr{"diagnostics-addl-probes.rds"}\hlstd{)}
\end{alltt}
\end{kframe}
\end{knitrout}

\subsection{Forecasting}

The codes in file \code{forecasts.R} perform all the forecasting computations.
In a directory containing \code{ebola.R}, \code{profiles.rds}, and \code{hosts}, a command like
\begin{verbatim}
mpirun -hostfile hosts -np 101 Rscript --vanilla forecasts.R
\end{verbatim}
will result in the execution of these computations.

Contents of the file \code{forecasts.R}:
\begin{knitrout}\scriptsize
\definecolor{shadecolor}{rgb}{0.969, 0.969, 0.969}\color{fgcolor}\begin{kframe}
\begin{alltt}
\hlkwd{require}\hlstd{(pomp)}
\hlkwd{require}\hlstd{(plyr)}
\hlkwd{require}\hlstd{(reshape2)}
\hlkwd{require}\hlstd{(magrittr)}
\hlkwd{options}\hlstd{(}\hlkwc{stringsAsFactors}\hlstd{=}\hlnum{FALSE}\hlstd{)}

\hlkwd{set.seed}\hlstd{(}\hlnum{988077383L}\hlstd{)}

\hlkwd{require}\hlstd{(foreach)}
\hlkwd{require}\hlstd{(doMPI)}
\hlkwd{require}\hlstd{(iterators)}

\hlkwd{source}\hlstd{(}\hlstr{"ebola.R"}\hlstd{)}

\hlstd{horizon} \hlkwb{<-} \hlnum{13}

\hlkwd{foreach} \hlstd{(}\hlkwc{country}\hlstd{=}\hlkwd{c}\hlstd{(}\hlstr{"SierraLeone"}\hlstd{),}\hlkwc{.inorder}\hlstd{=}\hlnum{TRUE}\hlstd{,}\hlkwc{.combine}\hlstd{=c)} \hlopt{%:%}
  \hlkwd{foreach} \hlstd{(}\hlkwc{type}\hlstd{=}\hlkwd{c}\hlstd{(}\hlstr{"raw"}\hlstd{,}\hlstr{"cum"}\hlstd{),}\hlkwc{.inorder}\hlstd{=}\hlnum{TRUE}\hlstd{,}\hlkwc{.combine}\hlstd{=c)} \hlopt{%do%}
  \hlstd{\{}
    \hlstd{M1} \hlkwb{<-} \hlkwd{ebolaModel}\hlstd{(}\hlkwc{country}\hlstd{=country,}\hlkwc{type}\hlstd{=type,}
                     \hlkwc{timestep}\hlstd{=}\hlnum{0.01}\hlstd{,}\hlkwc{nstageE}\hlstd{=}\hlnum{3}\hlstd{,}\hlkwc{na.rm}\hlstd{=}\hlnum{TRUE}\hlstd{)}
    \hlstd{M2} \hlkwb{<-} \hlkwd{ebolaModel}\hlstd{(}\hlkwc{country}\hlstd{=country,}\hlkwc{type}\hlstd{=}\hlstr{"raw"}\hlstd{,}
                     \hlkwc{timestep}\hlstd{=}\hlnum{0.01}\hlstd{,}\hlkwc{nstageE}\hlstd{=}\hlnum{3}\hlstd{,}\hlkwc{na.rm}\hlstd{=}\hlnum{TRUE}\hlstd{)}
    \hlkwd{time}\hlstd{(M2)} \hlkwb{<-} \hlkwd{seq}\hlstd{(}\hlkwc{from}\hlstd{=}\hlnum{1}\hlstd{,}\hlkwc{to}\hlstd{=}\hlkwd{max}\hlstd{(}\hlkwd{time}\hlstd{(M1))}\hlopt{+}\hlstd{horizon,}\hlkwc{by}\hlstd{=}\hlnum{1}\hlstd{)}
    \hlstd{M3} \hlkwb{<-} \hlkwd{ebolaModel}\hlstd{(}\hlkwc{country}\hlstd{=country,}\hlkwc{type}\hlstd{=}\hlstr{"raw"}\hlstd{,}
                     \hlkwc{timestep}\hlstd{=}\hlnum{0.01}\hlstd{,}\hlkwc{nstageE}\hlstd{=}\hlnum{3}\hlstd{,}\hlkwc{na.rm}\hlstd{=}\hlnum{TRUE}\hlstd{)}
    \hlkwd{time}\hlstd{(M3)} \hlkwb{<-} \hlkwd{seq}\hlstd{(}\hlkwc{from}\hlstd{=}\hlkwd{max}\hlstd{(}\hlkwd{time}\hlstd{(M1))}\hlopt{+}\hlnum{1}\hlstd{,}\hlkwc{to}\hlstd{=}\hlkwd{max}\hlstd{(}\hlkwd{time}\hlstd{(M1))}\hlopt{+}\hlstd{horizon,}\hlkwc{by}\hlstd{=}\hlnum{1}\hlstd{)}
    \hlkwd{timezero}\hlstd{(M3)} \hlkwb{<-} \hlkwd{max}\hlstd{(}\hlkwd{time}\hlstd{(M1))}
    \hlkwd{list}\hlstd{(M1,M2,M3)}
    \hlstd{\}} \hlkwb{->} \hlstd{models}
\hlkwd{dim}\hlstd{(models)} \hlkwb{<-} \hlkwd{c}\hlstd{(}\hlnum{3}\hlstd{,}\hlnum{2}\hlstd{,}\hlnum{1}\hlstd{)}
\hlkwd{dimnames}\hlstd{(models)} \hlkwb{<-} \hlkwd{list}\hlstd{(}\hlkwd{c}\hlstd{(}\hlstr{"fit"}\hlstd{,}\hlstr{"det.forecast"}\hlstd{,}\hlstr{"stoch.forecast"}\hlstd{),}
                         \hlkwd{c}\hlstd{(}\hlstr{"raw"}\hlstd{,}\hlstr{"cum"}\hlstd{),}\hlkwd{c}\hlstd{(}\hlstr{"SierraLeone"}\hlstd{))}

\hlstd{noexport} \hlkwb{<-} \hlkwd{c}\hlstd{(}\hlstr{"models"}\hlstd{)}

\hlcom{## Weighted quantile function}
\hlstd{wquant} \hlkwb{<-} \hlkwa{function} \hlstd{(}\hlkwc{x}\hlstd{,} \hlkwc{weights}\hlstd{,} \hlkwc{probs} \hlstd{=} \hlkwd{c}\hlstd{(}\hlnum{0.025}\hlstd{,}\hlnum{0.5}\hlstd{,}\hlnum{0.975}\hlstd{)) \{}
  \hlstd{idx} \hlkwb{<-} \hlkwd{order}\hlstd{(x)}
  \hlstd{x} \hlkwb{<-} \hlstd{x[idx]}
  \hlstd{weights} \hlkwb{<-} \hlstd{weights[idx]}
  \hlstd{w} \hlkwb{<-} \hlkwd{cumsum}\hlstd{(weights)}\hlopt{/}\hlkwd{sum}\hlstd{(weights)}
  \hlstd{rval} \hlkwb{<-} \hlkwd{approx}\hlstd{(w,x,probs,}\hlkwc{rule}\hlstd{=}\hlnum{1}\hlstd{)}
  \hlstd{rval}\hlopt{$}\hlstd{y}
  \hlstd{\}}

\hlstd{starts} \hlkwb{<-} \hlkwd{c}\hlstd{(}\hlkwc{Guinea}\hlstd{=}\hlstr{"2014-01-05"}\hlstd{,}\hlkwc{Liberia}\hlstd{=}\hlstr{"2014-06-01"}\hlstd{,}\hlkwc{SierraLeone}\hlstd{=}\hlstr{"2014-06-08"}\hlstd{)}

\hlstd{cl} \hlkwb{<-} \hlkwd{startMPIcluster}\hlstd{()}
\hlkwd{registerDoMPI}\hlstd{(cl)}

\hlstd{bake} \hlkwb{<-} \hlkwa{function} \hlstd{(}\hlkwc{file}\hlstd{,} \hlkwc{expr}\hlstd{) \{}
  \hlkwa{if} \hlstd{(}\hlkwd{file.exists}\hlstd{(file)) \{}
    \hlkwd{readRDS}\hlstd{(file)}
    \hlstd{\}} \hlkwa{else} \hlstd{\{}
      \hlstd{val} \hlkwb{<-} \hlkwd{eval}\hlstd{(expr)}
      \hlkwd{saveRDS}\hlstd{(val,}\hlkwc{file}\hlstd{=file)}
      \hlstd{val}
      \hlstd{\}}
  \hlstd{\}}

\hlkwd{readRDS}\hlstd{(}\hlstr{"profiles.rds"}\hlstd{)} \hlopt{%>%}
  \hlkwd{ddply}\hlstd{(}\hlopt{~}\hlstd{country}\hlopt{+}\hlstd{type}\hlopt{+}\hlstd{model,subset,loglik}\hlopt{>}\hlkwd{max}\hlstd{(loglik)}\hlopt{-}\hlnum{6}\hlstd{,}
        \hlkwc{select}\hlstd{=}\hlopt{-}\hlkwd{c}\hlstd{(conv,etime,loglik.se,nfail.min,nfail.max,profile))} \hlkwb{->} \hlstd{mles}

\hlstd{mles} \hlopt{%>%} \hlkwd{melt}\hlstd{(}\hlkwc{id}\hlstd{=}\hlkwd{c}\hlstd{(}\hlstr{"country"}\hlstd{,}\hlstr{"type"}\hlstd{,}\hlstr{"model"}\hlstd{),}\hlkwc{variable.name}\hlstd{=}\hlstr{'parameter'}\hlstd{)} \hlopt{%>%}
  \hlkwd{ddply}\hlstd{(}\hlopt{~}\hlstd{country}\hlopt{+}\hlstd{type}\hlopt{+}\hlstd{model}\hlopt{+}\hlstd{parameter,summarize,}
        \hlkwc{min}\hlstd{=}\hlkwd{min}\hlstd{(value),}\hlkwc{max}\hlstd{=}\hlkwd{max}\hlstd{(value))} \hlopt{%>%}
  \hlkwd{subset}\hlstd{(parameter}\hlopt{!=}\hlstr{"loglik"}\hlstd{)} \hlopt{%>%}
  \hlkwd{melt}\hlstd{(}\hlkwc{measure}\hlstd{=}\hlkwd{c}\hlstd{(}\hlstr{"min"}\hlstd{,}\hlstr{"max"}\hlstd{))} \hlopt{%>%}
  \hlkwd{acast}\hlstd{(country}\hlopt{~}\hlstd{type}\hlopt{~}\hlstd{model}\hlopt{~}\hlstd{parameter}\hlopt{~}\hlstd{variable)} \hlkwb{->} \hlstd{ranges}

\hlstd{mles} \hlopt{%>%} \hlkwd{ddply}\hlstd{(}\hlopt{~}\hlstd{country}\hlopt{+}\hlstd{type}\hlopt{+}\hlstd{model,subset,loglik}\hlopt{==}\hlkwd{max}\hlstd{(loglik),}\hlkwc{select}\hlstd{=}\hlopt{-}\hlstd{loglik)} \hlopt{%>%}
  \hlkwd{mutate}\hlstd{(}\hlkwc{k}\hlstd{=}\hlkwd{round}\hlstd{(k,}\hlnum{4}\hlstd{),}\hlkwc{rho}\hlstd{=}\hlkwd{round}\hlstd{(rho,}\hlnum{4}\hlstd{),}\hlkwc{R0}\hlstd{=}\hlkwd{round}\hlstd{(R0,}\hlnum{4}\hlstd{),}\hlkwc{E_0}\hlstd{=}\hlnum{3}\hlopt{*}\hlkwd{round}\hlstd{(E_0}\hlopt{/}\hlnum{3}\hlstd{))} \hlopt{%>%}
  \hlkwd{unique}\hlstd{()} \hlopt{%>%}
  \hlkwd{arrange}\hlstd{(country,type,model)} \hlkwb{->} \hlstd{mles}

\hlcom{### DETERMINISTIC MODELS}

\hlkwd{bake}\hlstd{(}\hlkwc{file}\hlstd{=}\hlstr{"forecasts_det.rds"}\hlstd{,\{}
  \hlkwd{foreach} \hlstd{(}\hlkwc{country}\hlstd{=}\hlkwd{c}\hlstd{(}\hlstr{"SierraLeone"}\hlstd{),}
           \hlkwc{.inorder}\hlstd{=}\hlnum{TRUE}\hlstd{,}\hlkwc{.combine}\hlstd{=rbind)} \hlopt{%:%}
    \hlkwd{foreach} \hlstd{(}\hlkwc{type}\hlstd{=}\hlkwd{c}\hlstd{(}\hlstr{"raw"}\hlstd{,}\hlstr{"cum"}\hlstd{),}\hlkwc{nsamp}\hlstd{=}\hlkwd{c}\hlstd{(}\hlnum{1000}\hlstd{,}\hlnum{3000}\hlstd{),}
             \hlkwc{.inorder}\hlstd{=}\hlnum{TRUE}\hlstd{,}\hlkwc{.combine}\hlstd{=rbind)} \hlopt{%do%}
    \hlstd{\{}

      \hlstd{params} \hlkwb{<-} \hlkwd{sobolDesign}\hlstd{(}\hlkwc{lower}\hlstd{=ranges[country,type,}\hlstr{'det'}\hlstd{,,}\hlstr{'min'}\hlstd{],}
                            \hlkwc{upper}\hlstd{=ranges[country,type,}\hlstr{'det'}\hlstd{,,}\hlstr{'max'}\hlstd{],}
                            \hlkwc{nseq}\hlstd{=nsamp)}

      \hlkwd{foreach}\hlstd{(}\hlkwc{p}\hlstd{=}\hlkwd{iter}\hlstd{(params,}\hlkwc{by}\hlstd{=}\hlstr{'row'}\hlstd{),}
              \hlkwc{.inorder}\hlstd{=}\hlnum{FALSE}\hlstd{,}
              \hlkwc{.combine}\hlstd{=rbind,}
              \hlkwc{.noexport}\hlstd{=noexport,}
              \hlkwc{.options.multicore}\hlstd{=}\hlkwd{list}\hlstd{(}\hlkwc{set.seed}\hlstd{=}\hlnum{TRUE}\hlstd{),}
              \hlkwc{.options.mpi}\hlstd{=}\hlkwd{list}\hlstd{(}\hlkwc{chunkSize}\hlstd{=}\hlnum{10}\hlstd{,}\hlkwc{seed}\hlstd{=}\hlnum{1568335316L}\hlstd{,}\hlkwc{info}\hlstd{=}\hlnum{TRUE}\hlstd{)}
              \hlstd{)} \hlopt{%dopar%}
        \hlstd{\{}
          \hlstd{M1} \hlkwb{<-} \hlstd{models[}\hlstr{"fit"}\hlstd{,type,country][[}\hlnum{1}\hlstd{]]}
          \hlstd{M2} \hlkwb{<-} \hlstd{models[}\hlstr{"det.forecast"}\hlstd{,type,country][[}\hlnum{1}\hlstd{]]}
          \hlstd{ll} \hlkwb{<-} \hlkwd{logLik}\hlstd{(}\hlkwd{traj.match}\hlstd{(M1,}\hlkwc{start}\hlstd{=}\hlkwd{unlist}\hlstd{(p)))}
          \hlstd{x} \hlkwb{<-} \hlkwd{trajectory}\hlstd{(M2,}\hlkwc{params}\hlstd{=}\hlkwd{unlist}\hlstd{(p))}
          \hlstd{p} \hlkwb{<-} \hlkwd{parmat}\hlstd{(}\hlkwd{unlist}\hlstd{(p),}\hlnum{20}\hlstd{)}
          \hlkwd{rmeasure}\hlstd{(M2,}\hlkwc{x}\hlstd{=x,}\hlkwc{times}\hlstd{=}\hlkwd{time}\hlstd{(M2),}\hlkwc{params}\hlstd{=p)} \hlopt{%>%}
            \hlkwd{melt}\hlstd{()} \hlopt{%>%}
            \hlkwd{mutate}\hlstd{(}\hlkwc{time}\hlstd{=}\hlkwd{time}\hlstd{(M2)[time],}
                   \hlkwc{period}\hlstd{=}\hlkwd{ifelse}\hlstd{(time}\hlopt{<=}\hlkwd{max}\hlstd{(}\hlkwd{time}\hlstd{(M1)),}\hlstr{"calibration"}\hlstd{,}\hlstr{"projection"}\hlstd{),}
                   \hlkwc{loglik}\hlstd{=ll)}
        \hlstd{\}} \hlopt{%>%}
        \hlkwd{subset}\hlstd{(variable}\hlopt{==}\hlstr{"cases"}\hlstd{,}\hlkwc{select}\hlstd{=}\hlopt{-}\hlstd{variable)} \hlopt{%>%}
        \hlkwd{mutate}\hlstd{(}\hlkwc{weight}\hlstd{=}\hlkwd{exp}\hlstd{(loglik}\hlopt{-}\hlkwd{mean}\hlstd{(loglik)))} \hlopt{%>%}
        \hlkwd{arrange}\hlstd{(time,rep)} \hlkwb{->} \hlstd{sims}

      \hlstd{ess} \hlkwb{<-} \hlkwd{with}\hlstd{(}\hlkwd{subset}\hlstd{(sims,time}\hlopt{==}\hlkwd{max}\hlstd{(time)),weight}\hlopt{/}\hlkwd{sum}\hlstd{(weight))}
      \hlstd{ess} \hlkwb{<-} \hlnum{1}\hlopt{/}\hlkwd{sum}\hlstd{(ess}\hlopt{^}\hlnum{2}\hlstd{)}
      \hlkwd{cat}\hlstd{(}\hlstr{"ESS det"}\hlstd{,country,type,}\hlstr{"="}\hlstd{,ess,}\hlstr{"\textbackslash{}n"}\hlstd{)}

      \hlstd{sims} \hlopt{%>%}
        \hlkwd{ddply}\hlstd{(}\hlopt{~}\hlstd{time}\hlopt{+}\hlstd{period,summarize,}\hlkwc{prob}\hlstd{=}\hlkwd{c}\hlstd{(}\hlnum{0.025}\hlstd{,}\hlnum{0.5}\hlstd{,}\hlnum{0.975}\hlstd{),}
              \hlkwc{quantile}\hlstd{=}\hlkwd{wquant}\hlstd{(value,}\hlkwc{weights}\hlstd{=weight,}\hlkwc{probs}\hlstd{=prob))} \hlopt{%>%}
        \hlkwd{mutate}\hlstd{(}\hlkwc{prob}\hlstd{=}\hlkwd{mapvalues}\hlstd{(prob,}\hlkwc{from}\hlstd{=}\hlkwd{c}\hlstd{(}\hlnum{0.025}\hlstd{,}\hlnum{0.5}\hlstd{,}\hlnum{0.975}\hlstd{),}
                              \hlkwc{to}\hlstd{=}\hlkwd{c}\hlstd{(}\hlstr{"lower"}\hlstd{,}\hlstr{"median"}\hlstd{,}\hlstr{"upper"}\hlstd{)))} \hlopt{%>%}
        \hlkwd{dcast}\hlstd{(period}\hlopt{+}\hlstd{time}\hlopt{~}\hlstd{prob,}\hlkwc{value.var}\hlstd{=}\hlstr{'quantile'}\hlstd{)} \hlopt{%>%}
        \hlkwd{mutate}\hlstd{(}\hlkwc{country}\hlstd{=country,}\hlkwc{type}\hlstd{=type)}
      \hlstd{\}}
  \hlstd{\})} \hlkwb{->} \hlstd{fc_tm}

\hlcom{### STOCHASTIC MODEL}

\hlkwd{bake}\hlstd{(}\hlkwc{file}\hlstd{=}\hlstr{"forecasts_stoch.rds"}\hlstd{,\{}
  \hlkwd{foreach} \hlstd{(}\hlkwc{country}\hlstd{=}\hlkwd{c}\hlstd{(}\hlstr{"SierraLeone"}\hlstd{),}
           \hlkwc{.inorder}\hlstd{=}\hlnum{TRUE}\hlstd{,}\hlkwc{.combine}\hlstd{=rbind)} \hlopt{%:%}
    \hlkwd{foreach} \hlstd{(}\hlkwc{type}\hlstd{=}\hlkwd{c}\hlstd{(}\hlstr{"raw"}\hlstd{,}\hlstr{"cum"}\hlstd{),}\hlkwc{nsamp}\hlstd{=}\hlkwd{c}\hlstd{(}\hlnum{200}\hlstd{,}\hlnum{200}\hlstd{),}
             \hlkwc{.inorder}\hlstd{=}\hlnum{TRUE}\hlstd{,}\hlkwc{.combine}\hlstd{=rbind)} \hlopt{%do%}
    \hlstd{\{}

      \hlstd{params} \hlkwb{<-} \hlkwd{sobolDesign}\hlstd{(}\hlkwc{lower}\hlstd{=ranges[country,type,}\hlstr{'stoch'}\hlstd{,,}\hlstr{'min'}\hlstd{],}
                            \hlkwc{upper}\hlstd{=ranges[country,type,}\hlstr{'stoch'}\hlstd{,,}\hlstr{'max'}\hlstd{],}
                            \hlkwc{nseq}\hlstd{=nsamp)}

      \hlkwd{foreach}\hlstd{(}\hlkwc{p}\hlstd{=}\hlkwd{iter}\hlstd{(params,}\hlkwc{by}\hlstd{=}\hlstr{'row'}\hlstd{),}
              \hlkwc{.inorder}\hlstd{=}\hlnum{FALSE}\hlstd{,}
              \hlkwc{.combine}\hlstd{=rbind,}
              \hlkwc{.noexport}\hlstd{=noexport,}
              \hlkwc{.options.multicore}\hlstd{=}\hlkwd{list}\hlstd{(}\hlkwc{set.seed}\hlstd{=}\hlnum{TRUE}\hlstd{),}
              \hlkwc{.options.mpi}\hlstd{=}\hlkwd{list}\hlstd{(}\hlkwc{chunkSize}\hlstd{=}\hlnum{1}\hlstd{,}\hlkwc{seed}\hlstd{=}\hlnum{1568335316L}\hlstd{,}\hlkwc{info}\hlstd{=}\hlnum{TRUE}\hlstd{)}
              \hlstd{)} \hlopt{%dopar%}
        \hlstd{\{}
          \hlstd{M1} \hlkwb{<-} \hlstd{models[}\hlstr{"fit"}\hlstd{,type,country][[}\hlnum{1}\hlstd{]]}
          \hlstd{M2} \hlkwb{<-} \hlstd{models[}\hlstr{"stoch.forecast"}\hlstd{,type,country][[}\hlnum{1}\hlstd{]]}
          \hlstd{pf} \hlkwb{<-} \hlkwd{pfilter}\hlstd{(M1,}\hlkwc{params}\hlstd{=}\hlkwd{unlist}\hlstd{(p),}\hlkwc{Np}\hlstd{=}\hlnum{2000}\hlstd{,}\hlkwc{save.states}\hlstd{=}\hlnum{TRUE}\hlstd{)}
          \hlstd{pf}\hlopt{$}\hlstd{saved.states} \hlopt{%>%} \hlkwd{tail}\hlstd{(}\hlnum{1}\hlstd{)} \hlopt{%>%} \hlkwd{melt}\hlstd{()} \hlopt{%>%}
            \hlkwd{acast}\hlstd{(variable}\hlopt{~}\hlstd{rep,}\hlkwc{value.var}\hlstd{=}\hlstr{'value'}\hlstd{)} \hlopt{%>%}
            \hlkwd{apply}\hlstd{(}\hlnum{2}\hlstd{,}\hlkwa{function} \hlstd{(}\hlkwc{x}\hlstd{) \{}
              \hlkwd{setNames}\hlstd{(}\hlkwd{c}\hlstd{(x[}\hlstr{"S"}\hlstd{],}\hlkwd{sum}\hlstd{(x[}\hlkwd{c}\hlstd{(}\hlstr{"E1"}\hlstd{,}\hlstr{"E2"}\hlstd{,}\hlstr{"E3"}\hlstd{)]),x[}\hlstr{"I"}\hlstd{],x[}\hlstr{"R"}\hlstd{]),}
                       \hlkwd{c}\hlstd{(}\hlstr{"S_0"}\hlstd{,}\hlstr{"E_0"}\hlstd{,}\hlstr{"I_0"}\hlstd{,}\hlstr{"R_0"}\hlstd{))\})} \hlkwb{->} \hlstd{x}
          \hlstd{pp} \hlkwb{<-} \hlkwd{parmat}\hlstd{(}\hlkwd{unlist}\hlstd{(p),}\hlkwd{ncol}\hlstd{(x))}
          \hlstd{pp[}\hlkwd{rownames}\hlstd{(x),]} \hlkwb{<-} \hlstd{x}
          \hlkwd{simulate}\hlstd{(M2,}\hlkwc{params}\hlstd{=pp,}\hlkwc{obs}\hlstd{=}\hlnum{TRUE}\hlstd{)} \hlopt{%>%}
            \hlkwd{melt}\hlstd{()} \hlopt{%>%}
            \hlkwd{mutate}\hlstd{(}\hlkwc{time}\hlstd{=}\hlkwd{time}\hlstd{(M2)[time],}
                   \hlkwc{period}\hlstd{=}\hlkwd{ifelse}\hlstd{(time}\hlopt{<=}\hlkwd{max}\hlstd{(}\hlkwd{time}\hlstd{(M1)),}\hlstr{"calibration"}\hlstd{,}\hlstr{"projection"}\hlstd{),}
                   \hlkwc{loglik}\hlstd{=}\hlkwd{logLik}\hlstd{(pf))}
        \hlstd{\}} \hlopt{%>%} \hlkwd{subset}\hlstd{(variable}\hlopt{==}\hlstr{"cases"}\hlstd{,}\hlkwc{select}\hlstd{=}\hlopt{-}\hlstd{variable)} \hlopt{%>%}
        \hlkwd{mutate}\hlstd{(}\hlkwc{weight}\hlstd{=}\hlkwd{exp}\hlstd{(loglik}\hlopt{-}\hlkwd{mean}\hlstd{(loglik)))} \hlopt{%>%}
        \hlkwd{arrange}\hlstd{(time,rep)} \hlkwb{->} \hlstd{sims}

      \hlstd{ess} \hlkwb{<-} \hlkwd{with}\hlstd{(}\hlkwd{subset}\hlstd{(sims,time}\hlopt{==}\hlkwd{max}\hlstd{(time)),weight}\hlopt{/}\hlkwd{sum}\hlstd{(weight))}
      \hlstd{ess} \hlkwb{<-} \hlnum{1}\hlopt{/}\hlkwd{sum}\hlstd{(ess}\hlopt{^}\hlnum{2}\hlstd{)}
      \hlkwd{cat}\hlstd{(}\hlstr{"ESS stoch"}\hlstd{,country,type,}\hlstr{"="}\hlstd{,ess,}\hlstr{"\textbackslash{}n"}\hlstd{)}

      \hlstd{sims} \hlopt{%>%} \hlkwd{ddply}\hlstd{(}\hlopt{~}\hlstd{time}\hlopt{+}\hlstd{period,summarize,}\hlkwc{prob}\hlstd{=}\hlkwd{c}\hlstd{(}\hlnum{0.025}\hlstd{,}\hlnum{0.5}\hlstd{,}\hlnum{0.975}\hlstd{),}
                     \hlkwc{quantile}\hlstd{=}\hlkwd{wquant}\hlstd{(value,}\hlkwc{weights}\hlstd{=weight,}\hlkwc{probs}\hlstd{=prob))} \hlopt{%>%}
        \hlkwd{mutate}\hlstd{(}\hlkwc{prob}\hlstd{=}\hlkwd{mapvalues}\hlstd{(prob,}\hlkwc{from}\hlstd{=}\hlkwd{c}\hlstd{(}\hlnum{0.025}\hlstd{,}\hlnum{0.5}\hlstd{,}\hlnum{0.975}\hlstd{),}
                              \hlkwc{to}\hlstd{=}\hlkwd{c}\hlstd{(}\hlstr{"lower"}\hlstd{,}\hlstr{"median"}\hlstd{,}\hlstr{"upper"}\hlstd{)))} \hlopt{%>%}
        \hlkwd{dcast}\hlstd{(period}\hlopt{+}\hlstd{time}\hlopt{~}\hlstd{prob,}\hlkwc{value.var}\hlstd{=}\hlstr{'quantile'}\hlstd{)} \hlopt{%>%}
        \hlkwd{mutate}\hlstd{(}\hlkwc{country}\hlstd{=country,}\hlkwc{type}\hlstd{=type)}
      \hlstd{\}}
  \hlstd{\})} \hlkwb{->} \hlstd{fc_if}

\hlkwd{ldply}\hlstd{(}\hlkwd{list}\hlstd{(}\hlkwc{stoch}\hlstd{=fc_if,}\hlkwc{det}\hlstd{=fc_tm),}\hlkwc{.id}\hlstd{=}\hlstr{'model'}\hlstd{)} \hlopt{%>%}
  \hlkwd{ddply}\hlstd{(}\hlopt{~}\hlstd{country,mutate,}
        \hlkwc{model}\hlstd{=}\hlkwd{factor}\hlstd{(model,}\hlkwc{levels}\hlstd{=}\hlkwd{c}\hlstd{(}\hlstr{"stoch"}\hlstd{,}\hlstr{"det"}\hlstd{)),}
        \hlkwc{date}\hlstd{=}\hlkwd{as.Date}\hlstd{(starts[}\hlkwd{unique}\hlstd{(}\hlkwd{as.character}\hlstd{(country))])}\hlopt{+}\hlnum{7}\hlopt{*}\hlstd{(time}\hlopt{-}\hlnum{1}\hlstd{))} \hlopt{%>%}
  \hlkwd{saveRDS}\hlstd{(}\hlkwc{file}\hlstd{=}\hlstr{'forecasts.rds'}\hlstd{)}

\hlkwd{closeCluster}\hlstd{(cl)}
\hlkwd{mpi.quit}\hlstd{()}
\end{alltt}
\end{kframe}
\end{knitrout}

\end{document}